\documentclass{article}

\usepackage{fourier}
\usepackage{amsmath}
\usepackage{amssymb}
\usepackage{soul}
\usepackage[dvipsnames,table]{xcolor}
\colorlet{excolor}{blue!45!green!45!black}
\colorlet{rdfcolor}{black} %{blue!90!black}
\colorlet{shadecolor}{black!10}
\colorlet{goodgreen}{green!70!black}
\colorlet{todocolor}{red}
\colorlet{brouilloncolor}{blue!50!black}
\colorlet{GBcolor}{Fuchsia!20!black}
\colorlet{GBxcolor}{Fuchsia!70!red!20!black}
\colorlet{MNcolor}{Lavender}
\colorlet{GCcolor}{Orange}
\definecolor{pencolor}{HTML}{ADDFAD}
\colorlet{pencolorSTR}{LimeGreen!50!black}
\definecolor{existencecolor}{HTML}{A1CAF1} %{8faef8} % #5789ff
\definecolor{narrativecolor}{HTML}{FAE7B5}
\definecolor{actioncolor}{HTML}{d9d9d9}
\definecolor{questioncolor}{HTML}{FFA07A}%{FF4F00}%{FF9933}%{F88379}
\definecolor{questionscolor}{HTML}{FFA07A}
\definecolor{datacolor}{HTML}{D2B48C}
\definecolor{pertainscolor}{rgb}{0.56, 0.27, 0.52}
\definecolor{answerscolor}{rgb}{1.0, 0.75, 0.0} % FFBF00
\definecolor{nuancescolor}{rgb}{0.9, 0.17, 0.31} %  	#E52B50 \definecolor{amaranth}{rgb}{0.9, 0.17, 0.31}
\definecolor{detailscolor}{rgb}{1.0, 0.75, 0.0} %  	#FFBF00 amber
\definecolor{instantiatescolor}{rgb}{1.0, 0.0, 0.5}
\definecolor{equatescolor}{rgb}{0.19, 0.55, 0.91} % bleudefrance
\colorlet{relatecolor}{black}
\usepackage{ragged2e}
\usepackage{multirow}
\usepackage{nameref}

\usepackage{amssymb}
\usepackage{pifont}

\usepackage{array}

\usepackage[includeheadfoot,margin=0.8in]{geometry}
\usepackage{marginnote}
  \reversemarginpar
\usepackage{lastpage,refcount}
\usepackage{fancyhdr}
\usepackage{tocloft}

\usepackage{graphicx}
\graphicspath{ {.} }

\newcommand{\mmmtype}[2][orange!60!black]{\textcolor{#1}{{\tt #2}}}
\def\pertains{\mmmtype[pertainscolor]{pertains}}
\def\relatesto{\mmmtype[relatecolor]{relatesTo}}
\def\obsolete{\mmmmark{obsolete}}

\newcommand{\mmmmark}[1]{\textcolor{purple!60!black}{{\tt #1}}}
\def\sharemark{\mmmmark{sharedWith}}
\newcommand{\sharedwith}[1]{\mmmmark{sharedWith:}\textcolor{purple!60!black}{#1}}
\def\syncmark{\mmmmark{syncWith}}

\def\subscribemark{\mmmmark{subscribedTo}}

\def\rewardmark{\mmmmark{rewarded}}

\newcommand{\mmmid}[1]{{\sc\scriptsize id=\ref*{#1}}}

\def\interfacelayer{interface layer}

\usepackage{caption}
\newcommand{\mmmtag}[1]{\textrm{$@$#1}}

\newcommand{\rdftag}[1]{\mmmtag{\textcolor{rdfcolor}{{#1}}}}

\newcommand{\todo}[1]{} %{{\color{todocolor}\small \ding{47} #1 }} % \reflectbox{\ding{47}}

\def\MMMapp{MMM application}

\newcommand{\GBxx}[1]{}

\def\equates{\mmmtype[equatescolor]{equates}}

\colorlet{furtherworkcolor}{black}
\newcommand{\furtherwork}[1]{#1}
\newcommand{\furtherworkmark}[0]{}
\usepackage{etoolbox}
\newcommand{\furtherworkcaption}[1]{#1}

\newcommand{\SMref}[1]{Suppl. Mat. \ref{#1}}

\newcommand{\landscape}{landscape} % MMM network

\newsavebox{\selvestebox}
\newenvironment{colbox}[1]
  {\newcommand\colboxcolor{#1}%
   \begin{lrbox}{\selvestebox}%
   \begin{minipage}{\dimexpr\columnwidth-2\fboxsep\relax}}
  {\end{minipage}\end{lrbox}%
   \begin{center}
   \colorbox[HTML]{\colboxcolor}{\usebox{\selvestebox}}
   \end{center}}

\usepackage{framed}

\newcounter{principle}

    \newenvironment{principle}[1]{
        \refstepcounter{principle}
        \begin{colbox}{FAE7B5} \vspace{7pt}
           \begin{center} \vspace{4pt}\centerline{Design Bias \#\theprinciple:  \textbf{#1}} \vspace{4pt}
            \begin{minipage}{0.95\textwidth}
            
        }%
        { 
        \end{minipage}\end{center}\vspace{2pt}
        \end{colbox}\smallskip
        }

    \newcounter{contribution}
    \newcommand{\contributionID}[1]{\refstepcounter{contribution}\thecontribution\label{#1}}

    \newcounter{bestpractice}
    \newenvironment{bestpractices}[1]{
        \refstepcounter{bestpractice}
        \begin{colbox}{d9d9d9} \vspace{7pt}
           \begin{center}\begin{minipage}{0.95\textwidth}

            \textit{Best practices for users:}%\#\thebestpractice.:  \textbf{#1}} %\vspace{9pt}
        }%
        { 
        \end{minipage}\end{center}\vspace{2pt}
        \end{colbox}\smallskip
        }

        \newcommand{\citedesignbias}[1]{Design Bias \ref{#1}}
    \usepackage{mdframed}
\usepackage{float}
\usepackage{enumitem}
\usepackage{hyperref}
\definecolor{links}{HTML}{288e47}%{2A1B81}
\hypersetup{colorlinks=true,urlcolor=links,urlbordercolor=links,linkcolor=goodgreen
}

\newcommand{\episodetitle}[1]{\title{\fontsize{14}{15}\selectfont\vspace{-1cm}\LARGE Proposal for an Organic Web\\\Large The missing link between the Web and the Semantic Web\\[3mm] {\sc #1}}}

\newcommand{\episodetitleS}[2]{\title{\fontsize{14}{15}\selectfont\vspace{-1cm}\LARGE Proposal for an Organic Web\\[3mm] \Large {\sc #1}\\[3mm] \Large #2}}

\usepackage{tikz}
\usetikzlibrary{shapes,snakes,shadows,arrows}
\usetikzlibrary{positioning}
\usetikzlibrary{calc,fit,matrix}
\usetikzlibrary{decorations.text,decorations.markings}

\tikzset{   
    existence/.style={fill=existencecolor, ellipse,align=center}, 
    question/.style={rectangle,fill=questioncolor,align=center},
    narrative/.style={fill=narrativecolor, rectangle,align=flush left}, 
    action/.style={rectangle,fill=actioncolor,align=center},
    datavalue/.style={fill=datacolor, rectangle,align=flush left}, 
    pen/.style={fill=pencolor, rectangle,align=flush left}, 
    edge/.style={font=\scriptsize,ultra thick,align=center,text=black},
    pertains/.style={edge, -latex, pertainscolor, text=black},
    relatesTo/.style={edge, -latex},
    answers/.style={edge, -latex, answerscolor,text=black},
    nuances/.style={edge, -latex, nuancescolor,text=black},
    details/.style={edge, -latex, detailscolor,text=black,font=\scriptsize},
    questions/.style={edge, -latex, questionscolor,text=black},
    relate/.style={edge, -, black},
    instantiates/.style={edge, -latex, instantiatescolor,text=black,font=\scriptsize},
    differsfrom/.style={edge, latex-latex, ForestGreen,text=black},
    pennedin/.style={edge, -latex, pencolor,text=black},
    equates/.style={edge, latex-latex, equatescolor,text=black},
    edgelabel/.style={font=\scriptsize,sloped},
    metadata/.style={draw=none,fill=none,anchor=south,inner sep = 0pt,font=\scriptsize},      
}

\episodetitle{Part I}
\author{M. Noual}
\date{2023}

\begin{document}
\maketitle
% \nocite{*}    

% ============================================================== %
% ========                    ABSTRACT                  ======== %
% ============================================================== %
\begin{abstract}
    \noindent
    A huge amount of information is produced in digital form. The Semantic Web  stems from the realisation that dealing efficiently with this production requires getting better at interlinking digital informational resources together \cite{berners2006linked}.
    % \href{https://www.w3.org/standards/semanticweb/data}{Linked Data} 
    % the web was a half-baked understanding of this, SW show maturity
    Its focus is on linking \textit{data}. Linking data isn't enough. Not all information produced is intended to be processed \textit{as data} per se. Most of the digital content produced today is unstructured (informal) text whose progressive semantics are only intended to be intelligible to humans. 
    The documents containing the information can themselves be interlinked  as if they were data. But links between granular  documents then only convey a shallow pre-defined semantics that ignores the rich progressive semantics expressed inside the documents. Dealing with traditional documents as if they were data, is bound to make suboptimal use of their contents, and arguably remains of limited utility.
    We need to provide infrastructural support for linking all sorts of informational resources including resources whose understanding and fine interlinking requires domain-specific human expertise.
    At times when many problems scale to planetary dimensions, it is essential to scale  coordination of information processing and information production,  
    without giving up on expertise and depth of analysis, nor forcing languages and formalisms  onto thinkers, decision-makers and innovators  that are only  suitable to some forms of intelligence. 
    I make  a proposal in this direction and in line with the idea of interlinking championed by the Semantic Web. 
\end{abstract}
\bigskip

This proposal is the result of a compilation of ideas contributed by the scientific and entrepreneur communities  over several years of discussions.
\bigskip

% ============================================================== %
% ========                 KEYWORDS                     ======== %
% ============================================================== %
\bigskip

% ============================================================== %
% ========               INPUT FILES                    ======== %
% ============================================================== %
\section{Introduction}\label{I-intro}

\subsection{Motivation}\label{motivation}

% ============================================================== %
% ========              BEGIN MOTIVATIONS               ======== %
% ============================================================== %
\def\motivations{% Need this for a subsequent figure
Most of the digital content produced today is "unstructured", meaning its semantics is not understood without the involvement of a human mind. Natural language processing techniques only extract a tiny proportion of the semantics of all unstructured content produced by humans (and now machines). Huge quantities of unstructured digital content are a problem. More unstructured content means more work for humans. The risks are: (i) that content be made suboptimal use of, by both humans and machines, and (ii) that  content stand in the way of humans communicating and working efficiently with each other. A primary motivation of my proposal is to address this problem and deal with unstructured digital content \textit{early} in its  life cycle. My aim is  to help preempt the production of digital content with low capacity to positively impact on and empower 
}
\motivations{}% end motivations extract 
human {entreprises} and with high capacity to stand in their way. This requires means to gauge the value of new pieces of information against the body of pre-existing information which Vannevar Bush famously referred to  as "\textit{the record}" \cite{VB}, as I do here too.
\medskip

% ============================================================== %
% =======               THE RECORD                       ======= %
% ============================================================== %
It isn't enough to evaluate pieces of information individually, we also need to take a step back and watch over humanity's global informational commons. 
The actual digitalisation of knowledge and of knowledge work is an opportunity  to materialise a well delineated version of the record. 
But amassing quantities of digital content is not enough to make the record manageable. Pieces of information in the record must be related with one another \textit{meaningfully} (cf \SMref{socks}). 
It isn't enough to have all theorems about, say, Boolean Automata Networks digitally recorded. To ensure that we make good use of all those theorems, and that we don't add repetitions of  knowledge that they already capture, we must also know, and document, how those theorems relate to each other: \textit{which ones are generalisations of which other ones, which ones are used in the proofs of others, which ones are equivalent to each other, which ones contradict each other\ldots{}} 
I contend that the record should honour the highest level of expertise of the humans who contribute knowledge to it.  More than just document the existence and nature of epistemic relations between theorems and other pieces of information, we should also {endeavour to \textit{highlight} the} finest known {details} of those relations -- 
e.g. \textit{how does theorem $t_1$ generalise theorem $t_2$, that is, what features of Boolean Automata Networks  does the generalisation relation operate on: the topology of the Boolean Automata Networks? their dynamics?  In what way is formalism $f_1$ equivalent to formalism $f_2$: mathematically? semantically? philosophically? What question do two results provide contradicting answers to? \ldots{}} \medskip

My proposal relies on design biases.  I document them in yellow boxes in the present article. The supplementary material~\ref{philo} accompanying this article, presents further fundamental philosophical biases underlying my proposal.

\begin{principle}{{Smart Networking}} 
    \label{expert-network}\label{smart}
    I wish to support an informational network whose \textit{structure}  reflects as best as possible the  depth of human expertise across the diversity of informational domains that get documented in the network.
\end{principle}

Design Bias \ref{smart} requires that expertise no longer be documented only \textit{inside} documents at the endpoints of the informational network's links. The links themselves should reflect expert insight. This disagrees with the documentation of links by/for outsiders.
\medskip

My proposal devises a solution to manage the record both locally and globally. It aims (1) to favour the systematic stripping down  of new digital content to the bare, useful minimum, defined in terms of what information the digital record already contains, and (2) to promote  care and detail in linking new digital content to  humanity's body of digital content. 
     
\subsection{Information}\label{information}

As suggested above, the focus of this proposal  is on \textit{unstructured} information meant to be consumed by humans rather than machines. \furtherwork{The proposal is nonetheless extended to deal marginally with the case of structured data. This extension will be presented in a subsequent article.}   
Until then, our focus is more specifically on textual information, for instance: pieces of science, of philosophical arguments, of geopolitical discussions, polemics, recipes, tutorials, stories\ldots{}
The solution proposed in this paper may also to some extent accommodate poems and videos,
% of cats 
but I leave this less argumentative kind of content aside for the moment to concentrate on textual information that invites discussion, nuance, questioning, updating, reformulating\ldots{} 
A future extension of the solution  to at least pictures (e.g. photos of whiteboards) will be important in my opinion. The solution should ultimately  accommodate the diversity of successful informational work practices and not impose cumbersome documentation efforts on thinkers, especially when less interfering documentation alternatives can be put into place (see Design Bias~\ref{experts}).
It remains that the extension of our solution to non-textual media can be taken care of later once I have mastered the solution for textual information. %  cf cold

% ============================================================== %
% ========               EXPERTS KNOW BEST              ======== %
% ============================================================== %

\begin{principle}{{Experts at work know best / Support what already  works }}\label{experts} 
    I wish to support humans in their practice of informational work. Many information workers, i.e., thinkers (e.g. scientist researchers) are already equipped to doing worthwhile informational work and are doing it well. 
    Different thinkers operate within different epistemic cultures, and their chains of thought follow different series of landmarks.
    It is essential for "informational welfare" (progress and quality of information) that existing informational know-how, cultures and habits be respected. 
    I wish to dedicate a solution to supporting what works well already\footnotemark. And in particular I wish to propose a solution that preserves the focus of experts on what they are successfully contributing.
\end{principle}

\footnotetext{I believe it is far more important and safe to  support what already functions well than to attempt relieving pain points of informational workers. Indeed, solutions brought  to informational workers  are bound to  emphasise certain of  their activities over others. Pain points of informational workers that are \textit{at the core} of informational work and that outsiders don't understand (e.g. difficulty in proving a conjecture) are best dealt with by the informational workers themselves. All the other pain points  relate to activities that are \textit{marginal} to the  informational work (e.g. self-marketing activities). Arguably, mitigating those pain points risks facilitating and thereby emphasising those marginal activities at the expense of the core informational work. Dynamics and balance of experts' activities should be protected from generic technologies built by outsiders. }

\citedesignbias{experts} means the solution aimed at here, is \textit{not} primarily a  solution in support of communication.
The Supplementary Material \ref{philo} accompanying the present article emphasises a distinction between supporting communication and supporting information (cf \SMref{SMcommunication}). 
\medskip

The present proposal emerged from a context of transdisciplinary scientific research involving formal sciences and life sciences.   Its motivating intention isn't so much to support the production of formal documents that help communicate ideas. It rather is to support the stage of informational work at which  ideas are shaped with words, sometimes  upstream from the work of formal documentation.
The emphasis   is \textit{less} on supporting use and reuse of informational resources (contrary to the \href{https://www.w3.org/TR/annotation-model/}{Web Annotation Data Model} \cite{WADM}), than it is on supporting \textit{the continual improvement, renewal and obsolescence of informational resources}. I expect enhanced use and reuse to naturally follow from \hyperref[smart]{smart networking} (as well as  from architectural choices presented in  \SMref{implementation}).

% ============================================================== %
% ========               DYNAMIC INFORMATION            ======== %
% ============================================================== %
\begin{principle}{{Continual Improvement and Updating of Information}} 
    \label{continual-improvement}
    I consider information as a process, something to do, until it no longer is relevant. A piece of information is neither definite nor free-standing.  Unlike data is not a given and unlike knowledge it is not established.  It is \textit{at its best} when it is subjected to continual improvement: when it is getting nuanced, detailed, challenged, (re)contextualised, updated\ldots{} to follow the world it describes as it changes. Eventually a piece of information becomes obsolete. 
    I wish to support information \textit{at its best} 
    and facilitate the obsolescence of unprocessable information. 
\end{principle}

Notably, \citedesignbias{continual-improvement} disagrees with  ridding  the record of low quality information. Low quality information is typically information that can be improved. See  \SMref{quality}.
\medskip

Following \citedesignbias{continual-improvement} and its emphasis on the \textit{dynamism} of information, the solution I propose to build is also not optimized for automated reasoning and inference, which require \textit{settling} on some knowledge base. \medskip

The Supplementary Material \ref{philo} further details assumed characteristics of the notion of information, providing epistemic foundation to this proposed solution.    
\subsection{Epistemic Glue}
\label{glue}

A base assumption of the proposed solution is that different pieces of information may be produced by different humans. The solution is to support \textit{collective} documentation.
Appreciating the way  in which individual pieces of information relate to each other is paramount to this project of organising the record and managing redundancy in it.   
Consider two individual pieces of information, $\mathcal{I}_1$ and $\mathcal{I}_2$. A legitimate question is: How are $\mathcal{I}_1$ and $\mathcal{I}_2$ related? Possible answers are:

\begin{tabular}{p{7cm}p{7cm}}
    \textit{\begin{itemize}
        \item They are not related at all.
        \item They contradict each other.
        \item They are about the same topic $\mathcal{T}$.
        \item They use the same term $\mathcal{T}$.
        \item They refer to the same object or concept $x$.
        \item They denote the same object or concept $x$. 
        \item They imply the same consequences.
        \item They answer the same question.
        \item They appear in the same book.
    \end{itemize}}
    &
    \textit{\begin{itemize}
        \item $\mathcal{I}_1$ answers $\mathcal{I}_2$.\footnote{I consider questions to be pieces of information: cf \SMref{questions}}
        \item $\mathcal{I}_1$ is an example of $\mathcal{I}_2$.
        \item $\mathcal{I}_1$ is a subclass of $\mathcal{I}_2$.
        \item $\mathcal{I}_1$ implies $\mathcal{I}_2$.
        \item $\mathcal{I}_1$ generalises $\mathcal{I}_2$.
        \item $\mathcal{I}_1$ sums up $\mathcal{I}_2$.
        \item $\mathcal{I}_1$ builds on $\mathcal{I}_2$.
        \item $\mathcal{I}_1$ is the title of $\mathcal{I}_2$.
        \item[] ~~~~$\vdots$
    \end{itemize}}
\end{tabular}

There is a great diversity of ways in which two independent pieces of information might relate to each other. Obviously not all relations are possible between any two pieces of information. And  some relations are harder to document than others -- e.g. it is harder to say that $\mathcal{I}_1$ and $\mathcal{I}_2$ refer to the same concept than to say that they use the same term. 
Note also that $\mathcal{I}_1$ and $\mathcal{I}_2$ may be related in more than one way, and that a relation between $\mathcal{I}_1$ and $\mathcal{I}_2$ is itself a piece of information. 
\medskip

I use the term "glue" to refer to information that relates pieces of information together. Because the record is a collective document, glue is necessary for it to have structure. Without glue, the record would  merely be a collection of independent resources, possibly organised into some sub-collections and categories defined by an arbitrary (central) entity. An important desirable property of glue is that it be \textit{generic} so the structure it gives to the record be \textit{domain-agnostic}. For instance, saying that $\mathcal{I}_1$ answers the question $\mathcal{I}_2$ is a generic, domain-agnostic way of gluing $\mathcal{I}_1$ and $\mathcal{I}_2$ together. Saying that $\mathcal{I}_1 \implies \mathcal{I}_2$ ($\mathcal{I}_1$ mathematically implies $\mathcal{I}_2$ as in $\neg \mathcal{I}_1 \vee \mathcal{I}_2$ holds) is not. \medskip

Not all relations provide the same "epistemic depth" of glue. For instance, understanding that $\mathcal{I}_1$ implies $\mathcal{I}_2$ is epistemically deeper than understanding that  $\mathcal{I}_1$ and $\mathcal{I}_2$ appear in the same book. 
Generally, let  $\mathcal{R}$ and $\mathcal{R}'$ be two relations (like the ones listed above) both between pieces of information $\mathcal{I}_1$ and $\mathcal{I}_2$. Informally, we say that  $\mathcal{R}$ is \textit{epistemically deeper} than  $\mathcal{R}'$ if understanding $\mathcal{R}$ \textit{leads to}   more understanding of $\mathcal{I}_1$ and $\mathcal{I}_2$  than does understanding $\mathcal{R}'$, or if  $\mathcal{R}$ \textit{comes from} more understanding  of $\mathcal{I}_1$ and $\mathcal{I}_2$. 
\medskip

The glue we're interested here must  be \textit{smart} (cf \citedesignbias{smart}). It must be generic without being shallow. We want to materialise a \hyperref[smart]{\textit{smartly}} networked version of the digital record. So we are interested in emphasising relations that are  epistemically deep. However, the digital record is a \textit{collective} work. A diversity of epistemic cultures, approaches, formalisms and languages need to be accommodated. Our solution must support the provision of glue in a diversity of formalisms, languages \textit{etc}. 
For experts to contribute glue, glue should not be tedious to contribute. 
Following \citedesignbias{experts},  a legal scientist who understands that  $\mathcal{I}_1$ legally implies $\mathcal{I}_2$ should \textit{not} have to understand the mathematical notion of implication (nor any other notion of implication for that matter) in order to document the legal relation between $\mathcal{I}_1$ and $\mathcal{I}_2$. 
Similarly, when documenting a mathematical relation between two theorems $T_1$ and $T_2$, a mathematician who knows how theorem $T_1$ generalises theorem $T_2$ should \textit{not}  be required to take the approach of a formal ontology modeller\footnote{The ontology modeller formally distinguishes between relations like \texttt{is-a}, \texttt{belongs-to}, \texttt{is-a-part-of} \cite{brachman1983and,burgun2001aspects,keet2008representing},  and understands the consequences of the Closed/Open World Assumptions   and the Unique Name Assumption \cite{hustadt1994we}. The mathematician like all other experts considered in this proposal who are in no need of an ontology, should not have to worry about any of that. None of these three assumptions are made in the context of this proposal. And no knowledge of formal ontological relations is required to participate in smart networking.}. 
An expertise in $\mathcal{R}$, $\mathcal{I}_1$ and $\mathcal{I}_2$ (all three are pieces of information) should be enough to document $\mathcal{R}$ between $\mathcal{I}_1$ and $\mathcal{I}_2$. No understanding of the repercussions of $\mathcal{R}$ beyond $\mathcal{I}_1$ and $\mathcal{I}_2$ should be required. In short, from the point of view of experts at work, documenting content as part of our smart networking solution should not be significantly different from documenting content today with current digital technologies that don't support smart networking (e.g. text editors). 
\medskip

There are many ways of "gluing" informational resources together (cf bullet points above). A popular way involves ontological and taxonomic commitments  (cf \SMref{ontological-commitment}). For instance, two independent  informational resources A and B (e.g. two statements, or two articles) may be related based on the fact that they both assume that a certain object  '\texttt{Bob}' exists and they both agree on what \texttt{Bob} really refers to, what kind of object it is, e.g. a '\texttt{person}' which is a type of '\texttt{mammal}' with a '\texttt{phone number}', equal to \texttt{01-23-45-67-89}. If resources A and B agree on the existence and meaning of '\texttt{Bob}', there is deep epistemic glue between them in the form of shared rigorous semantics. Checking  that the '\texttt{Bob}' of resource A is exactly the same as the '\texttt{Bob}' of resource B can be very demanding. Often, correspondences across ontologies and taxonomies, between concepts used to define an object (e.g. \texttt{person}, \texttt{mammal}, \texttt{phone number}) are not trivial to establish \cite{DIALLO21}. 
I propose to avoid this difficulty altogether  and  \textit{not} restrict  the glue we consider here in the way formal ontologies and taxonomies restrict it to semantic relations. 
A founding hypothesis of my proposal is that good information feeds on a diversity of epistemic glue and that the glue itself must be challengeable like any other piece of information is. Thus, my proposal  neither {assumes}, {imposes} nor even aims at semantic homogeneity. I propose   to allow for 
%assumed and established  
ontological and taxonomic commitments to be documented explicitly, questioned and discussed like any other piece of information.
Even without a  guarantee that resources A and B rigorously agree on the existence and meaning of '\texttt{Bob}', A and B may still be glued together by  the \textit{explicit (motivated) assumption} that  they do agree, or simply by a question asking whether they agree\footnotemark{}. 
\footnotetext{This suggests an alternative to having to build trust in the content of the record. Rather than try to make the information trustworthy, we encourage anyone who has a doubt to express that doubt so that it can be addressed. Of course, for this, efficient redundancy management is key.}
An obvious consequence  is that  the digital version of the record that I propose to structure with glue, isn't expected to have global semantic consistency. It can't serve as a functional representation of  the world\footnotemark{}. 
The aim here is rather to support "safe passage" between  different documented perspectives on the world. 
\medskip

\footnotetext{This is desirable. A model doesn't scale well. A collectively built, all-encompassing model would have limited use for the huge majority of humans who continually need to change perspectives, deal with uncertainty, ambiguity \textit{etc}.}

When gluing a new piece of information $\mathcal{I}$ to existing pieces of information $\mathcal{I}_1,\mathcal{I}_2,\ldots \mathcal{I}_n$ already in the record,  $\mathcal{I}$ should gain in precision because of  the epistemically relevant context provided by the glue to $\mathcal{I}_1,\mathcal{I}_2,\ldots \mathcal{I}_n$. Conversely, $\mathcal{I}_1,\mathcal{I}_2,\ldots \mathcal{I}_n$ should also gain from being completed, nuanced, circumscribed \textit{etc} by the glue to $\mathcal{I}$. Otherwise, the relations between  $\mathcal{I}$ and $\mathcal{I}_1,\mathcal{I}_2,\ldots \mathcal{I}_n$ should highlight the inappropriateness of $\mathcal{I}$. Glue generally gives information on information\footnote{\textit{Not} meta-information -- cf \SMref{metainformation}}.
More glue should   make the record more manageable because it should increase the ease with which we can epistemically relate any two pieces of information in it and thus jointly deal with them. 
It should help take into account pieces of information that come from heterogeneous sources.  Individual pieces of information that make sense individually shouldn't make less sense when glued to other pieces of the record.  They shouldn't degrade the record either. 
On the contrary, more information should be beneficial to the record, or it should be easy to see that it isn't.

\def\yes{ \ding{52}}
\def\nope{\ding{56}}
\def\WWWcomparedWithSW{
    \begin{table}[h!]
\begin{tabular}{|p{10cm}|>{\centering\arraybackslash}p{2.8cm}|>{\centering\arraybackslash}p{2.8cm}|}
    \cline{2-3}
    \multicolumn{1}{c|}{}& \textbf{The Web} & \textbf{The Semantic Web}\\\hline
    Is designed primarily for \ldots{} & humans & machines  \\\hline
    Allows anyone to contribute without  technical skills & \yes & \nope\\\hline
    Doesn't limit expressivity, accepts any textual information expressed in  any formalism or  language, at any level of formalisation, including healthily challenging and nuancing information  & \yes %\par but infomation tends to be encapsulated in documents 
     & \nope  \\\hline
    Supports   \hyperref[glue]{epistemic glue} in between individual resources & \nope\par  hyperlinks only & \yes\par \textit{semantic} glue  \\\hline
    Is designed to network any granularity of informational resources & \nope\par mostly documents & \yes  \\\hline
    Natively allows to annotate any granularity of information & \nope & \yes{\par {cf RDFS  comments}}
    \\\hline
    Natively allows to interlink annotations like any other informational resource & \nope  & \nope \\\hline
\end{tabular}
\caption{ 
    There is room for an intermediary web whose content, like the Web's,  is primarily for humans to consume, and whose structure, like the Semantic Web's, is "\hyperref[smart]{smart}" (cf \citedesignbias{smart}) 
    and "\hyperref[glue]{epistemically deep"} (cf Section~\ref{glue}).  
    }
\end{table}
}

\subsection{There is a missing link between the Web and the Semantic Web.}\label{missing}

\WWWcomparedWithSW

Traditional human-centric information networks -- e.g. the Web, the network of inter-cited academic publications -- are not \hyperref[smart]{smart} epistemically structured networks. They are \textit{infra}structural networks taking care of the logistics of humans' informational activities. 
Information in these networks is mostly   \textit{inside} the documents at the endpoints of the network links. The links between documents are epistemically shallow.
\medskip

The Web is missing a smart epistemic backbone. The Semantic Web proposes to materialise a semantic one, in the aim of opening the wealth of Web content up to automatic processing. 
It defines standards like RDF (the Resource Description Framework) \cite{decker2000semantic,manola2004rdf}  that support  strong {semantic} gluing of informational resources. The gluing operates at a finer  granularity than the Web's cross-document  hyperlinking.  The shift in  granularity 
is consequential:  it unlocks possibilities of  collective documentation and enhanced resource sharing (data  can be reused in multiple RDF triples by multiple authors)\footnotemark{}.
\medskip

\footnotetext{RDF  is a standard data model  that expresses information  by way of  "semantic triples". A semantic triple consists in a subject, a predicate and an object which altogether  express a statement.  Each of the three components of a triple are individually addressable. Atomic pieces of information  can thus be reused in multiple triples by multiple authors.  "\texttt{Bob}" can be the subject and object of multiple statements. "\texttt{enjoys}" can be the predicate of multiple statements, linking together multiple resources acting as subjects and objects.
}

The Semantic Web only concerns a minor proportion of the information that is of interest to humans. Its standards are designed to make information machine-understandable in order to support automatic reasoning over it. 
Most information that is of interest to humans is expressed by humans, for humans%
\footnote{
    The common denominator for humans is natural language, not any specific formal language. 
    Even when humans (e.g. theoretical computer scientists at work) endeavour to produce highly formalised information, they spend most of their work time navigating \textit{between} levels of formalisation, rather than thinking committedly in one particular formalism. As most humans don't speak in rigorous RDF triples, populating the Semantic Web with  human produced information would require intermediary entities savvy of SW standards. The translation of a piece of information into SW standards is rarely worth  the cost and effort. Arguably, there also is a danger in handing the translation of expert information over to non-specialists. Profound domain-specific knowledge is best documented first-hand by the domain experts. }. 
It doesn't need to be made machine-understandable and is therefore usually not worth  being formalised into SW  standards.
When a researcher proves a new theorem $T$ there rarely is a  pressing  incentive, if any, to make a machine understand $T$. There is usually one for the researcher's human peers to understand $T$. 
The sort of information that the Semantic Web is primarily designed to deal with is  \textit{metadata} (cf \SMref{metainformation}).
Metadata is  essential for transferring rudimentary semantics over to the machines 
\cite{greenberg2003metadata}. 
But unlike the expression and proof of a theorem $T$, metadata describing $T$ (e.g. the authorship, the creation date of $T$) is not reflective of the depth and nuance of  understanding that domain experts  have of $T$.  So we can't make a \hyperref[smart]{smart} network out of metadata. 
Also, the SW standards are designed to constructively collect information as part of coherent ontological models.
But our interest here is supporting informational progress (cf Design Bias \ref{continual-improvement}), which can happen in a variety of ways,
including  through the  deconstruction  of established models. 
Human produced information  expresses doubts, questions, hypotheses \textit{etc} and sometimes challenges the best models actually in service. A web looser than the Semantic Web is needed to interlink and structure this information.
\medskip 

I propose to materialise \textbf{an intermediary   web} -- namely, the  \textbf{"the MMM"} -- geared towards supporting human reasoning and its organised documentation\footnote{The Semantic Web's  vision is to support automated reasoning by providing machines with  collectively built knowledge in the right formalisation. Arguably, a reasonable intermediary step to organising information for machines is to do it for humans (cf \SMref{machines}).}. 
{MMM} stands for \textbf{Mutual Mutable Medium}, meaning \textit{collective dynamic document} or \textit{record}\footnote{"Mutual" in MMM replaces "Worldwide" in WWW. The idea is to go from a paradigm where information is meant to be distributed to everyone, to a paradigm where information emerges from small scale invested relationships. }. 
\medskip

The intermediary MMM web is to  \textbf{co-exist with the original Web}, possibly even interface with it.
Theoretically, anything expressible  in text on the Web can be expressed on the MMM with no obligation for translation or reformulation. In practice, software interfaces may exist (for instance Web browser extensions comparable to \href{https://web.hypothes.is/}{hypothes.is} \cite{perkel2015annotating,hypothesis}, see also Fig.\,\ref{browser-plugin} in \SMref{interface-layer}) to copy or to reference information from the Web onto the MMM where information benefits from  epistemic interlinking.  \medskip

The intermediary MMM web is also to  \textbf{relay the Semantic Web}. Without serving the same purpose as the Semantic Web nor conveying the same kind of content, the MMM  is to offer some compatibility with the standards of the Semantic Web. 
A mapping  between the MMM and the RDF data models  will be provided in  a follow-up article.
Once mapped into the MMM, data looses some of its ability to be efficiently processed by machines. %
Automatically populating the MMM with Semantic Web resources may nonetheless have some advantages. It may favour the reuse of  resources because it brings the resources to humans, exposing them to human scrutiny and annotations, and making humans systematically aware of the existing  terminologies and taxonomies that are epistemically relevant to their work. It may help mitigate the amount of redundant and diverging new terminology.
Conversely, interfacing the Semantic Web with the MMM can enable more systematic generation and update of formal ontologies. 
The MMM data model can %act as "proto-taxonomical" structure.
be leveraged in the capacity of "proto-taxonomical" structure for smart systematic documenting and informing of  ontology design decisions \cite{ontotip,Rector04}.

\subsection{Requirements}\label{MMMreqs}

To materialise a version of the digital record that has the desirable properties discussed  in Section \ref{I-intro}, I propose to organise information as per a pre-defined data model. The data model  is called the \textbf{MMM data model} or \textbf{MMM format}. It is formally introduced in the next section. Here let us first  list some basic requirements for the MMM format.
\medskip

{N.B.}: This proposal is \textbf{\textit{not} a solution for organising already archived information}. It aims instead at supporting humans as they work on updating and renewing information. In other terms, the  solution is to resemble an upgrade on the concept of whiteboard more than it is to resemble an upgrade on the concept of library. 

\begin{enumerate}[leftmargin=*,label=R\arabic*]

    \item  \textbf{Inclusivity and  informalism-friendliness}: 
    \label{req-inclusivity}\label{req-informal}  
    It must be possible and easy for a human user to contribute anything expressible as text in natural language. 
    We shouldn't have to bother users  with unwelcome formalisation exercises. It must be possible to contribute content to the smart network, without knowledge of any formal language. 
  
    \item \textbf{Epistemic glue}:
    \label{req-glue}
    It must be easy for users to document links between their contributions and pre-existing ones. It must be possible for them to  make explicit the epistemic relationship between them. 

    \item \textbf{Recursive annotation}:
    \label{req-recursive}
    It must be possible to question, nuance, challenge, detail any contribution. 
    Generally, it must be possible to comment on/annotate any contribution, including annotations themselves, as well as links between contributions and links between annotations. 

    \item  \textbf{Minimal metadata}:
    \label{req-metadata}
    The amount of metadata 
    associated to each contribution must be kept minimal, and  for strictly administrative purposes    (cf \SMref{metadata}). Metadata should not be needed to assess the quality of contributions.

    \item \textbf{Reformulation}: 
    \label{req-reformulate}
    It must be possible to contribute to the record by adding a reformulation of an existing contribution (this proposal does \ul{\textit{not}} aim at finding a unique canonical expression of each piece of information).

    \item \textbf{Intelligible collective documentation}:
    \label{req-collective}
    It must be possible for independent users to contribute to the record without consulting each other, even when they document closely related information. The record should not decrease in quality.  It must be possible for the  set of theirs contributions     to constitute an intelligible collective document whose  smart structure allows navigating meaningfully from contribution to contribution. 
    
    \item \textbf{Contribution types}:
    \label{req-types}  \label{req-questions} 
    It must be easy to distinguish between different types of contributions. 
    In particular it must be easy to distinguish contributions that are questions from contributions that are not (cf \SMref{questions}). Generally:
    \begin{enumerate}
        \item %\textbf{Easy choice}:
        \label{easychoice} \label{easy}\label{req-easy} 
        The semantics of contribution types must be intuitive. It must be easy for contributors to know which type to assign  to their contributions. 

        \item %\textbf{Limited number of types}:
        The number of different contribution types must be kept small.  To assign a type to their contributions, users must not need to learn a long list of types. 

        \item  The set of different contribution types must be stable. Types must be pre-defined\footnote{This means that we need to have a good set of types from the start to limit the need for future evolutions. I claim to have a basis for this starting set (cf Section \ref{MMMformat}). Methodical and diverse testing of this set needs to be performed beyond what has already been accomplished before writing this proposal.  }. Users must not need to keep up to speed with evolving definitions. 
       
        \item  %\textbf{Generic epistemic purpose}:
        \label{req-epistemic} 
        The semantics of contribution types must be  generic (domain-independent). A contribution's type must convey the basic epistemic role or purpose of the contribution   (e.g. is the contribution a question, a statement or something else?)  
        The genericness of types must maintain bridges across domains of expertise.   

        \item %\textbf{Loose interpretation}:
        \label{req-interpretation} 
        The semantics of contribution types must be loose. There must be room for interpretation and specification of the meaning of a type. 
        It must be possible for users to make slightly different uses of the same  %have slightly different interpretations of what epistemic roles are conveyed by a given  
        type\footnote{Just like how biologists and mathematicians slightly diverge on  what kind of information goes in an article introduction. But they can still read each others' articles without being thrown off by  what they find in the introduction.}.
        \label{universal-types}\label{req-universal}
        Users from different domains  must still  use the same contribution type in \textit{relatable} ways\footnote{Just like how  a mathematician and a biologist use the concepts  "introduction", "abstract", and "title" in  relatable ways (cf \SMref{balance}).} (cf requirement \ref{req-epistemic}).     
        \label{smartspecification} 
        It must thus be possible for contributors to \textit{narrow down} the generic epistemic purpose conveyed by a type. For instance having assigned the type "question" to a contribution, it must be possible  for the contributor to specify that the contribution is a \textit{rhetorical} question or that it is a \textit{confirmation seeking} question.\smallskip

        \item \label{req-lazy}\label{easygoing}%\textbf{Tolerant Typing:} 
        % \textbf{Easy-Going.} 
        It must be easy for contributors to assign a \textit{default} type to their contributions when they don't want to bother choosing a more epistemically meaningful type. It must be possible for contributors to use our solution without leveraging its structuring opportunities.

        \item %\textbf{Ontological Commitments}:
        \label{req-onto} 
        It must be  easy to highlight the ontological commitments underlying a contribution. For instance the question "\textit{What are genes made of?}" makes the tacit ontological assumption that genes exist (cf \SMref{namingthings}). It must be easy to make a contribution whose purpose is  to highlight this about this question.
       \medskip
    
        \begin{center}
        \begin{tikzpicture}[]
            \node[question] at (0, 0)   (q1) {What are genes made of?};
            \node[left=7cm, existence] (n2) {Genes};
            \node[below right=0.6cm and 1cm of n2, question] (q2) {Are genes even real?};
            \draw[pertains] (n2) -- (q1)  node[midway,yshift=0pt] {\color{pertainscolor}Characterising\\ ontological commitement} ;
            \draw[questions] (q2) -- (n2)  node[edgelabel,midway,yshift=4pt] {\color{questionscolor}challenges} ;
        \end{tikzpicture}
        \end{center}\medskip
      
    \end{enumerate}
           
\end{enumerate}

Because of requirement \ref{req-informal} and also because of \citedesignbias{experts} ("Experts at work know best"),  the MMM data model must be very flexible by design.  This means that there may often be several  ways to document the same piece of information in MMM format. As a consequence, 
the definition of the MMM format needs to be relayed by a collection of \textit{best practices for users}. Some are mentioned below. Best practices may be promoted through the design of MMM editors and other tools   for contributing content in MMM format  (cf  \SMref{interface-layer}).\medskip

The primary version of the MMM data model introduced in Section \ref{MMMformat} is expected to need some minor tweaking.  
It is essential that it remain small and simple.  
The MMM format must strike a balance between  freedom and constraint in the exercise of manual documentation (cf \SMref{balance}). 
Possible necessary future adaptations of the MMM format should be weary of maintaining that balance. 
The definition of the MMM data model is  exclusively motivated by \textit{practical} reasons.
There is no epistemological theory behind its definition.
Any modification to the MMM data model must be done circumspectly to address practical needs only (rather than to be exhaustive and demonstrate a certain form coherence).

\section{Definition of the MMM format}   \label{MMMformat} \label{format}

In the sequel, I use the symbol $\mathbb{S}$ to denote the set of strings of characters {possibly including \href{https://www.markdownguide.org/basic-syntax/}{markdown} symbols}, and I use the symbol $\mathbb{D}$ to denote the set of dates. 

% ============================================================== %
% ========                LANDSCAPE                     ======== %
% ============================================================== %
\subsubsection{Landscape}\label{landscape}

A {MMM network}, a.k.a. "\textbf{\landscape{}}" consists of objects called "\textbf{landmarks}". Exactly one of those objects is a special landmark called the "\textbf{pit}", denoted by $\bot$. All other landmarks in a \landscape{}  ${\bf N}$  are \textbf{contributions} $ c\in {\bf C}$ belonging to the set $ {\bf C}\subset {\bf N}={\bf C} \cup \{\bot\}$ of contributions. Contributions have attributes that I  list in next paragraphs \S\ref{IDs} -- \S\ref{marks}.
    
\begin{center}\begin{tikzpicture}
        \node[existence]    (landscape) {landscape};
        \node[right=2cm, existence] (landmark) {landmark};
        % landmarks
        \node[right=1cm of landmark, existence] (contribution) {contribution};
        \node[above=0.3cm  of contribution, existence] (pit) {The pit $\bot$};
        \draw[pertains] (landmark) -- (landscape)  node[pos=0.3,yshift=3pt,sloped,pertainscolor] {$\in$};
        \draw[instantiates] (pit) -- (landmark)  node[pos=0.3,yshift=4pt,sloped,instantiatescolor] {is a};
        \draw[instantiates] (contribution) -- (landmark)  node[pos=0.3,yshift=4pt,sloped,instantiatescolor] {is a};
    \end{tikzpicture}
\end{center}

\subsection{Main Contribution Attributes}\label{main-attributes}

Contributions convey information through their three main attributes, namely, their label, their type, and their tags. 

\begin{center}\begin{tikzpicture}
    \node[existence]    (contribution) {contribution};
    \node[left=2cm of contribution, existence] (type) {type};
    \node[above=0.1cm of type, existence] (label) {label};
    \node[below=0.1cm  of type, existence] (tags) {tags};
    \node[right=2cm  of contribution, existence,inner sep=0pt] (other) {Other (metadata)\\ attributes};
    \draw[pertains] (label) -- (contribution)  node[pos=0.4,yshift=4pt,sloped,pertainscolor] {is an attribute of};
    \draw[pertains] (type) -- (contribution) ;
    \draw[pertains] (tags) -- (contribution) ;
    \draw[pertains] (other) -- (contribution) ;
\end{tikzpicture}
\end{center}

% ============================================================== %
% ========              LABELS                          ======== %
% ============================================================== %
\subsubsection{Labels}\label{labels}

Contributions $c\in {\bf C}$ have \textbf{labels}. For now, we consider labels are taken from the set  $\mathbb{S}$  of character strings of arbitrary length. Labels can be empty. And some types of contributions (namely \hyperref[bidir]{bidirectional edges}) can have multiple labels. Labels satisfy requirement \ref{req-inclusivity}.
\medskip

There is no limit on the length of a contribution label. An entire book could be copied into the label of a contribution. 

\begin{bestpractices}{}
    Keep labels short. Prefer to decompose a long text into multiple contribution labels.
\end{bestpractices}

% ============================================================== %
% ========              TYPES                           ======== %
% ============================================================== %
\subsubsection{Types}\label{types}

A contribution has  an abstract type   and a concrete type (a.k.a. a type and a subtype). There are five different sets of \textbf{abstract contribution types}, namely \textit{(i)} the set  $\bf{V}$ of  {vertex} types (cf \S\ref{vertices} below), \textit{(ii)} the set $\bf{P}$ of {pen} types  (cf \S\ref{pens}),   and the set $\bf{E}$ of  edge types (cf \S\ref{edges}) which comprises \textit{(iii)} the set $\bf{E_A}$ of {adirectional edge} types (cf \S\ref{adir}), \textit{(iv)} the set $\bf{E_U}$ of {unirectional edge} types (cf \S\ref{unidir}), and \textit{(v)} the set $\bf{E_B}$ of {bidirectional edge} types (cf \S\ref{bidir}). The set of abstract types is denoted ${\bf T}^{\textsc{ab}}$. It is equal to 
${\bf T}^{\textsc{ab}}={\bf V} \cup {\bf P} \cup{\bf E}={\bf V} \cup {\bf P}\cup{\bf E_A}\cup {\bf E_U}\cup {\bf E_B}$.\medskip

The abstract type of a contribution is specified by a \textbf{{concrete type}}. Examples of concrete types of vertex contributions are the $\mmmtype{question}$ and $\mmmtype{narrative}$ subtypes. Examples of concrete types of edges are the \pertains{} and \equates{} subtypes. 
The different  concrete types are formally introduced in subsequent paragraphs of this section: \S\ref{vertices} -- \S\ref{pens}. 
\medskip
    
\begin{figure*}
    \centering\begin{tikzpicture}
        \node[existence] (contribution) {contribution};
        % contributions
        \node[right=1cm  of contribution, existence] (pen) {pen};
        \node[above=1.5cm of pen.west, anchor=west, existence] (vertex) {vertex};
        \node[below= 2cm of pen.west, anchor=west, existence] (edge) {edge};
        % edges
        \node[right=1cm of edge, existence,inner sep=0] (uni) {unidirectional\\ edge};
        \node[above=1.8cm of uni.west, anchor=west, existence,inner sep=0] (a) {adirectional\\ edge};
        \node[below=2.5cm of uni.west, anchor=west, existence,inner sep=0] (bi) {bidirectional\\ edge};
        % vertices
        \node[right=1cm of vertex, existence] (question) {\mmmtype[questioncolor]{question}};
        \node[above=0.6cm of question.west, anchor=west, existence] (narrative) {\mmmtype[narrativecolor]{narrative}};
        \node[below=0.6cm of question.west, anchor=west, existence] (existence) {\mmmtype[existencecolor!50!black]{existence}};
        %
        % adirectional edges
        \node[right=1cm of a, existence] (relate) {\mmmtype[relatecolor]{relate}};
        % unidirectional edges
        \node[below=1.7cm of relate.west,anchor=west, existence] (pertains) {\mmmtype[pertainscolor]{pertains}};
        \node[above=0.5cm of pertains.west, anchor=west, existence] (instantiates) {\mmmtype[instantiatescolor]{instantiates}};
        \node[below=0.5cm of pertains.west, anchor=west, existence] (answers) {\mmmtype[answerscolor]{answers}};
        \node[below=0.5cm of answers.west, anchor=west, existence] (questions) {\mmmtype[questionscolor]{questions}};
        \node[below=0.5cm of questions.west, anchor=west, existence] (nuances) {\mmmtype[nuancescolor]{nuances}};
        \node[above=0.5cm of instantiates.west, anchor=west, existence] (relatesTo) {\mmmtype[relatecolor]{relatesTo}};
        %  bidirectional edges  
        \node[below=4cm of relate.west, anchor=west,, existence] (equates) {\mmmtype[equatescolor]{equates}};
        \node[below=0.5cm of equates.west, anchor=west, existence] (differsFrom) {\mmmtype[ForestGreen]{differsFrom}};
        \draw[instantiates] (vertex) -- (contribution);
        \draw[instantiates] (pen) -- (contribution) ;
        \draw[instantiates] (edge) -- (contribution);
        \draw[instantiates] (a) -- (edge);
        \draw[instantiates] (uni) -- (edge);
        \draw[instantiates] (bi) -- (edge);
        \draw[instantiates] (narrative) -- (vertex);
        \draw[instantiates] (question) -- (vertex);
        \draw[instantiates] (existence) -- (vertex);
        \draw[instantiates] (relate) -- (a);
        \draw[instantiates] (questions) -- (uni);
        \draw[instantiates] (answers) -- (uni);
        \draw[instantiates] (nuances) -- (uni);
        \draw[instantiates] (instantiates) -- (uni);
        \draw[instantiates] (pertains) -- (uni);
        \draw[instantiates] (relatesTo) -- (uni);
        \draw[instantiates] (equates) -- (bi);
        \draw[instantiates] (differsFrom) -- (bi);
    \end{tikzpicture}
\end{figure*}

Types generally satisfy requirement \ref{req-types} about contribution types. Abstract types are however merely infrastructural while concrete types convey structuring epistemic information in agreement with requirements \ref{req-epistemic} -- \ref{easygoing} (satisfaction of requirement \ref{easychoice} is to be further supported by UI application code).  Importantly, despite concrete types being epistemically structuring,  plenty of room is left for most of them (especially concrete edge types) to be interpreted with some flexibility (cf the \href{https://gitlab.com/MMM-Mat/mmm/-/blob/master/Graphml\%20files/interpretations.graphml}{\texttt{interpretations.graphml}} file \cite{MMMgraphml}).
\medskip

The set of concrete types is denoted ${\bf T}^{\textsc{co}}$. It is equal to ${\bf T}^{\textsc{co}}={\bf T}_{\bf V}\cup {\bf T}_{\bf P}\cup {\bf T}_{\bf E}={\bf T}_{\bf V}\cup {\bf T}_{\bf P}\cup{\bf T}_A\cup {\bf T}_U\cup {\bf T}_B$. \medskip

% ============================================================== %
% ========              TAGS                            ======== %
% ============================================================== %
\subsubsection{Tags}\label{tags}

Contributions $c\in {\bf C}$ are associated a possibly empty \textbf{set of tags}. The tag set associated to a contribution is often empty. 
By convention for now we expect that a tag is a string that starts with the character '$@$'. We denote by $\mathbb{S}_{@}$ the set of character strings in which tags are taken from. 
\medskip
    
\begin{figure}[H]
    \centering\begin{tikzpicture}
        \node[existence]    (a) {Animal};
        \node[left=10.5cm, existence] (c) {Cat};
        \draw[relatesTo] (c) -- (a)  node[midway,yshift=0pt] {is a \\ \mmmtag{\url{http://www.w3.org/1999/02/22-rdf-syntax-ns\#type}}};
    \end{tikzpicture}

    \caption{A \relatesto{} edge (cf \S\ref{unidir}) labelled "is a" and tagged with the URI of the RDF schema definition of  "RDF type" \cite{manola2004rdf}. }
    \label{RDF-tag}
\end{figure}
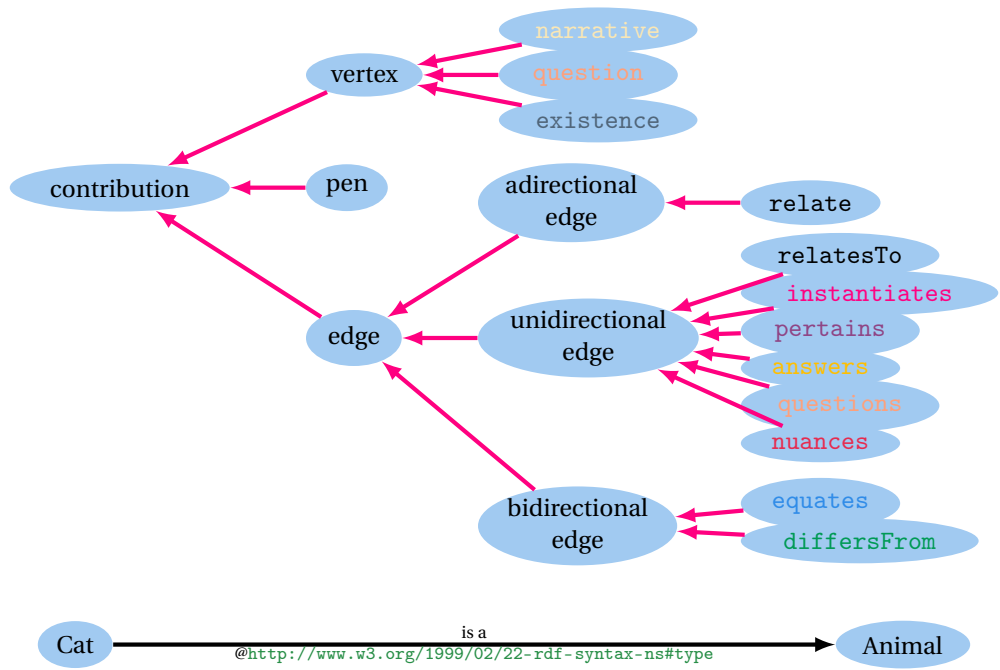

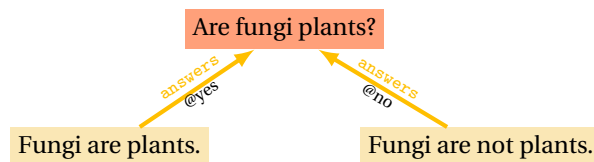
\begin{figure}[H]
    \centering\begin{tikzpicture}
        \node[question]    (q) {Are fungi plants?};
        \node[below left=1.3cm and 1cm, narrative] (yes) {Fungi are plants.};
        \draw[answers] (yes) -- (q)  node[edgelabel,midway,yshift=0pt] {\mmmtype[answerscolor]{answers} \\ \mmmtag{yes}};
        \node[below right=1.3cm and 1cm, narrative] (no) {Fungi are not plants.};
        \draw[answers] (no) -- (q)  node[edgelabel,midway,yshift=0pt] {\mmmtype[answerscolor]{answers} \\ \mmmtag{no}};
    \end{tikzpicture}

    \caption{A closed {question}  and two statement answers each linked to the question by an edge of type \mmmtype{answers} appropriately tagged.}
    \label{yesnotag}
\end{figure}

Like labels, tags are for enriching concrete types. Tags are used instead or in addition to labels when the full semantics of a contribution is specified somewhere else than in the contribution's label and than in the contribution's concrete type, e.g. when it is specified in an external resource (cf Fig \ref{RDF-tag}). {Tags are typically URIs or standardised  keywords, e.g.  the $@$yes and $@$no tags typically specify the meaning of an \mmmtype{answers} edge incoming a closed \mmmtype{question} contribution (cf Fig. \ref{yesnotag}). 
}  
\medskip
    
Pervasive tags  may eventually inform an update of the set of predefined concrete MMM contribution types -- provided they represent fundamental epistemic concepts that are not already represented in the set of predefined concrete  types. It however is essential to keep the set of concrete types (listed in the sequel) small to satisfy requirement \ref{req-types}.

% ============================================================== %
% ========                   METADATA                   ======== %
% ============================================================== %
\subsection{Metadata Attributes}\label{metadata-attributes}
The remaining {five} MMM landmark attributes that we introduce next constitute "\textbf{metadata}". 
For the sake of simplicity, except the identifier, we will ignore metadata attributes in the definition and in the examples of MMM contributions given later on.\medskip

Metadata attributes are kept minimal in agreement with requirement \ref{req-metadata}.

\begin{center}\begin{tikzpicture}
    \node[existence]    (contribution) {contribution};
    \node[left=1cm of contribution, existence,inner sep=0pt] (epistemic) {epistemic\\ attributes};
    \node[right=1cm  of contribution, existence,inner sep=0pt] (metadata) {metadata\\ attributes};
    \node[left=1cm of epistemic, existence] (type) {type};
    \node[above=0.6cm of type.west,anchor=west, existence] (label) {label};
    \node[below=0.6cm  of type.west,anchor=west, existence] (tags) {tags};
    \node[right=1cm of metadata, existence,inner sep=0] (author) {authorship list};
    \node[above=0.4cm of author.west,anchor=west, existence] (id) {id};
    \node[below=0.4cm of author.west,anchor=west, existence,inner sep=0] (status) {status (public/private)};
    \node[below=0.4cm of status.west,anchor=west, existence,inner sep=0] (marks) {marks
    (hidden,shared,obsolete)};
    \node[below=0.4cm of marks.west,anchor=west, existence,inner sep=0] (timestamp) {timestamp};
    \draw[pertains] (epistemic) -- (contribution) ;
    \draw[pertains] (metadata) -- (contribution) ;
    \draw[pertains] (label) -- (epistemic) ;
    \draw[pertains] (type) -- (epistemic) ;
    \draw[pertains] (tags) -- (epistemic) ;
    \draw[pertains] (id) -- (metadata) ;
    \draw[pertains] (author.west) -- (metadata) ;
    \draw[pertains] (status.west) -- (metadata) ;
    \draw[pertains] (marks.west) -- (metadata) ;
    \draw[pertains] (timestamp.west) -- (metadata) ;
\end{tikzpicture}
\end{center}

% ============================================================== %
% ========                   IDs                        ======== %
% ============================================================== %
\subsubsection{Identifiers}\label{IDs}

All landmarks $x\in {\bf N}$ in a \landscape{} ${\bf N}={\bf C} \cup \{\bot\}$ have a unique \textbf{identifier} taken in a certain set of identifiers denoted here by ${\bf I}$. Those could be {standard (namespace based) UUIDs (universally unique identifiers). A hash function applied to just the \hyperref[labels]{label} and  \hyperref[types]{type} of a contribution (see \S\ref{labels} and \S\ref{types} below) may facilitate the identification of duplicates (cf \S\ref{merging}). Involving a user identifier in the hash may also be required to support the digital signature of contributions. 
\furtherwork{Future research will explore how to define  identifiers of MMM contributions so as to facilitate search in the MMM space.}}
\medskip

Whatever the \landscape{} ${\bf N}$, the identifier of the pit is always the same because there is only one pit. The pit is the first collective landmark of the intermediary MMM web. 
Let us assume for now, in the examples below, that  ${\bf I}=\mathbb{N}$ and that the identifier of the pit is 0. 

% ============================================================== %
% ========               AUTHORSHIP                     ======== %
% ============================================================== %
\subsubsection{Authorship}
\label{authorship}

Contributions are associated a possibly empty, grow-only \textbf{set of authorships} in the power set $\mathcal{P}({\bf A})$ of the set $\bf A$ of authorships. Authorships are taken in the set  ${\bf A}=\mathcal{P}(\mathbb{S}) \times \mathbb{D}$. 
An \textit{authorship} $a\in {\bf A}$ is a pair comprised of a list of author names and a date. For instance the following is an authorship: %\checkrequirement{req-metadata}
$$
(\underbrace{\{\textrm{"Jane Smith"},\textrm{"Ivy Li"},\textrm{"Amari Doe"}\}}_{\textrm{team of authors}}, \underbrace{13/08/2023}_{\textrm{timestamp}}) \in {\bf A}.
$$
Contributions can be assigned several authorships, and usually they must be assigned at least one. 
The following example of an authorship set contains three authorships:
$$
\{~
\underbrace{{(\{\textrm{"Anne Martin"}\}, 10/10/2022)}}_{\textrm{one authorship}}, \underbrace{{(\{\textrm{"Jane Smith"},\textrm{"Ivy Li"},\textrm{"Amari Doe"}\}, 13/08/2023)}}_{\textrm{another authorship}}, \underbrace{{(\{\textrm{"Al B."}\}, 12/2023)}}_{\textrm{another}} ~\}\in \mathcal{P}({\bf A}).
$$
 
\begin{bestpractices}{}
Authorships are primarily to recognise the humans who have \textit{recorded} a contribution. When a contribution is a quote from the work of a different human, use additional contributions (cf \S\ref{referencing}) to make the source explicit. The late Alan Turing, for instance, should never appear as part of an authorship of an MMM contribution.
\end{bestpractices}

Recognition of intellectual activity is mentioned in \S\ref{trickling} below. 
 \medskip

The pit landmark has no authorship set attribute.

% ============================================================== %
% ========                 STATUS                       ======== %
% ============================================================== %
\subsubsection{Status}\label{status}

As will be detailed in the sequel, MMM landscapes are to be collectively built and distributed. Individual MMM contributions are stored locally on users' machines. They may also be shared.
\medskip

A contribution $c\in {\bf C}$ has a \textbf{status}. By default, the status of a contribution is {private} and local. If $c$ is private, a possibility is that $c$ is privately shared with  one or several groups of users who  have access right to $c$. A private contribution that is shared with groups $g_1,\ldots,g_n$ of users,  has its status attribute set to \sharedwith{$g_1,\ldots,g_n;R$} where $R$ is the reference to a sharing contract, cf \S\ref{sharing}.  If $c$ is private and shared with no-one, it has default status \mmmmark{local}. A contribution can also be public, meaning that \textit{any} user has access right to it.
Public contributions have their status attribute set to \mmmmark{public}.  When a contribution's status attribute is set to \mmmmark{public}, its label, type and tag attributes become immutable.
\medskip

Let $s$ and $s'$ be two contribution statuses. We define an  order $\leq$ on contribution statuses. We write $s\leq s'$ when either one of the following conditions is satisfied:
\begin{itemize}
    \item $s=$\mmmmark{local}, or 
     \item  $s'=$\mmmmark{public}, or 
     \item  $s=$\sharedwith{$g_1,\ldots,g_n;R$}, $s'=$\sharedwith{$g'_1,\ldots,g'_m;R'$},  $\bigcup g_i\subseteq \bigcup g'_i$ and $R'$ is no more constraining than $R$.
\end{itemize}
To downgrade (resp. upgrade) status $s$ is to replace $s$ with status $s'$ where $s'\neq s$ abd $s'\leq s$ (resp. $s\leq s'$). Downgrading a contribution status is forbidden.  A contribution's status can only be upgraded. 
\medskip

The pit landmark's status is \mmmmark{public}.

% ============================================================== %
% ========                MARKS                         ======== %
% ============================================================== %
\subsubsection{Marks}\label{marks}

Contributions may  be marked by any number of marks. Marks can be \textit{ad hoc} custom marks, e.g.: \mmmmark{archived}, \mmmmark{hidden}, \mmmmark{dim}, \mmmmark{highlighted}, \mmmmark{folder}, \mmmmark{unpublishable}. 
There also are predefined marks, e.g.: \mmmmark{new} (meaning unread), \obsolete{} (cf \S\ref{obsoleting}), \syncmark{} (cf \S\ref{synchronising}), \subscribemark{} (cf \S\ref{subscribing}), \rewardmark{} (cf \S\ref{trickling}). 
Marks may have parameters. For instance, the \syncmark{} mark is parametrised by the list of devices $d_1,\ldots,d_n$ that the contribution is meant to be copied to.  
\medskip

Marks are mostly for internal house-keeping, in contrast to the status which universally characterises a contribution. The   meaning of a mark and existence  is usually specific to one user or to a team of users.  For instance,  one user or team may choose to hide contribution $c$ marked as \mmmmark{hidden}, while other users may neither mark nor hide $c$, and others may mark it as \mmmmark{hidden} but not hide it. Also in contrast to the status, marks are usually locally revocable by the user.

% ============================================================== %
% ========                TIMESTAMPS                    ======== %
% ============================================================== %
\subsubsection{Timestamps}\label{timestamps} % Local/logical

A contribution $c\in {\bf C}$ is associated a timestamp corresponding to the date at which a user first encounters $c$.  If the user is the creator of $c$, the timestamp is the date of creation of $c$. Otherwise, it is  the date at which the user receives $c$ from another user. Defined this way, timestamp attributes allow ordering contributions into a timeline of granular events, namely contribution appearances (creation or reception) which are the events we propose to emphasise. The definition of the timestamp attribute of MMM contributions will need to be refined to support finer timelines also accounting for modifications of the attributes of existing contributions. 

\subsection{Landmarks}

We have defined the main attributes of MMM landmarks. Now we specify the different kinds of landmarks, and especially the different kinds of contributions.
% ============================================================== %
% ========             CONTRIBUTIONS                    ======== %
% ============================================================== %
\subsubsection{Contributions}\label{contributions}

For the sake of simplicity in the following sections we ignore the metadata of contributions -- i.e., {the authorship set, status, mark set and timestamp  attributes}.\medskip

A contribution is an object from ${\bf C}\subset{\bf I}\times \mathbb{S} \times {\cal P}(\mathbb{S}_@) \times {\bf T}^{\textsc{ab}} $ 
comprised of an \hyperref[IDs]{identifier} in ${\bf I}$, a \hyperref[labels]{label} in $\mathbb{S}$, a \hyperref[tags]{tag set} in  ${\cal P}(\mathbb{S}_@)$ and an abstract type in ${\bf T}^{\textsc{ab}}$.
Abstract types define the following sets of contributions:
\begin{itemize}
    \item The  set ${\bf C}_{\bf V} \subset {\bf I}\times \mathbb{S} \times {\cal P}(\mathbb{S}_@) \times {\bf V}$ of "\textbf{vertex contributions}" a.k.a. "vertices" a.k.a. "nodes" 
    \item The set ${\bf C}_{\bf P} \subset {\bf I}\times \mathbb{S} \times {\cal P}(\mathbb{S}_@) \times {\bf P} $ of "\textbf{pen contributions}" a.k.a. "pens" %:  ${\bf C}_{\bf P} \subset {\bf I}\times \mathbb{S} \times {\bf P} \times {\cal P}(\mathbb{S}_{\#}) \times {\cal P}({\bf A})$
    %  (whose  type is $ {\bf P}$).
    \item The set ${\bf C}_{\bf E}  \subset {\bf I}\times \mathbb{S} \times {\cal P}(\mathbb{S}_@) \times {\bf E} $ of "\textbf{edge contributions}" a.k.a. "edges" a.k.a. "links" %: ${\bf C}_{\bf E} \subset {\bf I}\times \mathbb{S} \times {\bf E} \times {\cal P}(\mathbb{S}_{\#}) \times {\cal P}({\bf A})$.
    %  (whose type is ${\bf E}$). The set  ${\bf C}_{\bf E}$ is 
    which is comprised of:
    \begin{itemize}
        \item The set ${\bf C_{E_A}}  \subset {\bf I}\times \mathbb{S} \times {\cal P}(\mathbb{S}_@) \times {\bf E_A}$ of "\textbf{adirectional edge contributions}" %: ${\bf C_{E_A}} \subset {\bf I}\times \mathbb{S} \times {\bf E_A} \times {\cal P}(\mathbb{S}_{\#}) \times {\cal P}({\bf A}) $.
        %  (whose type is ${\bf E_A}$).
        \item The set ${\bf C_{E_U}}  \subset {\bf I}\times \mathbb{S} \times {\cal P}(\mathbb{S}_@) \times {\bf E_U}$ of "\textbf{unidirectional edge contributions}" %: ${\bf C_{E_U}} \subset {\bf I}\times \mathbb{S} \times {\bf E_U} \times {\cal P}(\mathbb{S}_{\#}) \times {\cal P}({\bf A}) $.
        %  (whose type is ${\bf E_U}$).
        \item The set ${\bf C_{E_B}} \subset {\bf I}\times \mathbb{S} \times {\cal P}(\mathbb{S}_@) \times {\bf E_B} $ of "\textbf{bidirectional edge contributions}" %: ${\bf C_{E_B}} \subset {\bf I}\times \mathbb{S} \times {\bf E_B} \times {\cal P}(\mathbb{S}_{\#}) \times {\cal P}({\bf A}) $. 
        % (whose type set is ${\bf E_B}$).
    \end{itemize}
\end{itemize}
The set of all contributions is 
${\bf C}= {\bf C}_{\bf V} \cup {\bf C}_{\bf P} \cup {\bf C}_{{\bf E}}  = {\bf C}_{\bf V} \cup {\bf C}_{\bf P} \cup {\bf C_{E_A}}  \cup {\bf C_{E_U}}  \cup {\bf C_{E_B}} $.\medskip

\begin{figure}[H]
    \begin{center}
    \includegraphics[scale=0.5,trim=7cm 0 7cm 0, clip]{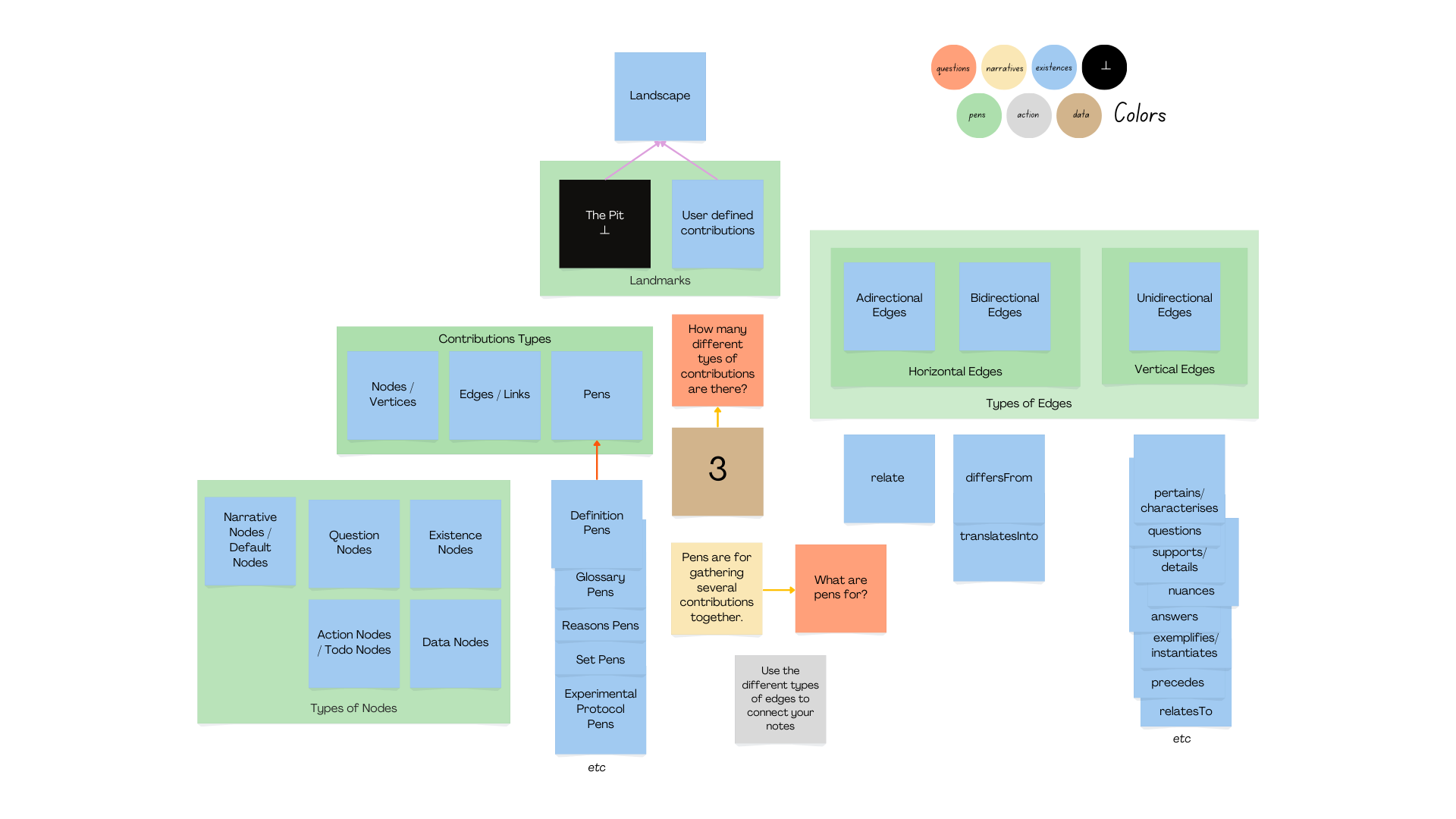}
    \end{center}
    \caption{Summary diagram of the MMM format definitions. }
\end{figure}

Visual design choices in the illustrations of this article are arbitrary. This proposal does not cover visualisation of the MMM formatted content. Different user interfaces can later offer different visualisations to accommodate different preferences.

% ============================================================== %
% ========                                              ======== %
% ============================================================== %
\subsubsection{Vertices}
\label{vertices}

A vertex contribution $c\in {\bf C}_{\bf V} \subset {\bf I}\times \mathbb{S} \times {\cal P}(\mathbb{S}_@)  \times {\bf V} $ is composed of an \hyperref[IDs]{identifier},  \textbf{a non-empty \hyperref[labels]{label}}, a \hyperref[tags]{tag set} in  ${\cal P}(\mathbb{S}_@)$ and \hyperref[types]{an abstract type} in ${\bf V} = {\bf T_V}\subset {\bf T}^{\textsc{ab}}$  equal to the concrete type. 
Vertices have one of {five} possible (abstract/concrete) types:

$${\bf T}_{\bf V}={\bf V}=\{\mmmtype{question}, \mmmtype{narrative},\mmmtype{existence},\mmmtype{action},\mmmtype{data}\}.$$

The  types of vertices that are the most central to our system are $\mmmtype{question}$ vertices, $\mmmtype{narrative}$ vertices  and  $\mmmtype{existence}$ vertices. Contrary to the other two types, redundancy management is to be severe on those central three types. We won't mind if there are several  $\mmmtype{data}$ vertices labelled "{\tt 42}". However, we will mind if there are several  $\mmmtype{question}$ vertices labelled "What colour is the sky?".\medskip

{Vertex labels \textit{cannot} be empty.}\medskip

Below  are examples of contributions that are vertices. I remind that visual choices made in the illustrations in this article are arbitrary. MMM documented information doesn't even need to be graphically represented. The MMM-JSON format for instance is enough to capture it.
\begin{itemize}
    \item $(\contributionID{qsky}, \textrm{"What colour is the sky?"}, \emptyset, \mmmtype{question})\in {\bf C}_{\bf V}$ 
    \hspace{6cm} %\hspace{7cm}
    \begin{tikzpicture}[overlay]
        % \node[question] at (0, 0)   (qsky)  {What\\ colour is\\ the sky?\nodepart{second} ID=\ref{qsky}};
        \node[question,yshift=3mm]   (qsky)  {What colour is the sky?};
        \node[metadata,below right=0mm and 0mm of qsky] {\mmmid{qsky}};
    \end{tikzpicture} 
    \item $(\contributionID{narrSkyBlue}, \textrm{"The sky is blue."}, \emptyset,\mmmtype{narrative})\in {\bf C_V}$
    \hspace{6cm}\begin{tikzpicture}[overlay]
        \node[narrative,yshift=3mm] (narrSkyBlue) {The sky is blue.};
    \node[metadata,below right=0mm and 0mm of narrSkyBlue] {\mmmid{narrSkyBlue}};
    \end{tikzpicture} 
    \item $(\contributionID{sky}, \textrm{"Sky"}, \emptyset,\mmmtype{existence})\in {\bf C_V}$  \hspace{7.8cm}\begin{tikzpicture}[overlay]
    \node[existence,yshift=2mm] (sky) {Sky};
    \node[metadata,below=1mm of sky] {\mmmid{sky}};
    \end{tikzpicture}
    \item  $(\contributionID{tobeblue}, \textrm{"To be blue"}, \emptyset,\mmmtype{existence})\in {\bf C_V}$ 
    \hspace{9cm}\begin{tikzpicture}[overlay]
     \node[existence,yshift=6mm] (tobeblue) {To be blue};
    \node[metadata,below right=0mm and 0mm of tobeblue] {\mmmid{tobeblue}};
    \end{tikzpicture}
    \item  $(\contributionID{blue}, \textrm{"Blue"}, \emptyset,\mmmtype{existence})\in {\bf C_V}$ %\hspace{6.6cm}
    \hspace{7.6cm}
    \begin{tikzpicture}[overlay]
     \node[existence,yshift=5mm] (blue) {Blue};
        \node[metadata,below=1mm  of blue] {\mmmid{blue}};
    \end{tikzpicture}

    \item   $(\contributionID{daytime}, \textrm{"the color  of a cloudless daytime sky"}, \emptyset,\mmmtype{existence})\in {\bf C_V}$ 
    \hspace{5.8cm}\begin{tikzpicture}[overlay]
     \node[existence,yshift=6mm,inner sep=0mm] (daytime) {the colour of a\\ cloudless daytime  sky};
        \node[metadata,below right=0mm and 0mm of daytime] {\mmmid{daytime}};
    \end{tikzpicture}
    \item  $(\contributionID{turquoise}, \textrm{"Turquoise"},\emptyset, \mmmtype{existence})\in {\bf C_V}$ \hspace{7.9cm}\begin{tikzpicture}[overlay]
      \node[existence,yshift=3mm] (turquoise) {Turquoise};
    \node[metadata,below right=0mm and 0mm of turquoise] {\mmmid{turquoise}};
        \end{tikzpicture}
    \item  $(\contributionID{bleu}, \textrm{"bleu"}, \emptyset,\mmmtype{existence})\in {\bf C_V}$ \hspace{7.7cm}\begin{tikzpicture}[overlay]
        \node[existence,yshift=3mm] (bleu) {bleu};
    \node[metadata,below=1mm of bleu] {\mmmid{bleu}};
    \end{tikzpicture}
    \item  $(\contributionID{white}, \textrm{"White"}, \emptyset,\mmmtype{existence})\in {\bf C_V}$ \hspace{8.5cm}\begin{tikzpicture}[overlay]
    \node[existence,yshift=3mm] (white) {White};
    \node[metadata,below=1mm of white] {\mmmid{white}};

    \end{tikzpicture}
    \item  $(\contributionID{colour}, \textrm{"Colour"}, \emptyset,\mmmtype{existence})\in {\bf C_V}$ 
    \hspace{7.2cm}
    \begin{tikzpicture}[overlay]
    \node[existence,yshift=3mm] (colour) {color};
    \node[metadata,below=1mm of colour] {\mmmid{colour}};
    \end{tikzpicture}

\item  $(\contributionID{datatrue}, \textrm{"true"},\{\mmmtag{boolean}\},\mmmtype{data})\in {\bf C_V}$ 
\hspace{8.2cm}
\begin{tikzpicture}[overlay]
\node[datavalue,yshift=3mm] (datatrue) {true};
\node[metadata,below=1mm of datatrue] {\mmmid{datatrue}};
\end{tikzpicture}

\item  $(\contributionID{data12}, \textrm{"12"}, \{\mmmtag{int}\},\mmmtype{data})\in {\bf C_V}$ 
\hspace{8.2cm}
\begin{tikzpicture}[overlay]
\node[datavalue,yshift=4mm] (data12) {true};
\node[metadata,below=1mm of data12] {\mmmid{data12}};
\end{tikzpicture}

\item  $(\contributionID{boilwater}, \textrm{"Boil water."}, \emptyset,\mmmtype{action})\in {\bf C_V}$ 
\hspace{7.2cm}
\begin{tikzpicture}[overlay]
\node[action,yshift=3mm] (boilwater) {Boil water.};
\node[metadata,below=1mm of boilwater] {\mmmid{boilwater}};
\end{tikzpicture}
\end{itemize}

Contributions with identifiers 5 to 9 above could be drawn from a glossary file listing/defining colours. Those contributions could be tagged $\{\mmmtag{colours} \}$ instead of $\emptyset$.  Contribution 3 could be tagged $\{\mmmtag{myWeatherVocab}, \mmmtag{some-standard-bird-ontology}\}$.

\begin{bestpractices}{}
Use \mmmtype{question} vertices for labels that end with a question mark. 
Use \mmmtype{narrative} as the default contribution
type when you don't want to bother determining what other type best suits your new contribution. But ideally, use \mmmtype{narrative} vertices for labels that end with a period, for instance, a statement, a series of statements, a theorem, a story\ldots{}. 
Use \mmmtype{existence} vertices for labels that typically don't end with any punctuation mark, naming or formulating something that matters to you, a concept or a property (e.g. "Sky", "Blue", "To be blue"). Use \mmmtype{action} vertices for labels that describe some action to perform (e.g. "Boil water").   Use \mmmtype{data} vertices for labels that express the value of a data point (e.g. "{\tt 42}", "{\tt true}", "{\tt 13/08/1991}").\medskip
\end{bestpractices}

The default contribution type \mmmtype{narrative}  satisfies requirement \ref{req-lazy} while 
the  vertex type \mmmtype{existence}  
satisfies requirement \ref{req-onto}.  
\bigskip

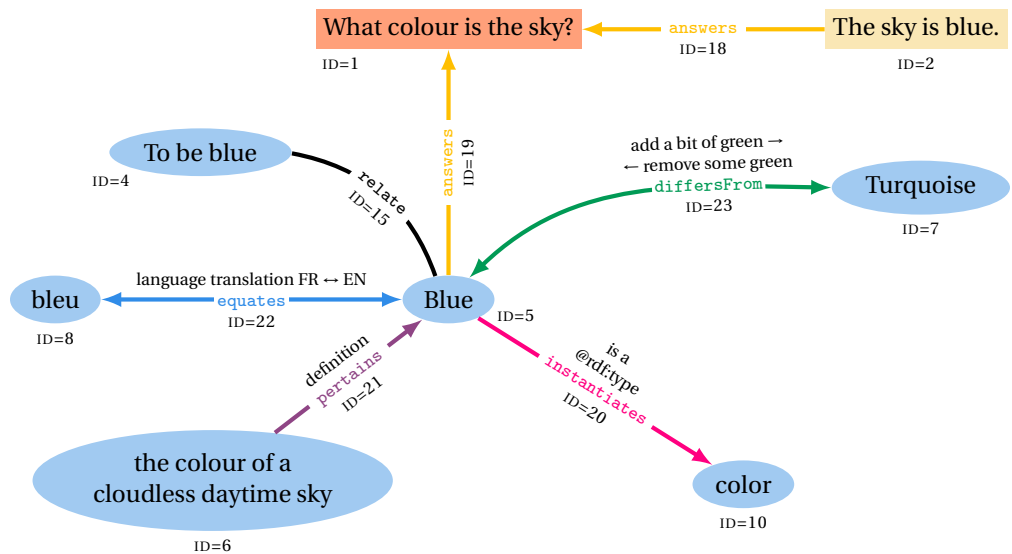
\begin{figure}[H]
    \centering
    \begin{tikzpicture}
        \node[question]   (qsky)  {What colour is the sky?};
        \node[metadata,below left=1mm and -6mm of qsky] {\mmmid{qsky}};
        \node[narrative,right=5cm] (narrSkyBlue) {The sky is blue.};
        \node[metadata,below=1mm of narrSkyBlue] {\mmmid{narrSkyBlue}};
        \node[existence,below=3cm of qsky] (blue) {Blue};
        \node[metadata,below right=-1mm and 2mm of blue] {\mmmid{blue}};
        \node[existence,above left=1.5cm and 2cm of blue] (tobeblue) {To be blue};
        \node[metadata,below left=1mm of tobeblue] {\mmmid{tobeblue}};
        \node[existence,below left=1.7cm and 1cm of blue] (daytime) {the colour of a \\cloudless daytime  sky};
        \node[metadata,below=1mm of daytime] {\mmmid{daytime}};
\node[existence,above right=1cm and 5cm of blue] (turquoise) {Turquoise};
\node[metadata,below=1mm of turquoise] {\mmmid{turquoise}};

\node[existence,left=4cm of blue] (bleu) {bleu};
\node[metadata,below=1mm of bleu] {\mmmid{bleu}};
\node[existence,below right=2cm and 3cm of blue] (colour) {color};
\node[metadata,below=1mm of colour] {\mmmid{colour}};
\path [draw,edge] (blue) to [bend right,edgelabel] node[yshift=-3pt,fill=white] (relate1){\mmmtype[black]{relate}\\\mmmid{relate1}} (tobeblue);
        % Narrative answer
        \draw[answers] (narrSkyBlue) -- (qsky)  node[edgelabel,midway,yshift=-4pt,fill=white] (narranswer) {\mmmtype[answerscolor]{answers}\\\mmmid{narranswer}} ;
        % Blue answer
        \draw[answers] (blue) -- (qsky)  node[edgelabel,midway,yshift=-4pt,fill=white] (blueanswer) {\mmmtype[answerscolor]{answers}\\\mmmid{blueanswer}} ;
        %  
        % isardftype
        \draw[instantiates] (blue) -- (colour)  node[edgelabel,midway,yshift=5pt,fill=white] (isardftype){is a\\\rdftag{rdf:type} \\\mmmtype[instantiatescolor]{instantiates}\\\mmmid{isardftype}} ;
        %
        % Blue definition
        \draw[pertains] (daytime) -- (blue)  node[edgelabel,midway,yshift=0pt,fill=white] (bluedefinition) {definition\\\colorbox{white}{\mmmtype[pertainscolor]{pertains}}\\\mmmid{bluedefinition}} ;
%
        % Blue <--> Turquoise           DIFFERSFROM
        \path [draw,differsfrom] (blue) to [bend left,in=165,edgelabel] node[yshift=5pt,pos=0.7] {add a bit of green $\rightarrow$\\$\leftarrow$ remove some green\\\colorbox{white}{\mmmtype[ForestGreen]{differsFrom}}\\ \mmmid{blueturquoise}} (turquoise.west);

        %
         % Blue <--> Bleu
         \draw[equates] (bleu) -- (blue)  node[edgelabel,midway,yshift=0pt] (bluebleu) {language translation FR $\leftrightarrow$ EN\\\colorbox{white}{\mmmtype[equatescolor]{equates}}\\\mmmid{ENFR}} ;
    \end{tikzpicture}
    \caption{Different concrete types of MMM edges. 
        As mentioned above (cf \S\ref{MMMreqs}), the concrete types of MMM edges (e.g. \mmmtype[equatescolor]{equates}, \mmmtype[Red]{instantiates}, \mmmtype[pertainscolor]{pertains}) have deliberately loose semantics. This proposal offers a starting set of edge types that is intended to remain small. \furtherworkcaption{Future modifications of this starting set may be needed to ensure the use of the different edge types is balanced.} 
        }
    \label{fig-examples}
\end{figure}
\furtherworkmark{}

Before I introduce the other types of MMM contributions, I recall that in agreement with Design Bias \ref{experts} and with requirement \ref{req-lazy}, the MMM format provides a default information container (namely the \mmmtype{narrative} node). {A busy/lazy user (using a lazy UI) doesn't need to know about, nor use anything else in order to document their notes in MMM format. They could possibly consider that one \mmmtype{narrative} corresponds one traditional document. I propose however to incentivise and facilitate the use of other MMM contributions.}

% ============================================================== %
% ========                                              ======== %
% ============================================================== %
\subsubsection{Edges}\label{edges}

There are three categories of edges: \hyperref[adir]{adirectional}, \hyperref[unidir]{unidirectional} and \hyperref[bidir]{bidirectional}. 
Contrary to that of vertices, the abstract type of edges contains more data than just the concrete type. For one, the abstract type specifies the endpoints of the edge.  Bidirectional edges are special. A bidirectional edge resembles two unidirectional edges in opposite direction. Each direction may have its own label and its own tag set.
\medskip

Contrary to node labels, edge labels can be empty.\medskip

Edge endpoints can be any kind of landmark: vertices, edges, pens and even the pit. This allows to satisfy requirement \ref{req-recursive}, while generally MMM edges satisfy requirements  \ref{req-glue} and \ref{req-collective}. An edge can't be one of its own two endpoints. An edge can however have another edge as one of its endpoints. We may have to limit the depth of the recursion\footnote{\label{edgerecursion}Say an ordinary  edge between two node contributions is of depth $0$. An edge of depth $n+1$ is an edge that has one of its endpoint that is an edge of depth $n$ and the other endpoint is of depth no greater than $n$. MMM landscapes might be easier to store and to navigate if we forbid edges of depth greater than a given maximum. I expect that very deep recursion is dispensable.  }. \medskip

\begin{bestpractices}{}
Make extensive use of directional edges.
Prefer contributing information in edge labels than in vertex labels (cf \S\ref{annotating}). \end{bestpractices}

{Concrete edge types play an important role in this proposal. They allow to \textit{roughly} sort  epistemic glue according to its intended purpose. Together with complementary edge information (conveyed by  tags and labels), concrete edge types allow aggregating information (cf \S\ref{aggregating}). }\medskip

Again let us insist on the flexibility of interpretation of concrete types of edges. It plays a central part in the intended universality of the MMM data structure.  Concrete types are meant to convey a form of common epistemic denominator like the concrete type \mmmtype{question}: most people can have some form of agreement on what a question is, if only they agree that it is not a statement. The same holds for concrete types of edges but the room for variations in interpretations of each concrete edge type is wider. For instance, the \equates{} edge introduced below can be interpreted as conveying a relation of synonymy between two linguistic terms. It can just as well be used to link two mathematically equivalent theorems expressed in different formalisms. \furtherwork{Future work will have to circumscribe the extent  of the variations in interpretations that is tolerated.}

% ============================================================== %
% ========                                              ======== %
% ============================================================== %
\subsubsection{Adirectional Edges}\label{adir}

There is only one concrete type of adirectional edge, namely the   \mmmtype{relate}  edge. 
The abstract type\footnotemark{} of an adirectional edge is a triple in ${\bf E_A} = {\bf I}\times{\bf I}\times{\bf T}_A = {\bf I}\times{\bf I}\times \{\mmmtype{relate}\}$. The first two components of an adirectional edge type are the identifiers of its endpoints. The order of these two components does not matter. 
\medskip

\footnotetext{The term "type" in "abstract type" is not ideal. For two edges to have the same abstract type, they need to share the same endpoints and the same concrete type.  Arguably, they need to  almost be the same edge. }

\begin{bestpractices}{}
    Use \mmmtype{relate} edges as default edges when you don't want to bother determining what other concrete edge type is more suitable. But ideally, don't use this type of edge.
\end{bestpractices} 
 
\begin{figure}[H]
    \centering
    \begin{tikzpicture}
        \node[existence] (blue) {Blue};
        \node[metadata,below=1mm of blue] {\mmmid{blue}};
        \node[existence,above left=1cm and 1cm of blue] (tobeblue) {To be blue};
        \node[metadata,below left=1mm of tobeblue] {\mmmid{tobeblue}};
        \node[existence,right=2cm of blue] (sky) {Sky};
        \node[metadata,below=1mm of sky] {\mmmid{sky}};
        \draw[relate] (sky) -- (blue)  node[midway] (relate4) {};
        \node[metadata,below=0mm  of relate4] {\mmmid{relate4}};
        \path [draw,edge] (blue) to [bend left] node (relate1){} (tobeblue);
        \node[metadata,left=1mm  of relate1] {\mmmid{relate1}};
        \path [draw,edge] (blue)  to [bend right]node[sloped,font=\scriptsize] (relate2){similar\\[1pt]\mmmid{relate2}} (tobeblue);
        \path [draw,edge] (blue)  to [bend left=60]node[] (relate3){} (relate1);
        \node[metadata,left=1mm  of relate3] {\mmmid{relate3}};       
    \end{tikzpicture}
    \caption{Adirectional edges of type \mmmtype{relate} (represented in black lines here). All other edge types should be preferred to this default edge type. Edges can have edges as endpoints. Their labels are empty or not. Above all edge labels are empty except one which is set to  "similar". }
    \label{fig-adir}
\end{figure}
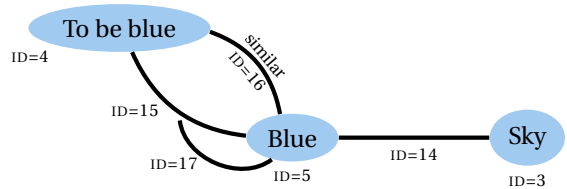

Here are some examples of contributions that are adirectional edges (see Fig. \ref{fig-adir}).  
\begin{itemize}
    \item  $(\contributionID{relate4}, \textrm{""},\emptyset,{\ref{blue},\ref{sky}, \mmmtype{relate}}) \in {\bf C_{E_A}}$ is a labelless  edge  between  vertices with IDs \ref{sky} and \ref{blue} introduced as examples in \S\ref{vertices}.
    \item     $(\contributionID{relate1}, \textrm{""},\emptyset,{\ref{blue},\ref{tobeblue}, \mmmtype{relate}})\in {\bf C_{E_A}}$  is another labelless  edge,  between  vertices with IDs \ref{tobeblue} and \ref{blue}.
    \item   $(\contributionID{relate2}, \textrm{""}, \emptyset,\ref{tobeblue}, \ref{blue},\mmmtype{relate})\in {\bf C_{E_A}}$ is an  edge labelled "similar" between the same two vertices as edge \ref{relate1}.   
\end{itemize}

As mentioned already, edges can have other landmarks than vertices as endpoints:
\begin{itemize}
    \item  $(\contributionID{relate3}, \textrm{""}, \emptyset,{\ref{relate1}, \ref{blue},\mmmtype{relate}})\in {\bf C_{E_A}}$ is an edge between edge with ID \ref{relate1} and vertex with ID \ref{blue}.
\end{itemize}

% ============================================================== %
% ========                                              ======== %
% ============================================================== %
\subsubsection{Unidirectional edges}\label{unidir}

The abstract type of an adirectional edge is a triple from ${\bf E_U} = {\bf I}\times{\bf I}\times{\bf T}_U$.  
The first (resp. second) component is the identifier of the edge's start point (resp. endpoint). 
The last component is the edge's concrete type taken in: % ${\bf T}_U $: 
\vspace{-7mm}

\begin{align*}
    {\bf T}_U  \supseteq&\{\mmmtype{answers}, \mmmtype{questions}, \pertains{},\mmmtype{instantiates},\\ & \mmmtype{nuances},\mmmtype{supports},\mmmtype{pennedIn},\mmmtype{precedes},\mmmtype{relatesTo}\}
\end{align*} 

The \mmmtype{relatesTo} edge is the default unidirectional edge that is to be used (sparingly) as a default edge similar to the adirectional \mmmtype{relate} edge.\medskip

The  \pertains{} edge is an ubiquitous type of contribution that we might consider renaming "\mmmtype{details}".
In agreement with requirement \ref{req-interpretation}, a \pertains{} edge from contribution $a$ to contribution $b$ can mean a number of things such as:  $a$ belongs to $b$, $a$   characterises  $b$,   $b$ involves $a$, $a$  is a detail of  $b$,  $b$ concerns $a$, $b$ refers to $a$, or $b$ is about  $a$\ldots{} \furtherwork{Work is actually being carried out to explore and circumscribe the relevant semantics of this edge type. I expect this edge type might need to be divided into two edge types, e.g. \pertains{} and \mmmtype{characterises}. }
Tying \pertains{} links between pieces of information and the topics/concepts they are about plays an important role in our solution {(cf \S\ref{parachutist})}. 
 \medskip

The \mmmtype{pennedIn} edge is the only MMM contribution that comes with a constraint of usage. 
{The endpoint of \mmmtype{pennedIn} edge can only be a pen contribution} (cf \S\ref{pens} below). The start point can be any type of contribution. In contrast to \pertains{} edges (and other edge types) which convey semantically structural information, \mmmtype{pennedIn} edges are rather for the meta coordination of MMM contributors.
We further discuss this special type of edge in \S\ref{mutable} below.
\medskip

Here are some examples of contributions that are unidirectional edges (see Fig. \ref{fig-examples}):
\begin{itemize}
    \item $(\contributionID{narranswer}, \textrm{""}, \emptyset, \ref{narrSkyBlue}, \ref{qsky},\mmmtype{answers})\in {\bf C_{E_U}}$ offers %\hyperref[vertices]
    {$\mmmtype{narrative}$ vertex \ref{narrSkyBlue}} as answer to %\hyperref[vertices]
    {$\mmmtype{question}$ vertex \ref{qsky}} (cf \S\ref{vertices}).
    \item $(\contributionID{blueanswer}, \textrm{""}, \emptyset,{\ref{blue},\ref{qsky}, \mmmtype{answers}})\in {\bf C_{E_U}}$ offers $\mmmtype{existence}$ vertex \ref{blue} as answer to $\mmmtype{question}$ vertex \ref{qsky}.
    \item  $(\contributionID{isardftype}, \textrm{"is a"}, \{\rdftag{rdf:type} \},\ref{blue},\ref{colour},  \mmmtype{instantiates} )\in {\bf C_{E_U}}$.
    \item  $(\contributionID{bluedefinition}, "\textrm{definition}", \emptyset,\ref{daytime},\ref{blue}, \pertains{})\in {\bf C_{E_U}}$.
\end{itemize}

% ============================================================== %
% ========                                              ======== %
% ============================================================== %
\subsubsection{Bidirectional edges}\label{bidir}

Formally, the abstract type of a bidirectional edge is a seven component uple  from ${\bf E_B} = {\bf I}\times{\bf I}\times{\bf T}_B\times \mathbb{S} \times \mathbb{S}\times {\cal P}(\mathbb{S}_{@}) \times {\cal P}(\mathbb{S}_{@}) $. 
The first two components are the identifiers of the edge's startpoint and endpoint. The third  component is the edge's concrete type taken in: % ${\bf T}_B $:
$${\bf T}_B\supseteq \{\equates{},\mmmtype{differsFrom}\}.$$  
The fourth (resp. fifth) component is the label of the edge specific to the direction start point $\to$ endpoint (resp. endpoint $\to$ start point). The sixth (resp. seventh) component is the tag list  specific to the direction start point $\to$ endpoint (resp. endpoint $\to$ start point). One, two or all three of the labels of a bidirectional edge can be empty. 
\medskip

Here are some examples of contributions that are bidirectional edges (see Fig. \ref{fig-examples}):
\begin{itemize}
    \item $(\contributionID{ENFR}, \underbrace{\textrm{"language translation"}}_{\textrm{main label}}, \emptyset,{\ref{blue}, \ref{bleu},\equates{},~~\underbrace{\textrm{"EN$\to$ FR"}}_{\textrm{label for dir 1}},~~\underbrace{\textrm{"FR$\to$EN"}}_{\textrm{label for dir 2}},\emptyset,\emptyset})\in {\bf C_{E_B}}$
   \item  $(\contributionID{blueturquoise}, \textrm{""}, \emptyset,{\ref{blue}, \ref{turquoise},\mmmtype{differsFrom},\textrm{"add a bit of  green"},\textrm{"remove some green"},\emptyset,\emptyset})\in {\bf C_{E_B}}$ 
\end{itemize}\medskip

Bidirectional \equates{} edges  are important as they allow connecting similar contributions that are not necessarily perfect duplicates. They support requirement \ref{req-reformulate}.

% ============================================================== %
% ========                PENS                          ======== %
% ============================================================== %
\subsubsection{Pens}\label{pens}

The abstract type of a pen is a 
couple taken from the set ${\bf P} ={\cal P}({{\bf I}})\times{\bf T}_{\bf P}$. 
The first component is a set of identifiers. The second component is the pen's concrete type from: 

$$ %{\bf T}_{\bf P}^p 
{\bf T}_{\bf P}  \supseteq \{\mmmtype{definition},\mmmtype{reasons},\mmmtype{conditions},\mmmtype{glossary},
\mmmtype{experimentalProtocol}, \mmmtype{measure}, \mmmtype{pointer},\mmmtype{document},  \mmmtype{default}\}  %\subsetneq {\bf T}_{\bf P}
.$$

% UNDIRECTED HYPERGRAPH
Pens are similar to edges in a  hypergraph. 
Let $p=(i,l,x,S,t)\in {\bf C}_{{\bf P}} $ be an arbitrary pen where $S\in {\cal P}({{\bf I}})$ is a set of landmark identifiers. Abusing language, for any landmark identifier $j\in S$, we say that the pen $p$ \textit{contains} the landmark $j$. 
A landmark can be contained in multiple pens, i.e., pens can overlap.
\medskip

Here are some examples of pen contributions:
\begin{itemize}
    \item  $(\contributionID{definitionpen}, \textrm{""},  \emptyset,\underbrace{\{\overbrace{\ref{blue}, \ref{daytime},\ref{bluedefinition}}^{\textrm{contents}}\},\mmmtype{definition}}_{\textrm{abstract type }\in\,{\bf P}} )\in {\bf C_P}$ is a \mmmtype{definition} type of pen. 
    
    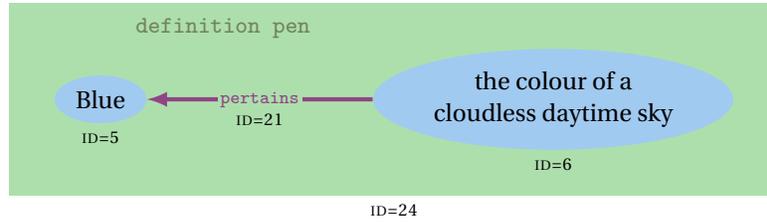
\begin{figure}[H]
        \centering
        \begin{tikzpicture}
            \node[existence] (blue) {Blue};
            \node[existence,right=3cm of blue] (daytime) {the colour of a\\ cloudless daytime  sky};
            \node [fill=pencolor, inner sep=6mm, fit={(blue) (daytime)}] (pen) {};
            \node[existence] (blue) {Blue};
            \node[metadata,below=1mm of blue] {\mmmid{blue}};
            \node[existence,right=3cm of blue] (daytime) {the colour of a\\ cloudless daytime  sky};
            \node[metadata,below=1mm of daytime] {\mmmid{daytime}};
            \draw[pertains] (daytime) -- (blue)  node[edgelabel,midway,yshift=-3pt] (bluedefinition) {\colorbox{pencolor}{\mmmtype[pertainscolor]{pertains}}\\\mmmid{bluedefinition}} ;
            \node[below left=1mm and 10mm  of pen.north,font=\small] {\mmmtype[pencolorSTR]{definition pen}};
            \node[metadata,below=1mm of pen] {\mmmid{definitionpen}};
        \end{tikzpicture}
        \caption{A \mmmtype{definition} pen used to specify that node \ref{daytime} not only characterises the concept "Blue" of node \ref{blue}, it defines it.  Not everyone has to agree with this definition. Someone else could include node \ref{blue} in a different \mmmtype{definition} pen. }
        \label{definition-pen}
        \begin{tikzpicture}
            \node[existence] (blue) {Blue};
            \node[existence,right=3cm of blue] (daytime) {the colour of a\\ cloudless daytime  sky};
            \node[existence,below=2cm of blue] (elec) {an electromagnetic wave\\ of wavelength\\ between 450 and 500 nm};
            \node [fill=pencolor, fill opacity=0.7, inner sep=6mm, fit={(blue) (daytime)}] (pen) {};
            \node [fill=pencolor, fill opacity=0.7, inner sep=4mm, fit={(blue) (elec)}] (penE) {};
            \node[existence] (blue) {Blue};
            \node[existence,right=3cm of blue] (daytime) {the colour of a\\ cloudless daytime  sky};
            \node[existence,below=2cm of blue] (elec) {an electromagnetic wave\\ of wavelength\\ between 450 and 500 nm};
            \draw[pertains] (daytime) -- (blue)  node[edgelabel,midway] (bluedefinition) {\colorbox{pencolor}{\mmmtype[pertainscolor]{pertains}}
            %\\\mmmid{bluedefinition}
            } ;
            \draw[pertains] (elec) -- (blue)  node[edgelabel,midway,pos=0.4] (bluedefinition) {{\mmmtype[pertainscolor]{\colorbox{pencolor!70}{pertai}\colorbox{pencolor}{ns}}}} ;
            \node[below left=1mm and 10mm  of pen.north,font=\small] {\mmmtype[pencolorSTR]{definition pen}};
            \node[left=-3mm  of penE.west,font=\small,rotate=90,xshift=25mm] {\mmmtype[pencolorSTR]{definition pen}};
        \end{tikzpicture}
    \end{figure}

    \item  $(\contributionID{defaultpen}, \textrm{""},  \{"@\textrm{colour names}"\},{\{\ref{blue}, \ref{turquoise},\ref{bleu},\ref{white},\ref{ENFR},\ref{blueturquoise}\},\mmmtype{default}} )\in {\bf C_P}$ is a \mmmtype{default}   pen tagged $"@\textrm{colour names}"$.
    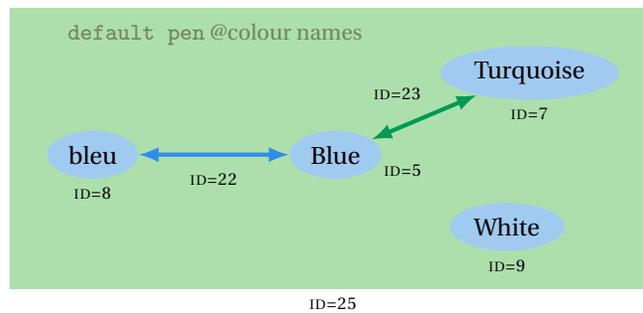
\begin{figure}[H]
        \centering
        \begin{tikzpicture}
            \node[existence] (blue) {Blue};
    \node[existence,above right=6mm and 1.3cm of blue] (turquoise) {Turquoise};
    \node[existence,left=2cm of blue] (bleu) {bleu};
    \node[existence,below right=5mm and 1.3cm of blue] (white) {White};
    \node [fill=pencolor, inner sep=5mm, fit={(blue) (bleu) (turquoise) (white)}] (pen) {};
         
            % Blue <--> Turquoise           DIFFERSFROM
            \path [draw,differsfrom] (blue) to        node (blueturquoise) {}     (turquoise);
            \node[metadata,above left=1mm and -1mm of blueturquoise] {\mmmid{blueturquoise}};
            %
             % Blue <--> Bleu
             \path [draw,equates] (blue) to   node (bluebleu) {}       (bleu);
             \node[metadata,below=1mm of bluebleu] {\mmmid{ENFR}};
    \node[existence] (blue) {Blue};
    \node[metadata,below right=-1mm and 2mm of blue] {\mmmid{blue}};
    \node[existence,above right=6mm and 1.3cm of blue] (turquoise) {Turquoise};
    \node[metadata,below=1mm of turquoise] {\mmmid{turquoise}};
    
    \node[existence,left=2cm of blue] (bleu) {bleu};
    \node[metadata,below=1mm of bleu] {\mmmid{bleu}};
    \node[existence,below right=5mm and 1.3cm of blue] (white) {White};
    \node[metadata,below=1mm of white] {\mmmid{white}};
    \node[below left=1mm and -5mm  of pen.north,font=\small] {\color{pencolorSTR}\mmmtype[pencolorSTR]{default pen} $@$colour names};
    \node[metadata,below=1mm of pen] {\mmmid{defaultpen}};
        \end{tikzpicture}
        \caption{A \mmmtype{default} pen for colour names, containing previously defined contributions.}
        \label{fig-colourpen}
    \end{figure}
    
\end{itemize}

Pens can contain any type of landmarks. Pen \ref{definitionpen} above for instance contains two vertices and an edge. Pens can also contain other pens.

% ============================================================== %
% ========                                              ======== %
% ============================================================== %
\subsubsection{The Pit}\label{pit}

The pit denoted $\bot$ is a very special kind of landmark in the landscape, the only landmark that is not a user contributed contribution. There only is one pit for all users. All users see the pit identified with the same identifier.  The pit represents absurdity. It plays an essential role in our quality management system. See \S\ref{censoring}.
\bigskip

We have defined the different kinds of landmarks and  their attributes. This concludes our definition of the MMM format. A JSON schema formalisation of the MMM format, namely JSON-MMM exists, cf \cite{MMMJSON}.  

\subsection{Areas \textit{etc}}

Now we look into sets of contributions. We already have seen in \S\ref{landscape} that a set of contributions including the pit constitutes a landscape.
% ============================================================== %
% ========               AREAS                          ======== %
% ============================================================== %
\subsubsection{Landscapes, Areas, and Territories}
\label{territory}\label{area} %\label{path}

A  \hyperref[landscape]{MMM \landscape{}}  ${\bf N}$   is  a set of landmarks, necessarily containing the pit landmark~$\bot$. \textbf{Areas} are a generalisation of  landscapes. An area is any set of landmarks not necessarily containing $\bot$.
\medskip

\textbf{The MMM} -- i.e. the structured digital version of the record   that I propose to materialise (cf \S\ref{motivation}) a.k.a. the intermediary epistemic web (cf \S\ref{missing}) -- is the reunion of all landscapes. It is  denoted ${\bf N}^\star$. 
Some areas  of  ${\bf N}^\star$ are private. Others are public. 
The reunion of all public landmarks  is called \textbf{the public  MMM} and denoted ${\bf N}^\star_p$.  \medskip

${\bf N}^\star$ and ${\bf N}^\star_p$ are \textit{collective} landscapes: their landmarks are contributed by different users and can be stored in a distributed manner.
Other smaller landscapes may also be  collective in that sense. \medskip

A special kind of landscape is a \textbf{territory}. A territory is  associated with a human user. It comprises $\bot$ and the set of contributions 
{that the user is acquainted with. Typically a territory is mostly stored locally, on the user's device(s), although parts of it may be stored remotely. The part that is stored locally is called the \textbf{local territory}. Users need not keep local copies of all landmarks they are acquainted with as  they may have little interest in some landmarks they have visited. Also,   they may trust other users to keep copies of the landmarks they are interested in and not store them themselves (cf \SMref{implementation}).}

\subsubsection{Paths}\label{path}
A MMM \textbf{path} is a series of MMM landmarks $l_1,l_2,\ldots,l_n$
such that for any even integer $i<n$, the contribution $l_i$ is an edge between landmark $l_{i-1}$ and landmark $l_{i+1}$. By default, edges in an MMM path don't need to all have the same direction. If they do, then we say the path is directed. Note that because edges can have edges as endpoints, a MMM path can have several successive edges. 

% ============================================================== %
% ========               MUTABLE PENS                   ======== %
% ============================================================== %
\subsubsection{Mutable Pens and Contributions}\label{mutable}

The set $S$ of contents of a pen $p$ is immutable (except for the local obsolescence mechanism, cf \S\ref{obsoleting}).
We can nonetheless define \textit{mutable} pens. Contrary to normal, immutable  pens defined in \S\ref{pens}, \textbf{mutable pens} aren't atomic contributions. They are \textit{sets} of contributions containing exactly one  (possibly empty) pen $p$ and  other contributions linked to $p$ by a \mmmtype{pennedIn} edge.  
The contents of the  mutable pen  are the contributions that are linked to the pen $p$  by a  \mmmtype{pennedIn} edge,  and if any, the contents of $p$. Contents can be added to a mutable pen. Using the obsoleting mechanism described in \S\ref{obsoleting}, contents can also be removed. 
Mutable pens are typically for delineating a mutable collaborative area of the landscape that plays the role of an editable document. Multiple users can include contents in a mutable pen.
Application code can offer the possibility to the user to remain "working inside" the mutable pen/document, meaning that all contributions created by the user during the work session are automatically linked with a \mmmtype{pennedIn} edge to the pen.   \medskip

\begin{figure}[H]
    \centering
    \begin{tikzpicture}
        \draw[pencolor,dashed, opacity=0.9] (0,0) rectangle ++(12,4.2);
        \node[pen,inner sep=6mm] (pen) at (6,3) {
            %\mmmtype[pencolorSTR]{default} 
            \hspace{1cm}Shopping List\hspace{1cm}};
        \node[existence,below=1cm of pen] (rice) {Rice};
        \node[existence,left=2cm of rice] (oranges) {oranges};
        \node[existence,right=2cm of rice] (bulbs) {Light bulbs};
     \path [draw,pennedin] (rice) to      node[fill=white,pos=0.4]  {\mmmtype[pencolor]{pennedin}}     (pen);
        \path [draw,pennedin] (oranges) to      node[fill=white]  {\mmmtype[pencolor]{pennedin}}     (pen);
        \path [draw,pennedin] (bulbs) to      node[fill=white]  {\mmmtype[pencolor]{pennedin}}     (pen);
\node[below left=-5mm and -22mm of pen.south west,font=\small] {\color{pencolorSTR}\mmmtype[pencolorSTR]{default pen}};
\node[font=\small,anchor=west, opacity=0.9] at (0.2,0.2) {\color{pencolorSTR}\mmmtype[pencolorSTR]{mutable pen}};
    \end{tikzpicture}
    \caption{{A mutable pen.   {UI application code} can visually represent atomic and mutable pens in a similar way although they are different objects. I recall that  this proposal is to deal with unstructured, typically disputable, information. Shopping lists are thus not typical content of interest to this proposal, but they make for simple illustrations.  }}
    \label{mutablepen}
\end{figure}
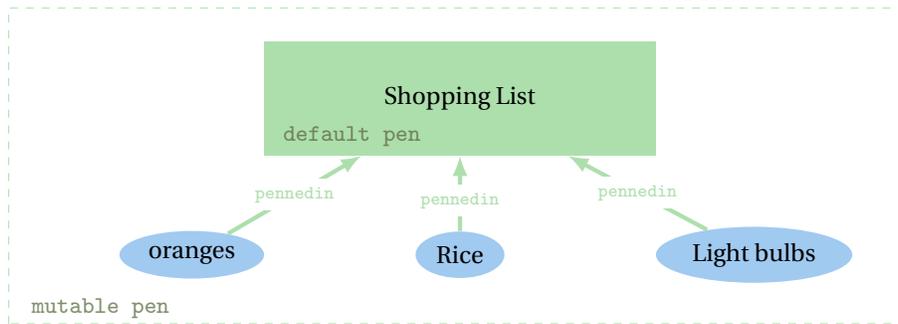

% ============================================================== %
% ========               MUTABLE CONTRIBUTIONS          ======== %
% ============================================================== %
By default, contribution labels and types are immutable. Pens together with the obsoleting mechanism described in \S\ref{obsoleting} can 
also
be used to define \textbf{mutable contributions}.  

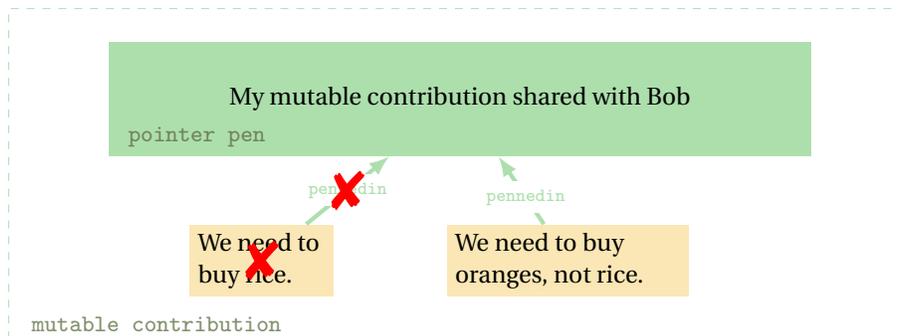
\begin{figure}[H]
    \centering
    \begin{tikzpicture}\draw[pencolor,dashed, opacity=0.9] (0,0) rectangle ++(12,4.4);
        \node[pen,inner sep=6mm] (pen) at (6,3.2) {
            \hspace{1cm}My mutable contribution shared with Bob\hspace{1cm}};
        \node[narrative,below left=0.9cm and -30mm of pen,text width=1.7cm] (rice) {We need to buy rice.};
        \node[narrative,right=15mm of rice,text width=3cm] (oranges) {We need to buy oranges, not rice.};
     \path [draw,pennedin] (rice) to      node[fill=white] (riceedge) {\mmmtype[pencolor]{pennedin}}     (pen);
        \path [draw,pennedin] (oranges) to      node[fill=white,pos=0.4]  {\mmmtype[pencolor]{pennedin}}     (pen);
\node[below left=-5mm and -22mm of pen.south west,font=\small] {\color{pencolorSTR}\mmmtype[pencolorSTR]{pointer pen}};
\node[text=red,font=\huge] at (rice) {\nope};
\node[text=red,font=\huge] at (riceedge) {\nope};
\node[font=\small,anchor=west, opacity=0.9] at (0.2,0.2) {\color{pencolorSTR}\mmmtype[pencolorSTR]{mutable contribution}};
    \end{tikzpicture}
    \caption{A mutable contribution. Contributions marked with {\color{red}\nope} are \mmmmark{obsolete}d contributions. They are not immediately deleted but eventually will (cf \S\ref{obsoleting}). Application code is required to handle marks and ensure desired  user experience in agreement with the sharing mechanisms presented in \S\ref{activities-sharing}.}
    \label{mutablecontribution}
\end{figure}

The MMM format isn't designed for real-time sharing of fine-grained information (see \S\ref{collaborating}). MMM contributions are meant to be final. The MMM is mostly an add-only space equipped with a slow obsoleting mechanism described in \S\ref{obsoleting}.   

\section{Landscape Based Activities}\label{activities}

Section \ref{format} above defined the MMM format.
Next, Section \ref{activities} here discusses how to use the elements of this format. Three kinds of landscape based activities are presented, namely, editing the landscape, exploring the landscape, and sharing  landmarks. 
The supplementary material \ref{implementation}  describes the technological infrastructure and tooling to assist with MMM landscape based activities. 
An essential part of the proposition is the  plan to support different user interfaces (UIs) in order to accommodate different epistemic approaches and interests towards information.

\subsection{Landscape Editing Activities}\label{activities-editing}
% use cases. 

% ============================================================== %
% ========                                              ======== %
% ============================================================== %
\subsubsection{Contributing }\label{contributing}

Contributing means recording one or several new contributions on a landscape. The most basic contributing actions are: posting a question using a $\mmmtype{question}$ node,  posting a vocabulary term using an $\mmmtype{existence}$ node, posting a statement or posting anything arbitrary  using a $\mmmtype{narrative}$ node. Other basic contributions are to post an edge -- e.g. a default $\mmmtype{relate}$ edge or an $\mmmtype{equates}$ edge -- between two existing nodes. Most contributing actions   involve multiple nodes and/or edges (cf \S\ref{annotating}). 
\medskip
 
\begin{bestpractices}{} In a context of collective documentation on a shared landscape like  ${\bf N}^\star_p$, the act of contributing should happen mainly through acts of "\hyperref[annotating]{annotation}" (see below \S\ref{annotating}). \end{bestpractices}

Contributions that are not annotations will typically contribute an isolated contribution or group of contributions. 

% ============================================================== %
% ========                                              ======== %
% ============================================================== %
\subsubsection{Annotating/Improving }\label{annotating}

\begin{principle}{Information as improving events}
\label{events}
Information is regarded here as 
\textit{events} that cause the record to improve. 
\end{principle}

Note that following Design Bias \ref{continual-improvement}, improving the record does not necessarily mean adding high quality information to it. It can mean exposing low quality information to nuance and detail. \medskip

To support \citedesignbias{events}, I propose to encourage the provision of new information in the form of  \textit{annotations} to pre-existing information (see also \SMref{quality} in relation to improving the quality of information on the record). A MMM annotation is a set of MMM contributions that involves at least one edge incident on a pre-existing contribution in ${\bf N}^\star$. 
To annotate contribution   $c\in {\bf C}$  means to add at least one edge to the landscape between $c$ and another  contribution.
Annotations   complete, nuance, challenge, support, detail, update \textit{etc} information that is already in the record ${\bf N}^\star$.
Here are some  annotation patterns that I expect to be useful:
\begin{enumerate}[leftmargin=*]
    \item Question a contribution, using the "\mmmtype[questionscolor]{questions}" type of edge: % {(cf Fig. \ref{question-a-contribution})} %\checkrequirement{req-questions}

    \begin{tikzpicture}  
        \node[narrative] (n) at (2,3.2) {The sky is blue.};
        \draw[help lines,step=40mm,gray!40] (0,0) grid (4,4);
        \node[text=gray] at (2,0.3) {\sc\small before};
        \begin{scope}[xshift=4.2cm]
        \node[narrative] (n) at (2,3.2) {The sky is blue.};
        \node[question]   (cqsky) at (2,1.2) {Is the sky \textit{always} blue?};
        \draw[help lines,step=40mm,gray!40] (0,0) grid (4,4);
        \draw[questions] (cqsky) -- (n)  node[midway,yshift=-4pt,fill=white] (narranswer) {\scriptsize \mmmtype[questionscolor]{questions}} ;
        \node[text=gray] at (2,0.3) {\sc\small after};
        \end{scope}
        \end{tikzpicture} 

    \item Answer a $\mmmtype{question}$ using the "\mmmtype[answerscolor]{answers}" type of edge: %{(cf Fig \ref{answer-a-question})}
   
        \begin{tikzpicture}  
            \node[question]   (qsky)  at (2,3.2)  {What colour is the sky?};
            \draw[help lines,step=40mm,gray!40] (0,0) grid (4,4);
            \node[text=gray] at (2,0.3) {\sc\small before};
            \begin{scope}[xshift=4.2cm]
            \node[question]   (qsky)  at (2,3.2)  {What colour is the sky?};
            \node[narrative]    (n) at (2,1.2) {The sky is blue.};
            \draw[help lines,step=40mm,gray!40] (0,0) grid (4,4);
            \draw[answers]  (n) -- (qsky)  node[midway,yshift=-4pt,fill=white] {\scriptsize \mmmtype[answerscolor]{answers}} ;
            \node[text=gray] at (2,0.3) {\sc\small after};
            \end{scope}
            \node[text=gray] at (8.8,0.3) {\sc\small or};
            \begin{scope}[xshift=9.4cm]
                \node[question]   (qsky)  at (2,3.2)  {What colour is the sky?};
                \node[existence]    (blue) at (2,1.2) {Blue};
                \draw[help lines,step=40mm,gray!40] (0,0) grid (4,4);
                \draw[answers]  (blue) -- (qsky)  node[midway,yshift=-4pt,fill=white] {\scriptsize \mmmtype[answerscolor]{answers}} ;
                \node[text=gray] at (2,0.3) {\sc\small after};
                \end{scope}
            \end{tikzpicture} 

    \item Pinpoint a notable concept or tricky term used in a contribution: %\checkrequirement{req-onto}
    
        \begin{tikzpicture}  
            \node[narrative] (n) at (2,3.2) {The sky is blue.};
            \draw[help lines,step=40mm,gray!40] (0,0) grid (4,4);
            \node[text=gray] at (2,0.3) {\sc\small before};
            \begin{scope}[xshift=4.2cm]
            \node[narrative] (n) at (2,3.2) {The sky is blue.};
            \node[existence]    (blue) at (2,1.2) {Blue};
            \draw[help lines,step=40mm,gray!40] (0,0) grid (4,4);
            \draw[pertains]  (blue) -- (n)  node[midway,yshift=-4pt,fill=white] {\scriptsize \mmmtype[pertainscolor]{pertains}} ;
            \node[text=gray] at (2,0.3) {\sc\small after};
            \end{scope}
            \end{tikzpicture} 
    % \end{figure}

    \item Define a term in a contribution (I recall that the ubiquitous "\mmmtype[pertainscolor]{pertains}" type of edge can be interpreted as meaning "characterises"): %\checkrequirement{req-onto}
    
        \begin{tikzpicture}  
            \node[narrative] (n) at (2,3.4) {The sky is blue.};
            \draw[help lines,step=40mm,gray!40] (0,0) grid (4,4);
            \node[text=gray] at (2,0.3) {\sc\small before};
            \begin{scope}[xshift=4.2cm]
            \draw[gray!40] (0,0) rectangle ++(9,4);
            \node[text=gray] at (4.5,0.3) {\sc\small after};
            \node[narrative] (n) at (2,3.4) {The sky is blue.};
            \node[existence]    (blue) at (1,1.5) {Blue};
            \node[existence,right=2cm of blue,font=\small] (elec) {an electromagnetic wave\\ of wavelength\\ between 450 and 500 nm};
            % pen
            \node [fill=pencolor, inner sep=1mm, fit={(blue) (elec)}] (pen) {};  
            \node[below left=1mm and 15mm  of pen.north,font=\small] {\mmmtype[pencolorSTR]{definition pen}};
            \node[existence]    (blue) at (1,1.5) {Blue};
            \node[existence,right=2cm of blue,font=\small] (elec) {an electromagnetic wave\\ of wavelength\\ between 450 and 500 nm};
            \draw[pertains] (elec) -- (blue)  node[edgelabel,midway,pos=0.4] (bluedefinition) {\colorbox{pencolor}{\mmmtype[pertainscolor]{pertains}}} ;
            \draw[pertains] (pen) |- (n)  node[midway,pos=0.25,fill=white] (bluedefinition) {\scriptsize \mmmtype[pertainscolor]{pertains}} ;            
            \end{scope}
            \end{tikzpicture} 

Note that as annotations may be documented independently by different users (cf requirement \ref{req-collective}), some form of redundancy is probable. Below the two \mmmtype[pertainscolor]{pertains} edges incoming the \mmmtype{narrative} node, don't mean exactly the same thing, but have some overlap in meaning. The edge outgoing the pen is  more specific, but also more disputable, than the edge outgoing the \mmmtype{existence} node. 

\begin{tikzpicture}  
    \draw[help lines,step=40mm,gray!0] (0,0) grid (4,4);
    \begin{scope}[xshift=4.2cm]
    \draw[gray!0] (0,0) rectangle ++(9,4);
    \node[narrative] (n) at (2,3.4) {The sky is blue.};
    \node[existence]    (blue) at (1,1.5) {Blue};
    \node[existence,right=2cm of blue,font=\small] (elec) {an electromagnetic wave\\ of wavelength\\ between 450 and 500 nm};
    % pen
    \node [fill=pencolor, inner sep=1mm, fit={(blue) (elec)}] (pen) {};  
    \node[existence]    (blue) at (1,1.5) {Blue};
    \node[existence,right=2cm of blue,font=\small] (elec) {an electromagnetic wave\\ of wavelength\\ between 450 and 500 nm};
    \draw[pertains] (elec) -- (blue)  ;
    \draw[pertains] (pen) |- (n)   ;   
    \draw[pertains]  (blue) -- ([xshift=2.3mm]n.south west);         
    \end{scope}
    \end{tikzpicture} 

    \item Ask a question about a term used in a contribution:
    
    \begin{tikzpicture}  
        \node[narrative] (n) at (2,3.4) {The sky is blue.};
        \draw[help lines,step=40mm,gray!40] (0,0) grid (4,4);
        \node[text=gray] at (2,0.3) {\sc\small before};
        \begin{scope}[xshift=4.2cm]
        \draw[gray!40] (0,0) rectangle ++(4,4);
        \node[text=gray] at (2,0.3) {\sc\small after};
        \node[narrative] (n) at (2,3.4) {The sky is blue.};
        \node[existence]    (blue) at (3.2,2.2) {Blue};
        \node[question,font=\small] (q) at (2,1) {What kind of blue\\ is the sky?};
        \draw[pertains] ([xshift=10mm]q.north west) -- ([xshift=10mm,yshift=1.7cm]q.north west) node[edgelabel,pos=0.4,yshift=1.5mm] {\mmmtype[pertainscolor]{pertains}}; 
        \draw[pertains] ([yshift=1mm,xshift=1pt]blue.west) -| ([xshift=-10mm]n.south east) node[pos=0.7] (nedge) {};
        \draw[pertains] ([yshift=-1mm,xshift=1pt]blue.west) -| ([xshift=-10.7mm]q.north east) ;
        \end{scope}
        \node[text=gray] at (8.8,0.3) {\sc\small or};
        \begin{scope}[xshift=9.4cm]
        \draw[gray!40] (0,0) rectangle ++(4,4);
        \node[text=gray] at (2,0.3) {\sc\small after};
        \node[narrative] (n) at (2,3.4) {The sky is blue.};
        \node[existence]    (blue) at (3.2,2.2) {Blue};
        \node[question,font=\small] (q) at (2,1) {What kind of blue\\ is the sky?};
        \draw[questions] ([xshift=10mm]q.north west) -- ([xshift=10mm,yshift=1.7cm]q.north west) node[edgelabel,pos=0.4,yshift=1.5mm] {\mmmtype[questionscolor]{questions}}; 
        \draw[pertains] ([yshift=1mm,xshift=1pt]blue.west) -| ([xshift=-10mm]n.south east) node[pos=0.7] (nedge) {};
        \draw[pertains] ([yshift=-1mm,xshift=1pt]blue.west) -| ([xshift=-10.7mm]q.north east) ;
         
        \end{scope}
    \end{tikzpicture} 

    Note again that the MMM datamodel being deliberately flexible, there can be several ways to document the same contribution. 

    \item Nuance a contribution using a "\mmmtype[nuancescolor]{nuances}" type of edge:

    \begin{tikzpicture}  
        \draw[gray!40] (0,0) rectangle ++(4,4);
        \node[text=gray] at (2,0.3) {\sc\small before};
        \node[narrative] (n) at (2,3.2) {The sky is blue.};
        \begin{scope}[xshift=4.2cm]
        \draw[gray!40] (0,0) rectangle ++(4,4);
        \node[text=gray] at (2,0.3) {\sc\small after};
        \node[narrative] (n) at (2,3.2) {The sky is blue.};
        \node[narrative,font=\small]   (nn) at (2,1.2) {The sky isn't always blue.\\ It can be white and black.};
        \draw[nuances] (nn) -- (n)  node[midway,yshift=-4pt,fill=white] {\scriptsize \mmmtype[nuancescolor]{nuances}} ;
        \end{scope}
    \end{tikzpicture} 

    \item Provide supporting details to a contribution:

    \begin{tikzpicture}  
        \draw[gray!40] (0,0) rectangle ++(4,4);
        \node[text=gray] at (2,0.3) {\sc\small before};
        \node[narrative] (n) at (2,3.2) {The sky is blue.};
        \begin{scope}[xshift=4.2cm]
        \draw[gray!40] (0,0) rectangle ++(4,4);
        \node[text=gray] at (2,0.3) {\sc\small after};
        \node[narrative] (n) at (2,3.2) {The sky is blue.};
        \node[narrative,font=\small]   (nn) at (2,1.2) {Blue light in the sky\\ scatters more.};
        \draw[details] (nn) -- (n)  node[midway,yshift=-4pt,fill=white] { \mmmtype[detailscolor]{supports}} ;
        \end{scope}
    \end{tikzpicture} 

    \item Give an example of the object of a contribution:
    
    \begin{tikzpicture}  
        \draw[gray!40] (0,0) rectangle ++(4,4);
        \node[text=gray] at (2,0.3) {\sc\small before};
        \node[narrative] (n) at (2,3.2) {Lots of things are blue.};
        % \node[existence] (n) at (2,3.2) {Blue things};
        %
        \begin{scope}[xshift=4.2cm]
        \draw[gray!40] (0,0) rectangle ++(4,4);
        \node[text=gray] at (2,0.3) {\sc\small after};
        \node[narrative] (n) at (2,3.2) {Lots of things are blue.};
        % \node[existence] (n) at (2,3.2) {Blue things};
        \node[existence]   (nn) at (2,1.2) {The sky};
        \draw[instantiates] (nn) -- (n)  node[midway,yshift=-4pt,fill=white] { \mmmtype[instantiatescolor]{instantiates}} ;
        \end{scope}
    \node[text=gray] at (8.8,0.3) {\sc\small or};
    \begin{scope}[xshift=9.4cm]
        \draw[gray!40] (0,0) rectangle ++(4,4);
        \node[text=gray] at (2,0.3) {\sc\small before};
        \node[existence] (n) at (2,3.2) {Colour};
        \begin{scope}[xshift=4.2cm]
        \draw[gray!40] (0,0) rectangle ++(4,4);
        \node[text=gray] at (2,0.3) {\sc\small after};
        \node[existence] (n) at (2,3.2) {Colour};
        \node[existence]   (nn) at (2,1.2) {Blue};
        \draw[instantiates] (nn) -- (n)  node[midway,yshift=-4pt,fill=white] { \mmmtype[instantiatescolor]{instantiates}} ;
        \end{scope}
    \end{scope}
    \end{tikzpicture} 

    \item Characterise  a contribution by its properties or contents using a \pertains{} edge: %, cf \S\ref{unidir}):
  
    \begin{tikzpicture}  
        \draw[gray!40] (0,0) rectangle ++(4,4);
        \node[text=gray] at (2,0.3) {\sc\small before};
        \node[existence] (n) at (2,3.2) {Blue};
        \begin{scope}[xshift=4.2cm]
        \draw[gray!40] (0,0) rectangle ++(4,4);
        \node[text=gray] at (2,0.3) {\sc\small after};
        \node[existence] (n) at (2,3.2) {Blue };
        \node[existence,font=\scriptsize]   (nn) at (2,1.2) {wavelength between\\ 450 and 500 nm};
        \draw[pertains] (nn) -- (n)  node[midway,yshift=-4pt,fill=white] { \mmmtype[pertainscolor]{pertains}} ;
        \end{scope}
        \node[text=gray] at (8.8,0.3) {\sc\small or};
        \begin{scope}[xshift=9.4cm]
        \draw[gray!40] (0,0) rectangle ++(4,4);
        \node[text=gray] at (2,0.3) {\sc\small before};
        \node[existence,font=\small,inner sep=0pt] (n) at (2,3.2) {MMM\\[-3pt]landscape};
        \begin{scope}[xshift=4.2cm]
        \draw[gray!40] (0,0) rectangle ++(4,4);
        \node[text=gray] at (2,0.3) {\sc\small after};
        \node[existence,font=\small,inner sep=0pt] (n) at (2,3.2) {MMM\\[-3pt]landscape};
        \node[existence,font=\small,inner sep=0pt]   (nn) at (2,1.2) {MMM\\[-3pt]landmark};
        \draw[pertains] (nn) -- (n)  node[midway,yshift=-4pt,fill=white] { \mmmtype[pertainscolor]{pertains}} ;
        \end{scope}
        \end{scope}
    \end{tikzpicture} 

    Please note again that concrete types of edges such as \pertains{} are not enough to specify the meaning conveyed by the edge. Meaning has to be specified through the edge's label or tag. In the left case above, the \pertains{} edge could be labelled or tagged "characterises" or "details". In the right case above, the \pertains{} edge could be labelled or tagged "involved in". 

    \item Reformulate or update a contribution $c$ \textit{without changing its meaning} using an \mmmtype[equatescolor]{equates} edge: %\checkrequirement{req-reformulate}
    
    \begin{tikzpicture}  
        \draw[gray!40] (0,0) rectangle ++(4,4);
        \node[text=gray] at (2,0.3) {\sc\small before};
        \node[narrative] (n) at (1.5,3.2) {The sky is ble.};
        \node[existence]    (blue) at (3.2,2.2) {Blue};
        \draw[pertains] ([yshift=1mm,xshift=2pt]blue.west) -| ([xshift=-5mm]n.south east) ;
        \begin{scope}[xshift=4.2cm]
        \draw[gray!40] (0,0) rectangle ++(4,4);
        \node[text=gray] at (2,0.3) {\sc\small after};
        \node[narrative] (n) at (1.5,3.2) {The sky is ble.};
        \node[existence]    (blue) at (3.2,2.2) {Blue};
        \draw[pertains] ([yshift=1mm,xshift=2pt]blue.west) -| ([xshift=-5mm]n.south east) ;
        \node[narrative] (nn) at (1.5,1.2) {The sky is blue.};
        % \draw[pertains] ([yshift=-1mm,xshift=2pt]blue.west) -| ([xshift=-6mm]nn.north east) ;
        \draw[equates] (nn) -- (n)  node[midway,fill=white] { \mmmtype[equatescolor]{equates}} ;
        \end{scope}
        \node[text=gray] at (8.8,0.3) {\sc\small or};
        \begin{scope}[xshift=9.4cm]
            \draw[gray!40] (0,0) rectangle ++(4,4);
            \node[text=gray] at (2,0.3) {\sc\small after};
            \node[narrative,opacity=0.6] (n) at (1.5,3.2) {The sky is ble.};
            \node[existence]    (blue) at (3.2,2.2) {Blue};
            \draw[pertains,opacity=0.6] ([yshift=1mm,xshift=2pt]blue.west) -| ([xshift=-5mm]n.south east) node[pos=0.7] (nedge) {};
            \node[narrative] (nn) at (1.5,1.2) {The sky is blue.};
            \draw[pertains] ([yshift=-1mm,xshift=2pt]blue.west) -| ([xshift=-6mm]nn.north east) ;
            \draw[equates] (nn) -- (n)  node[midway,fill=white] { \mmmtype[equatescolor]{equates}} ;
            \node[text=red,font=\huge] at (n) {\nope};
            \node[text=red,font=\huge] at (nedge) {\nope};
        \end{scope}
    \end{tikzpicture}

    Typically, the \equates{} edge can be labelled with the reason of the reformulation -- e.g. "corrected a typo". Application code can facilitate this for the user. If the new formulation is to \textit{replace} the original formulation as is the case in versioning (cf \S\ref{versioning}), edges incoming the old version have to be duplicated and redirected towards the new. In the example on the right above, {\color{red}\nope} marks \mmmmark{obsolete}d contributions that will eventually be deleted. Because MMM contributions are meant to be shared, old versions may be in use by other users and can't be immediately deleted locally. The \equates{} edge is necessary to maintain the link between old and new version until all users have been made aware of the replacement of the old by the new.  

    \item Red-flag a contribution (see details given in \S\ref{redflag}):

    \begin{tikzpicture}  
        \draw[gray!40] (0,0) rectangle ++(4,4);
        \node[text=gray] at (2,0.3) {\sc\small before};
        \node[question] (q) at (2,3.5) {What colour is the sky?};
        \node[narrative,font=\small] (n) at (2,1.2) {Chemtrails are for mind\\ and weather control.};
        \draw[answers]  ([xshift=10mm]n.north west) -- ([xshift=10mm,yshift=16mm]n.north west)  node[midway,yshift=-4pt,fill=white] {\scriptsize \mmmtype[answerscolor]{answers}} ;
        \begin{scope}[xshift=4.2cm]
        \draw[gray!40] (0,0) rectangle ++(4,4);
        \node[text=gray] at (2,0.3) {\sc\small after};
        \node[question] (q) at (2,3.5) {What colour is the sky?};
        \node[narrative,font=\small] (n) at (2,1.2) {Chemtrails are for mind\\ and weather control.};
        \draw[answers]  ([xshift=10mm]n.north west) -- ([xshift=10mm,yshift=16mm]n.north west)  node[midway] (a) {} ;
        \node[star,star points=12,fill=black, text=white,inner sep=1pt,right=15mm of a] (pit)  {$\bot$};
        \draw[equates]  (a) -- (pit)  node[midway,yshift=3pt] (a) {\mmmtype[equatescolor]{equates}} ;
        \end{scope}
    \end{tikzpicture} 

    \item Reference a contribution: \label{referencing}

    \begin{tikzpicture}  
        \draw[gray!40] (0,0) rectangle ++(4,4);
        \node[text=gray] at (2,0.3) {\sc\small before};
        \node[narrative,font=\scriptsize] (n) at (2,3.2) {Blue light in the sky\\ scatters more.};
        \begin{scope}[xshift=4.2cm]
        \draw[gray!40] (0,0) rectangle ++(4,4);
        \node[text=gray] at (2,0.3) {\sc\small after};
        \node[narrative,font=\scriptsize] (n) at (2,3.2) {Blue light in the sky\\ scatters more.};
        \node[narrative,font=\scriptsize]   (nn) at (2,0.8) {\href{https://scijinks.gov/blue-sky/}{https://scijinks.gov/blue-sky/}};
        \draw[details] (nn) -- (n)  node[midway,xshift=-6mm] { \mmmtype[detailscolor]{supports}} ;
        \end{scope}
        \node[text=gray] at (8.8,0.3) {\sc\small or};
        \begin{scope}[xshift=9.4cm]
            \draw[gray!40] (0,0) rectangle ++(4,4);
            \node[text=gray] at (2,0.3) {\sc\small before};
            \node[narrative,font=\scriptsize,text width=3.6cm] (n) at (2,2.8) {Gases and particles in Earth's atmosphere scatter sunlight in all directions. Blue light is scattered more than other colours because it travels as shorter, smaller waves. This is why we see a blue sky most of the time. };
            \begin{scope}[xshift=4.2cm]
            \draw[gray!40] (0,0) rectangle ++(4,4);
            \node[text=gray] at (2,0.3) {\sc\small after};
            \node[narrative,font=\scriptsize,text width=3.6cm] (n) at (2,2.8) {Gases and particles in Earth's atmosphere scatter sunlight in all directions. Blue light is scattered more than other colours because it travels as shorter, smaller waves. This is why we see a blue sky most of the time. };
            \node[existence,font=\scriptsize,inner sep=1pt,text width=2.5cm,text height=1em]   (nnex) at (2,0.8) {};
            \node[font=\scriptsize]   (nn) at (2,0.8) {\href{https://scijinks.gov/blue-sky/}{https://scijinks.gov/blue-sky/}};
            \draw[pertains] (nn) -- (n)  node[pos=0.5,xshift=-6mm,yshift=-1mm] { \mmmtype[pertainscolor]{pertains}} ;
            \end{scope}
        \end{scope}
    \end{tikzpicture} 
    
    An author  documented in the authorship set of a contribution $c$, isn't necessarily the original author of the text documented in $c$'s label. Alice may document in the MMM that contribution $c$ is \mmmtype[detailscolor]{supported} by reference $r$. The author of  $r$ might not agree with Alice on that. In the case pictured on the right above, the \pertains{} edge links  reference $r$ to  a quote extracted from the referenced document. In this case, the \pertains{} edge conveys the meaning that the quote has the property of coming from the document referenced by $r$: the quote is \textit{characterised} by its source.
%
    
    % , including an edge linking a bibliographical reference, 

    \begin{tikzpicture}  
        \node[text width= 9cm] at (0,3.4) {In contrast, in the case pictured on the right, the  \pertains{} edge conveys the fact that the quote is a part of the referenced document.  };
        \node[text width= 9cm] at (0,1.7) {By convention, MMM edges used for linking a contribution to its reference can be labelled or tagged "reference". UI application code can promote the documentation of references in standard formats such as {APA, MLA, bibtex}.};
        \begin{scope}[xshift=4.9cm]
            \draw[gray!40] (0,0) rectangle ++(4,4);
            \node[text=gray] at (2,0.3) {\sc\small before};
            \node[narrative,font=\scriptsize,text width=3.6cm] (n) at (2,2.8) {Gases and particles in Earth's atmosphere scatter sunlight in all directions. Blue light is scattered more than other colours because it travels as shorter, smaller waves. This is why we see a blue sky most of the time. };
            \begin{scope}[xshift=4.2cm]
            \draw[gray!40] (0,0) rectangle ++(4,4);
            \node[text=gray] at (2,0.3) {\sc\small after};
            \node[narrative,font=\scriptsize,text width=3.6cm] (n) at (2,2.8) {Gases and particles in Earth's atmosphere scatter sunlight in all directions. Blue light is scattered more than other colours because it travels as shorter, smaller waves. This is why we see a blue sky most of the time. };
            \node[existence,font=\scriptsize,inner sep=1pt,text width=2.5cm,text height=1em]   (nnex) at (2,0.8) {};
            \node[font=\scriptsize]   (nn) at (2,0.8) {\href{https://scijinks.gov/blue-sky/}{https://scijinks.gov/blue-sky/}};
            \draw[pertains] (n) -- (nn)  node[pos=0.3,xshift=-6mm,yshift=-1mm] { \mmmtype[pertainscolor]{pertains}} ;
            \end{scope}
        \end{scope}
    \end{tikzpicture}

\end{enumerate}
\bigskip

Annotation is recursive in the sense that any contribution that takes part in an annotation can itself be annotated.\medskip
% \checkrequirement{req-recursive}

\begin{itemize}[leftmargin=*]
    \item[]

    \begin{tikzpicture}  
        \draw[gray!40] (0,0) rectangle ++(4,4);
        \node[text=gray] at (2,0.3) {\sc\small before};
        \node[narrative,font=\scriptsize] (n) at (2,3.2) {Blue light in the sky\\ scatters more.};
        \node[narrative,font=\scriptsize]   (nn) at (2,0.8) {\href{https://scijinks.gov/blue-sky/}{https://scijinks.gov/blue-sky/}};
        \draw[details] (nn) -- (n)  node[midway,xshift=-6mm] { \mmmtype[detailscolor]{supports}} ;
        \begin{scope}[xshift=4.2cm]
            \draw[gray!40] (0,0) rectangle ++(7,4);
            \node[text=gray] at (3.5,0.3) {\sc\small after};
            \node[narrative,font=\scriptsize] (n) at (2,3.2) {Blue light in the sky\\ scatters more.};
            \node[narrative,font=\scriptsize]   (nn) at (2,0.8) {\href{https://scijinks.gov/blue-sky/}{https://scijinks.gov/blue-sky/}};
            \draw[details] (nn) -- (n)  node[midway,xshift=-6mm] (e) { \mmmtype[detailscolor]{supports}} ;
            \node[narrative,text width=2.5cm,font=\scriptsize,right=2cm of e] (nn)  {The  NOAA SciJinks website is more precise. };
            \draw[nuances] (nn) -- (e)  node[midway,yshift=1mm]  { \mmmtype[nuancescolor]{nuances}} ;
        \end{scope}
    \end{tikzpicture} 
\end{itemize}

The  \ref*{referencing} annotation patterns listed above  are suggestions. I recall that except for the \mmmtype{pennedIn} edge, there are no grammar rules for assembling MMM contributions in a landscape. 
I propose that user interfaces of MMM editing tools  be designed to facilitate the  annotation patterns listed above. They may  provide shortcuts to  some annotating patterns, and possibly emphasise some patterns more than others depending on what annotations are more typical to a given user segment served by the given UI. 
\bigskip

% ============================================================== %
% ========                METADATA                      ======== %
% ============================================================== %
% \subsubsection{A preliminary remark on information and meta-data}
\label{metadata-in-activity}

The metadata  of a MMM contribution (id, authorship set, mark set, status, timestamp) can't be annotated. MMM edges refer to the main informational payload of their contribution endpoints. And the main informational payload of a contribution is expressed in the contribution's label, type and possibly tag set. 
Consider the node with ID 2 given as an example in \S\ref{vertices} above. 
While you can comment on the statement "The sky is blue." given by the label of the node, you can't comment on the fact that the ID of that contribution is 2. Similarly, while you can  comment on the fact that edge with ID 12 has concrete type $\mmmtype[relatecolor]{relate}$ (see for instance \S\ref{redflag}) you can't comment on the fact that its documented contributor is Anne Martin. 
\medskip

\begin{bestpractices}{} If you wish to use some metadata as information, then document it explicitly as information: on the MMM, this means  document it  as a contribution which has its own metadata.
\end{bestpractices}

% ============================================================== %
% ========               IMPLANT                        ======== %
% ============================================================== %
\subsubsection{Implanting }\label{implanting}

Implanting means contributing edges between a new contribution and pre-existing contributions on the landscape. It is comparable to annotating. Annotating is aimed at improving pre-existing content by way of new content. Implanting is aimed at using old content to  provide context and visibility  to new contributions (cf \SMref{implantation}).\medskip

\begin{bestpractices}{} 
Implant much and {well}. Poorly implanted contributions will have less visibility as the  pathways leading to them are rare and/or at risk  of being \hyperref[redflag]{red-flagged} and \hyperref[obsoleting]{obsoleted}. 
\end{bestpractices}

Implanting is central to my proposal. In the distributed MMM network, the better a contribution is implanted (not just the \textit{more} it is), the more \hyperref[finding]{visible}  it is likely to be: the more likely it is that a peer will discover it by following a path to it (cf \S\ref{finding}). Implantation is also key to redundancy management. The better implanted a new contribution is, the easier it is to identify how the information it conveys relates to other recorded information and possibly overlaps some of it. Before a contribution $c$ is recorded for good on the landscape, and before it propagates to other users' territories, contributions in the neighbourhood of $c$ where $c$ is implanted can help determine the value that $c$ adds to the area. Good, early implantation can also spare authors  the effort of documenting atomic pieces of information that have already been documented by them or by someone else. Authors can concentrate on the added value they bring. They need only link that added value to the already documented relevant information that has been found, offering all the epistemic glue they have to favour good implantation. Incentives for authors are discussed in  \cite{fumier}. 
\medskip

Because they favour connectedness of the overall MMM network, implantation and annotation, also contribute to a \textit{desirable} form of redundancy. See \S\ref{translating} below.

% ============================================================== %
% ========              BRIDGING                        ======== %
% ============================================================== %
\subsubsection{Bridging}\label{bridging}

Bridging two contributions (or areas) $c$ and $c'$ means contributing an edge between them (epistemically gluing them together) or contributing several contributions which together materialise a path between  $c$ and $c'$. \medskip

Bridging is particularly important when $c$ and $c'$ are initially "distant" on the MMM  (cf \S\ref{measuring}), i.e., when no epistemic relation has been documented between them, not even an indirect one. Following \citedesignbias{experts}, I mentioned that the MMM solution is intended to address information problems primarily rather than to address communication problems.  Communication solutions  help get messages across geographical distance (e.g. powerful vocal cords, emailing \textit{etc}) and across language barriers (e.g. hand waving, dictionaries, automatic translators). But  interlocutors in the same room speaking the same language sometimes still  struggle to understand each other because of different epistemic cultures, viewpoints, mindsets \textit{etc}. Increasing the likelihood of an epistemic bridge being documented between the interlocutors' MMM territories may help.
\medskip

Any mechanism that increases the connectedness of the MMM network favours bridging. % between distant contributions. 

% ============================================================== %
% ========              DOCUMENTING                     ======== %
% ============================================================== %
\subsubsection{Documenting }\label{documenting}\label{atomic-documenting}

Traditional documents like scientific articles contain multiple pieces of information. And unless they amount to unexplained bullet point lists, traditional documents also provide  epistemic links between the pieces of information they contain. Information in scientific articles is typically organised in sections, subsections \textit{etc} (cf \SMref{balance}). In the MMM, a traditional self-standing document can be recorded in one piece, e.g. as the label of  a $\mmmtype{narrative}$ node. Or it can be decomposed into several pieces conveyed by a network of  several interlinked MMM contributions.
% (cf \SMref{atomic}).
\medskip  

\begin{bestpractices}{} Decompose the content that you want to document into atomic contributions well interlinked with each other (rather than dump it as a monolithic text in a single contribution). Limit the amount of content you convey through each single contribution and make extensive use of meaningful MMM edges.
\end{bestpractices}

% The traditional self-standing document translates into the MMM as an epistemic network of contributions .

Optionally a MMMified document can be delineated using a \hyperref[pens]{pen} -- typically a \textit{mutable} pen to support collaborative editing (cf \S\ref{mutable}).

\begin{figure}[H]
    \centering\scriptsize
    \begin{tikzpicture}
        % \draw[pencolor,dashed, opacity=0.9] (0,0) rectangle ++(12,4.2);
        % General title
        \node[pen,inner sep=6mm,font=\scriptsize,align=center] (genpen) at (4,3) {Proposal for an Organic Web\\ The missing link between the Web and the Semantic Web};
            % \node[pen,fill=red, opacity=0.5,inner sep=6mm,font=\scriptsize,align=center] (genppen) at (6,3) {            Proposal for an Organic Web\\ The missing link between the Web and the Semantic Web};
        \node[font=\scriptsize,anchor=west] at ([xshift=2pt,yshift=-2mm]genpen.north west) {\color{pencolorSTR}\mmmtype[pencolorSTR]{document pen}}; 
        % Part I
        \node[anchor=west,pen,inner sep=6mm,align=center,below =12mm of genpen,xshift=-10mm, text width=14mm] (pen) at (6,3) {\sc Part I};
        \node[font=\scriptsize,anchor=west] at ([xshift=2pt,yshift=-2mm]pen.north west) {\color{pencolorSTR}\mmmtype[pencolorSTR]{document pen}}; 
        % pen --> genpen 
        \draw[pennedin] (pen.west) -| ([xshift=-6mm,yshift=11mm]pen.west)  node[pos=0.7,xshift=-7mm] {\mmmtype[pencolor]{pennedin}};
        %
        % Part II
        \node[anchor=west,pen,inner sep=6mm,font=\scriptsize,align=center,right=15mm of pen, text width=14mm] (pen2)  {\sc Part II};
        \node[font=\scriptsize,anchor=west] at ([xshift=2pt,yshift=-2mm]pen2.north west) {\color{pencolorSTR}\mmmtype[pencolorSTR]{document pen}}; 
        \node[existence, above right=0.6cm and -2.1cm of pen2] (onto) {formal ontologies\\ {\large$\longleftrightarrow$} MMM};
        \node[existence, above right=1cm and 0.1cm of pen2] (search) {formal ontologies $\approx$\\ "backbone"\\ for MMM search};
        \node[question, above right=0.3cm and 0.3cm of pen2] (ident) {MMM identifiers?};
        \draw[pennedin] (onto) -- (pen2);
        \draw[pennedin] (search) -- (pen2);
        \draw[pennedin] (ident) -- (pen2);
        % pen2 --> genpen 
        \draw[pennedin] (pen2.west) -| ([xshift=-6mm,yshift=11mm]pen2.west)  ; 
        \node[existence,below=0.8cm of pen] (implant) {Implantation};
        \node[existence,left=3.5cm of implant,yshift=7mm] (m) {Motivation};
        \node[existence,right=3.3cm of implant] (c) {Continual\\ Improvement};
        %    
        % IMPLANTATION xxx
        \node[narrative,below=0.6cm of implant,font=\tiny,text width= 4.8cm,opacity=1]   (it)  {Implanting means contributing edges between a new contribution and pre-existing contributions on the landscape. It is the counterpart of annotating. Annotating is aimed at improving pre-existing content by way of new content. Implanting is aimed at using old content to  provide context and visibility to new contributions.%\vspace{-3mm}
        % %\begin{bestpractices}{} 
        % \textit{Best practices for users}
        % Implant much and {well}. Poorly implanted contributions will have less visibility as the  pathways leading to them are rare and/or at risk  of being {red-flagged} and {obsoleted}. \vspace{-3mm}\end{bestpractices}
        } ;
        \draw[details] (it) -- (implant) ;
        \node[action,below=0.4cm of it,font=\tiny,text width= 4.8cm,opacity=0.9]   (bpit)  {
        \textit{Best practices for users:}
        Implant much and {well}. Poorly implanted contributions will have less visibility as the  pathways leading to them are rare and/or at risk  of being {red-flagged} and {obsoleted}.     } ;
        \draw[details] (bpit) -- (it) ;
        % MOTIVATIONS
        \node[narrative,below=0.6cm of m,font=\tiny,text width= 4.8cm,opacity=0.8]   (mt)  {\motivations{}[\ldots]} ;
        \draw[details] (mt) -- (m) ;
        % CONTINUAL IMPROVEMENT
        \node[narrative,below=0.6cm of c,font=\tiny,text width= 4.8cm,opacity=0.8]   (ct)  { We consider information as a process, something to do, until it no longer is relevant. A piece of information is neither definite nor free-standing.  Unlike data is not a given and unlike knowledge it is not established.  It is \textit{at its best} when it is subjected to continual improvement: when it is getting nuanced, detailed, challenged, (re)contextualised, updated\ldots{} But eventually it becomes obsolete. 
        We wish to support information \textit{at its best} in this sense,  and facilitate the obsolescence of unprocessable information. } ;
        \draw[details] (ct) -- (c) ;
        \path [draw,pennedin] (implant) to  (pen);
        \path [draw,pennedin] (m) to        (pen);
        \path [draw,pennedin] (c) to        (pen);
        % \path [draw,pennedin] (implant) to      node[fill=white,pos=0.4]  {\mmmtype[pencolor]{pennedin}}     (pen);
        % \path [draw,pennedin] (m) to      node[fill=white]  {\mmmtype[pencolor]{pennedin}}     (pen);
        % \path [draw,pennedin] (c) to      node[fill=white]  {\mmmtype[pencolor]{pennedin}}     (pen);
        %
        % MUTABLE PEN ALL AROUND
        \node [draw=pencolor,dashed, opacity=0.9, inner sep=5mm, fit={(genpen) (mt) (pen) (pen2) (c) (bpit) (ct)}] (mutpen) {};
        \node[font=\scriptsize,anchor=west,opacity=0.9] at ([xshift=2pt,yshift=-2mm]mutpen.north west) {\color{pencolorSTR}\mmmtype[pencolorSTR]{mutable pen}}; 
    \end{tikzpicture}
    \caption{An "MMMification" of a traditional linear text document. N.B.: the MMM format isn't designed for  MMMifying existing documents. It rather is conversely meant for doing the informational work in preparation of  a new document  composition (cf \S\ref{drafting} and \SMref{rendering}).  }
    \label{fig-documenting}
    % \caption{{A mutable pen.   {UI application code} can visually represent atomic and mutable pens in a similar way although they are different objects. }}
\end{figure}
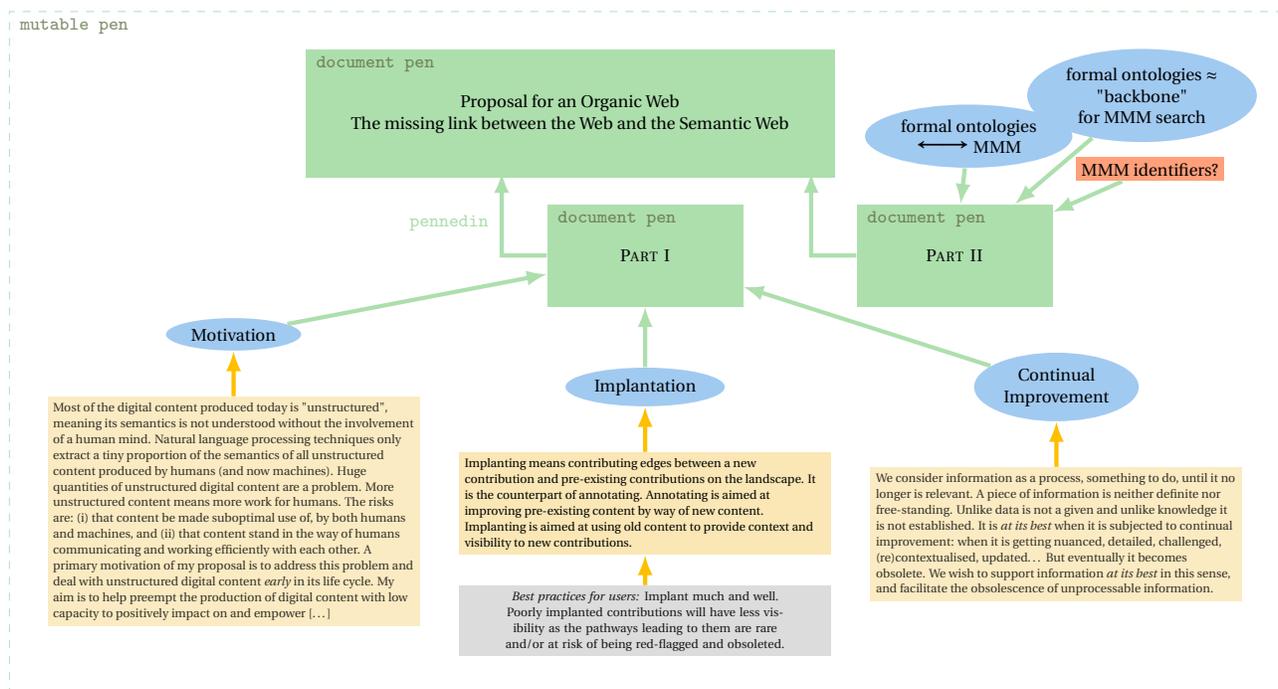

% ============================================================== %
% ========                                              ======== %
% ============================================================== %
\subsubsection{Disapproving and Red-Flagging}\label{censoring}\label{disapproving}

A contribution can be deemed of low quality -- i.e., disapproved of -- for exactly two reasons:
\begin{enumerate}
    \item The contribution is poorly positioned in the landscape. The edges between it and other landmarks are absurd. For instance the contribution is off-topic. Or it is on-topic, but it is implanted with the wrong kind of edge,  e.g. it is linked with an $\mmmtype{answers}$ edge to a $\mmmtype{question}$ node while not providing an answer to that specific question. 
    % eg what colour  to doi 
    \item The contribution is well positioned, but its label is of poor quality. For instance the contribution is a  $\mmmtype{narrative}$ node conveying a statement that someone considers untrue.
    % eg white to colour of daytime cloudless sky
\end{enumerate}

I propose to manage the two situations  differently, and neither by resorting  to censorship.
%
% We propose to manage the first situation  through \hyperref[redflag]{red-flagging} -- see  below. 
%
The second situation is a case where we especially don't want to make the low quality contribution less visible on the MMM. My conviction is that mistakes, misinformation, disinformation \textit{etc} are unavoidable if not normal on a collaborative information space (cf \SMref{errors}).  
\medskip

% ============================================================== %
% ========                 RED-FLAG                     ======== %
% ============================================================== %
\label{redflag}

Red-flagging  is an annotation pattern for dealing  with the first situation listed above. Red-flagging a contribution  $c\in {\bf C}$ consists in recording  an \equates{}  edge between  $c$  and  $\bot$.
Typically, red-flagging applies on contributions  that are edges. Red-flagging an edge conveys the idea that the edge's endpoints should not be linked in the way the edge says they are. \medskip

\begin{tikzpicture}  
    \draw[gray!40] (0,0) rectangle ++(4,4);
    \node[text=gray] at (2,0.3) {\sc\small before};
    \node[question] (q) at (2,3.5) {What colour is the sky?};
    \node[narrative,font=\small] (n) at (2,1.2) {We need to buy rice.};
    \draw[answers]  ([xshift=10mm]n.north west) -- ([xshift=10mm,yshift=18mm]n.north west)  node[midway,yshift=-4pt,fill=white,pos=0.35] {\scriptsize \mmmtype[answerscolor]{answers}} ;
    \begin{scope}[xshift=4.2cm]
    \draw[gray!40] (0,0) rectangle ++(4,4);
    \node[text=gray] at (2,0.3) {\sc\small after};
    \node[question] (q) at (2,3.5) {What colour is the sky?};
    \node[narrative,font=\small] (n) at (2,1.2) {We need to buy rice.};
    \draw[answers]  ([xshift=10mm]n.north west) -- ([xshift=10mm,yshift=18mm]n.north west)  node[midway] (a) {} node[midway,yshift=-4pt,fill=white,pos=0.35] {\scriptsize \mmmtype[answerscolor]{answers}} ;
    \node[star,star points=12,fill=black, text=white,inner sep=1pt,right=15mm of a] (pit)  {$\bot$};
    \draw[equates]  (a) -- (pit)  node[midway,yshift=3pt] (a) {\mmmtype[equatescolor]{equates}} ;
    \end{scope}
\end{tikzpicture} 
\hfill
\begin{tikzpicture}  
    \draw[gray!40] (0,0) rectangle ++(4,4);
    \node[text=gray] at (2,0.3) {\sc\small before};
    \node[question, text width=3.3cm,font=\small] (q) at (2,3.3) {How is RNA involved in vaccines?};
    \node[narrative, text width=3.3cm,font=\small] (n) at (2,1.2) {5G chips in vaccines causes people to spontaneously combust.};
    \draw[answers]  ([xshift=10mm]n.north west) -- ([xshift=10mm,yshift=11mm]n.north west)  node[midway,yshift=-4pt,fill=white,pos=0.5] {\scriptsize \mmmtype[answerscolor]{answers}} ;
    \begin{scope}[xshift=4.2cm]
    \draw[gray!40] (0,0) rectangle ++(4,4);
    \node[text=gray] at (2,0.3) {\sc\small after};
    \node[question, text width=3.3cm,font=\small] (q) at (2,3.3) {How is RNA involved in vaccines?};
    \node[narrative, text width=3.3cm,font=\small] (n) at (2,1.2) {5G chips in vaccines causes people to spontaneously combust.};
    \draw[answers]  ([xshift=10mm]n.north west) -- ([xshift=10mm,yshift=11mm]n.north west)  node[midway] (a) {}  ;
    \node[star,star points=12,fill=black, text=white,inner sep=1pt,right=15mm of a] (pit)  {$\bot$};
    \draw[equates]  (a) -- (pit)  node[midway,yshift=3pt] (a) {\mmmtype[equatescolor]{equates}} ;
    \end{scope}
\end{tikzpicture} 
\medskip

Consider for instance the example on the right above. A \mmmtype{question} $q$ asks about RNA mechanisms involved in vaccines. A \mmmtype{narrative} contribution $n$ is provided as an  answer  to $q$ through the \mmmtype{answers} edge $e$ connecting $n$ to $q$. Narrative $n$ makes no mention of RNA molecules. Whatever the value of the statement $n$ makes, 
whether it is true or false, whether we agree with $n$ or not, it is easy to see that $n$ is not an answer to $q$ because an answer to $q$ has to involve the notion of RNA. So independently of what one may think of $n$, one can safely red-flag  edge $e$  (not $n$ itself). 
This operation leaves all three contributions $q$, $n$, and $e$ unchanged. The only change to the landscape is the addition of a new \equates{} edge from $e$  to $\bot$. This new edge contribution can be labelled "An answer to the question must mention RNA mechanisms." or simply "Off topic" to specify the reason for the red-flagging of $e$.

\begin{bestpractices}{}
    Document the reason for the red-flagging of a contribution  in one of the labels of the  \equates{} edge linking it to $\bot$.
\end{bestpractices}

In the example above the \mmmtype{narrative} contribution $n$ about 5G doesn't need to be red-flagged. Justifying a red-flagging of   $n$ itself is more demanding than justifying the red-flagging of $e$. If $n$ could be standing alone disconnected of other contributions, and then "Off topic" wouldn't apply.  

\medskip

% N.B.: Red-flagging a contribution leaves that contribution unchanged. \medskip

\begin{bestpractices}{} {Avoid red-flagging a well positioned contribution} especially if it is of low quality. If it is of low quality, then \hyperref[annotating]{annotate it}: link new contributions to it that explicit in what way it is of low quality. In particular, use $\mmmtype[nuancescolor]{nuance}$ and $\mmmtype[questionscolor]{questions}$ edges abundantly to nuance and question the contribution. Only red-flag  poorly positioned contributions. 
\end{bestpractices}

Application code can deal with red-flagged contributions differently depending on the desired user experience. One UI could hide all contributions that have been red-flagged at least once by the community of users. This would mean that the 5G contribution above wouldn't be reachable from the question about vaccines.  Another UI could wait for the contribution to have been red-flagged 5 times (5 different authorships for all \equates{} edges between $n$ and $\bot$). Another UI could highlight red-flagged contributions. 

% ============================================================== %
% ========                 OBSOLETE                     ======== %
% ============================================================== %
\subsubsection{Obsoleting}\label{obsoleting} 

Obsoleting a contribution starts with  \hyperref[marks]{marking} it as \mmmmark{obsolete}. Different MMM editing tools may deal with obsolete contributions slightly differently. I describe the main idea. \medskip

The end result of obsolescence is the deletion of the contribution from the user's local database state. 
There may be several copies of a contribution $c$ in the distributed MMM network. 
By default, \mmmmark{obsolete} contributions are not propagated. If Alice has marked contribution $c$ as \mmmmark{obsolete}, and if Bob has no copy of $c$ in his own local database, then Bob won't get $c$ from Alice. However, if Charlie already has a copy of $c$, then Charlie may be notified of Alice's obsolescence of $c$.
%   \mmmmark{obsolete}d copy of $c$ may be shared with Charlie so that Charlie becomes aware that 
% among peers. They may be ignored altogether in communication among peers. Or, only the the \mmmmark{obsolete} mark of a contribution $c$ may be shared to peers who have a copy of $c$. 
So  \mmmmark{obsolete} contributions tend to disappear for good from the MMM, although no single user can be responsible for the definite disappearance of a shared contribution. 
% Without a centralised storage solution, users would be the sole custodians of contributions. 
\medskip
% Clients don't have to store their entire diff history while offline, and then, on reconnection, send the whole batch to their peers for a very expensive merge. 

When a contribution $c$ is marked as \mmmmark{obsolete}, all edges incident on it are also marked as \mmmmark{obsolete}. The edge recursion mentioned in \S\ref{edges} (see footnote \ref{edgerecursion} on page \pageref{edgerecursion}) means that obsolescence  cascades down series of incident edges. We say that we \textbf{recursively obsolete} $c$.\medskip

Obsolete contributions are not immediately  deleted. They remain "in limbo" for a customisable amount of time.\medskip

Suppose contribution $c$ is marked as \mmmmark{obsolete} on Alice's territory. Suppose that Bob has the three contributions  $c$, $c'$ and edge $e$ between $c$ and $c'$ stored on his territory. Neither of these three contributions is marked as \mmmmark{obsolete}  on Bob's territory. And suppose Alice synchronises with Bob. % connects with Bob to get his input. 
% Suppose that Alice connects to Bob who has three contributions to propose to Alice:  $c$, $c'$ and edge $e$ between $c$ and $c'$. Neither $c$, $c'$ nor $e$ is marked as \mmmmark{obsolete} on Bob's disk. 
Having $c$ in limbo on Alice's side allows dealing appropriately with Bob's input to Alice's territory. The fact that Bob's input is linked to a piece of information that Alice has already obsoleted may be used as a pretext to filter out everything Bob has linked to $c$. In this case Alice neither gets $c$, $c'$ nor $e$ from Bob. 

\begin{center}
\begin{tikzpicture}  
    \node[narrative,opacity=0.6] (n) at (2,3.4) {The sky is blue.};
    % \node[metadata,below right=0mm and 0mm of n] {\mmmid{narrSkyBlue}};
    \draw[help lines,step=40mm,gray!40] (0,0) grid (4,4);
    \node[text=gray] at (2,0.3) {\sc\small Alice's territory};
    \node[text=red,font=\huge] at (n) {\nope};
    \begin{scope}[xshift=4.2cm]
    \draw[gray!40] (0,0) rectangle ++(4,4);
    \node[text=gray] at (2,0.3) {\sc\small Bob's territory};
    \node[narrative] (n) at (2,3.4) {The sky is blue.};
    % \node[metadata,below right=0mm and 0mm of n] {\mmmid{narrSkyBlue}};
    \node[narrative,font=\small]   (nn) at (2,1.2) {Blue light in the sky\\ scatters more.};
        \draw[details] (nn) -- (n)  node[midway,yshift=-4pt,fill=white] { \mmmtype[detailscolor]{supports}} ;
    \end{scope}
\end{tikzpicture} 
\end{center}

Alice may also on the contrary decide to take $c$ out of the limbo because if Bob is still interested in $c$ she decides she may also be after all. 
Another possibility is that Alice obsoleted $c$ because she replaced $c$ with $c"$. If  there is an \equates{} edge connecting $c$ to $c"$, it suggests that $c$ and $c''$ are "epistemically equivalent", i.e., they convey the same idea.  Alice might then  be interested in Bob's annotation of $c$, provided it is redirected onto $c"$ as in the figure below:

\begin{center}
    \begin{tikzpicture}  
        \draw[gray!40] (0,0) rectangle ++(8.2,4);
        \node[text=gray] at (4.1,0.3) {\sc\small Alice's territory};
        \node[narrative,opacity=0.6] (n) at (2.2,3.4) {The sky is blue.};
        \node[narrative,font=\small]   (nn) at (2,1.2) {Blue light in the sky\\ scatters more.};
        \node[text=red,font=\huge] at (n) {\nope};
        \node[narrative,font=\small,text width=3.6cm]   (nnn) at (5.7,2.2) {The sky is blue during the day in the absence of clouds.};
        \draw[equates,opacity=0.6] (nnn) |- (n)  node[pos=0.7,yshift=4pt] (e) { \mmmtype[equatescolor]{equates}} ;
        \node[text=red,font=\huge] at ([xshift=-3mm,yshift=6mm]nnn.north) {\nope};
        %
        % Redirected edge :
        \draw[details] (nn.east) -| (nnn.south)  node[pos=0.3,yshift=-5pt] { \mmmtype[detailscolor]{supports}} ;
    \end{tikzpicture} 
    \end{center}

The limbo period allows to gradually adjust to the planned disappearance of a contribution. The idea is similar to "tombstones" in CRDTs \cite{sun2018real}. {Strategies need to be devised to decide when to end the limbo period of contributions and discard them for good to avoid unnecessary accumulation of obsolete material \cite{bieniusa2012optimized}.} % 
Definitely deleting  a contribution too early in the MMM {is not as serious as deleting a tombstone too early in a typical, fine-grained CRDT.} Deleting  an \mmmmark{obsolete} MMM contribution $c$ too early  doesn't prevent the system from functioning correctly. Without $c$, the landscape still makes sense. The redundancy management might however not be as good. Deleting an obsolete contribution early  increases the risk that the deletion be revoked. 
Suppose Alice's copy of "The sky is blue." and the incident \equates{} edge  are deleted for good  before synchronisation with Bob is made. When later Alice does synchronise with Bob, Bob's copy of "The sky is blue."  is shared with Alice as if Alice never had had a copy of it before.

\begin{center}
    \begin{tikzpicture}  
        \draw[gray!40] (0,0) rectangle ++(8.2,4);
        \node[text=gray] at (4.1,0.3) {\sc\small Alice's territory};
        \node[narrative] (n) at (2.2,3.4) {The sky is blue.};
        \node[narrative,font=\small]   (nn) at (2.2,1.2) {Blue light in the sky\\ scatters more.};
        \node[narrative,font=\small,text width=3.6cm]   (nnn) at (5.7,2.4) {The sky is blue during the day in the absence of clouds.};
        \draw[details] (nn) -- (n)  node[midway,yshift=-4pt,fill=white] { \mmmtype[detailscolor]{supports}} ;
    \end{tikzpicture} 
    \end{center}

Limbo periods are periods during which a user's disinterest for a contribution is remembered and can be shared.
\medskip

The main reason for obsoleting a contribution is that the contribution is no longer \textit{useful}. An \mmmmark{obsolete} contribution is not "epistemically inconvenient". Its presence in a landscape doesn't make the landscape less valuable. However, an  \mmmmark{obsolete} contribution, since it no longer is relevant to a particular user, becomes  \textit{visually} inconvenient to that user. 
\medskip

Because obsolescence is local, a user can't retract a contribution  that she has already shared. Once a contribution $c$ labelled "\textit{The cat is out of the box.}" has propagated, it can't be deleted from the MMM by any single user. If the author of $c$ regrets her publication, the only things she can do is  (1) try to propagate her obsoleting of $c$, and (2) direct attention away from $c$ or change the perception of $c$ by adding annotations around $c$ (she could also red-flag her own contribution, but that might not be advantageous to her in some contexts). 

% ============================================================== %
% ========                 DRAFTING                     ======== %
% ============================================================== %
\subsubsection{Drafting, Substituting and Versioning}\label{drafting}
    
The MMM can be used as a drafting medium: a place to organise ideas before deciding a linear order to present those ideas in a traditional document. 
MMM contributions are not meant to undergo significant changes after their creation.
Contributions represent atomic units of work. Replacing a contribution by a new contribution is a workaround the definiteness of contributions. %It should be used with caution. \medskip

\begin{bestpractices}{} Avoid leaving unfinished contributions at the end of a work session.
\end{bestpractices}

As before, in the figures below,  {\color{red}\nope} marks \mmmmark{obsolete}d contributions that will eventually be deleted.
    
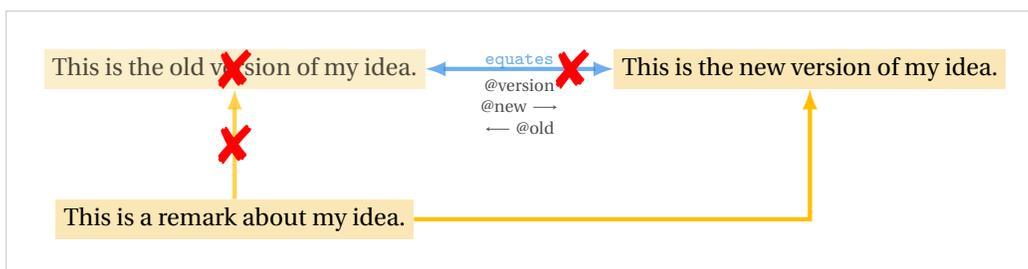
\begin{figure}[H]
    \centering
\begin{tikzpicture}  
    \node[narrative,opacity=0.7] (n) at (1.5,3.2) {This is the old version of my idea.};
    \node[narrative,right=2.5cm of n] (nn) {This is the new version of my idea.};
    \draw[equates,opacity=0.7] (nn) -- (n)  node[midway,yshift=1mm] (e) {\mmmtype[equatescolor]{equates}} ;
    \node[text width= 2cm,font=\scriptsize,align=center,opacity=0.7] (ee) at ([yshift=-6mm]e) {$@$version\\ $@$new $\longrightarrow$\\$\longleftarrow$ $@$old};
    \node[narrative]    (blue) at (1.5,1.2) {This is a remark about my idea.};
    \draw[details,opacity=0.7] (blue) -- (n) node[midway] (be) {} ;
    \draw[details] (blue) -| (nn)  ;
    \node [draw=gray!40, inner sep=5mm, fit={(n) (nn) (ee) (blue)}] {};
    \node[text=red,font=\huge] at (n) {\nope};
    \node[text=red,font=\huge,xshift=7mm,yshift=-1mm] at (e) {\nope};
    \node[text=red,font=\huge] at (be) {\nope};
\end{tikzpicture} 
\caption{Versioning. An idea is reformulated. The old version is obsoleted and {appropriately} linked to the new version.}
    \label{reformulate}
\end{figure}
    
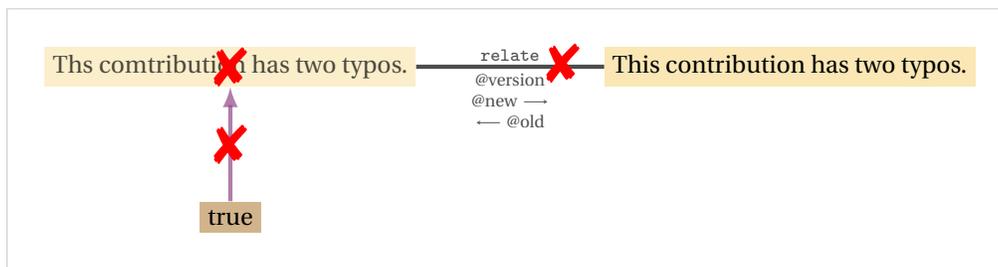
\begin{figure}[H]
    \centering
\begin{tikzpicture}  
    \node[narrative,opacity=0.7] (n) at (1.5,3.2) {Ths comtribution has two typos.};
    \node[narrative,right=2.5cm of n] (nn) {This contribution has two typos.};
    \draw[relate,opacity=0.7] (nn) -- (n)  node[midway,yshift=1.5mm] (e) {\mmmtype[relatecolor]{relate}} ;
    \node[text width= 2cm,font=\scriptsize,align=center,opacity=0.7] (ee) at ([yshift=-6mm]e) {$@$version\\ $@$new $\longrightarrow$\\$\longleftarrow$ $@$old};
    \node[datavalue]    (blue) at (1.5,1.2) {true};
    \draw[pertains,opacity=0.7] (blue) -- (n) node[midway] (be) {} ;
    \node [draw=gray!40, inner sep=5mm, fit={(n) (nn) (ee) (blue)}] {};
    \node[text=red,font=\huge] at (n) {\nope};
    \node[text=red,font=\huge,xshift=7mm,yshift=-1mm] at (e) {\nope};
    \node[text=red,font=\huge] at (be) {\nope};
\end{tikzpicture} 
\caption{Versioning when the old and new versions of a contribution are not epistemically equivalent.  The vertical pertains edge incoming the old version can't simply be redirected to the new version.   }
    \label{tricky-versioning}
\end{figure}

Rather than make contributions evolve over time, I advise the following   default strategy for replacing a contribution $c$ with a new version $c'$ of $c$: 

\begin{itemize}
    \item Make a copy $c'$ of $c$ that is a brand-new contribution with a different identifier than $c$.
    \item Let the user change whatever attribute she needs to change in $c'$.
    \item Link $c$ and $c'$ using an  \equates{} edge.  
    The new \equates{} edge can be tagged "$@$version" and optionally labelled with the 
    reason for the replacement of the old version by the new. 
    In some cases, the old and new versions might be too different to justify an \equates{} link. Another type of link should be preferred, possibly a \mmmtype{relate} link (cf Fig. \ref{tricky-versioning}). 
   
    \item Recursively obsolete $c$.
    
    \item If $c$ and $c'$ are epistemically equivalent (e.g. if $c'$ corrects a typo in $c$) and have been linked with an \equates{} edge, then make copies (with different identifiers) of all edges incident on $c$, replacing the $c$ endpoints of these edges with $c'$ endpoints. If $c$ and $c'$ aren't epistemically equivalent, leave it to the user (or possibly to a trained AI)
    %machine learning model) 
    to decide how to redirect edges from $c$ to $c'$. Because the obsolescence around $c$ is recursive, recursive redirection might be necessary. 
    \item Add $c'$ and new incident edges to all pens to which $c$ and old edges belong. 
\end{itemize}

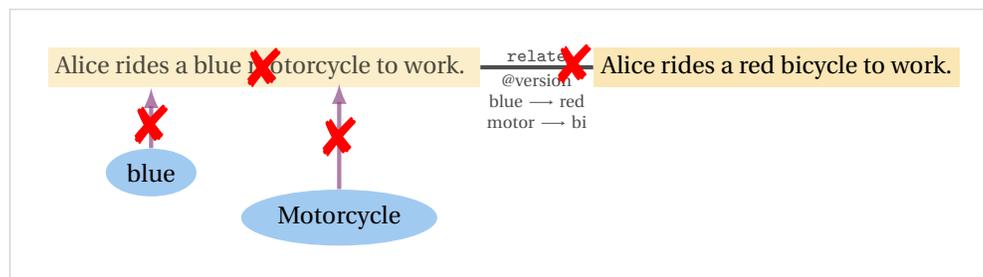
\begin{figure}[H]
    \centering
\begin{tikzpicture}  
    % n
    \node[narrative,opacity=0.7] (n) at (1.5,3.2) {Alice rides a blue motorcycle to work. };
    \node[text=red,font=\huge] at (n) {\nope};
    % n'
    \node[narrative,right=1.5cm of n] (nn) {Alice rides a red bicycle to work.};
    % n --- n' 
    \draw[relate,opacity=0.7] (nn) -- (n)  node[midway,yshift=1.5mm] (e) {\mmmtype[relatecolor]{relate}} ;
    \node[text width= 2cm,font=\scriptsize,align=center,opacity=0.7] (ee) at ([yshift=-6mm]e) {$@$version\\ blue $\longrightarrow$ red\\motor $\longrightarrow$ bi};
    \node[text=red,font=\huge,xshift=5mm,yshift=-1mm] at (e) {\nope};
    % Blue
    \node[existence]    (blue) at (0,1.8) {blue};
    \draw[pertains,opacity=0.7] (blue.north) -- ([yshift=8mm]blue.north) node[pos=0.4] (be) {} ;
    \node[text=red,font=\huge] at (be) {\nope};
    % cycle
    \node[existence]    (cycle) at (2.5,1.2) {Motorcycle};
    \draw[pertains,opacity=0.7] (cycle.north) -- ([yshift=14mm]cycle.north) node[midway] (ce) {} ;
    \node[text=red,font=\huge] at (ce) {\nope};
    % \draw[details] (blue) -| (nn)  ;
    %
    \node [draw=gray!40, inner sep=5mm, fit={(n) (nn) (ee) (blue) (cycle)}] {};
    %
    
    % yyy
    
\end{tikzpicture} 
\caption{Another case of versioning in which the old and new versions of the contribution aren't epistemically equivalent. Again the vertical pertains edge incoming the old version can't simply be redirected to the new version.   }
    \label{tricky-versioning-2}
\end{figure}

% ============================================================== %
% ========                 VERSIONING                   ======== %
% ============================================================== %
\label{versioning}

\begin{bestpractices}{} To modify an existing MMMified document $D$ -- i.e. a document documented in the MMM as a network of interlinked contributions (cf Fig. \ref{fig-documenting}) -- first locate  the exact pieces of information (contributions) in $D$ that your modification applies to. If you  want to specify one or several of those pieces of information, then preferably annotate them (cf \S\ref{annotating}). If you want to delete them then 
 {obsolete} them (cf \S\ref{marks}).
And if you want to replace them 
then record their new versions  and link them appropriately to the old using an \equates{} link as detailed above. 
\end{bestpractices}

% ============================================================== %
% ========                 MERGE                        ======== %
% ============================================================== %
\subsubsection{Merging}\label{merging}

Merging on the MMM only concerns contributions that are "epistemically equivalent", i.e., that convey the same idea. A contribution labelled "\textit{The Earth is round.}" will never be merged nor replaced with a contribution labelled "\textit{The Earth is flat.}" nor even with a contribution labelled "\textit{The Earth is roundish.}".
The MMM remains a mostly add-only system. The atomic unit of information that can be added to, and obsoleted from, the landscape is a MMM contribution.  MMM contributions don't disappear from the landscape because they are replaced \textit{per se}, but because they are \textit{in themselves} no longer useful. 
\medskip

For the sake of simplicity I ignore mark sets in this section, although a merging operation for mark sets can 
be defined. I assume an order can be defined on MMM contribution identifiers, for instance using timestamps.\medskip

% ================ order on contributions ====================== %

Let $c$ and $c'$ be two MMM contributions, respectively with identifiers $i$ and $i'$, with identical labels $l=l'$, with tag sets $x$ and $x'$, with identical types $t=t'$, with authorship sets $a$ and $a'$, and with statuses $s$ and $s'$.
%, and with mark sets $m$ and $m'$. 
I define the order $\preceq$ on contributions so that $c\preceq c'$ holds whenever all the following holds:  $i\leq i'$, $l=l'$,  $x\subseteq x'$, $t=t'$, $a\subseteq a'$,   and  $s\leq s'$.
\medskip

% ==================== contribution merge ====================== %

I define function $\mathtt{m}: {\bf C}\times {\bf C} \to {\bf C}$ such that for any two contributions $c$, $c'\in {\bf C}$ that share the same label and type, 
$\mathtt{m}(c,c') \in {\bf C}$ is the contribution whose identifier is $\textrm{max}\{i,i'\}$, whose label is $l=l'$, whose tag set is $x\cup x'$, whose type is $t=t'$, whose authorship set is $a \cup a'$, and whose status is $\textrm{max}\{s,s'\}$. 
Contribution $\mathtt{m}(c,c')=c\vee c'$ is 
the join 
%$c\vee c'$  
of $c$ and $c'$. The set of contributions that share the same label and type 
is a join-semilattice partially ordered by $\preceq $.
\medskip

% ======================== different IDs ====================== %

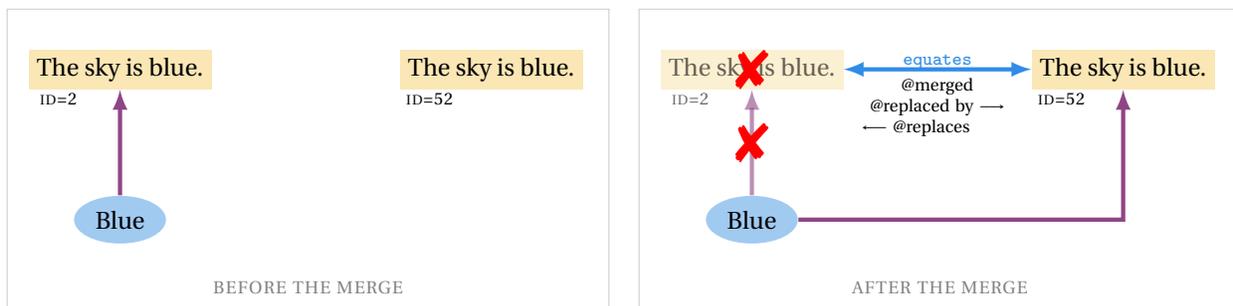
\begin{figure}[H]
    \centering
\begin{tikzpicture}  
    \draw[gray!40] (0,0) rectangle ++(8,4);
    \node[text=gray] at (4,0.3) {\sc\small before the merge};
    \node[narrative] (n) at (1.5,3.2) {The sky is blue.};
    \node[metadata,below=1pt of n.south west,xshift=4mm] {\mmmid{narrSkyBlue}};
    \node[narrative,right=2.5cm of n] (nn) {The sky is blue.};
    \node[metadata,below=1pt of nn.south west,xshift=4mm] {{\sc\scriptsize id=52}};
    \node[existence]    (blue) at (1.5,1.2) {Blue};
    \draw[pertains] (blue) -- (n) ;
    \begin{scope}[xshift=8.4cm]
        \draw[gray!40] (0,0) rectangle ++(8,4);
        \node[text=gray] at (4,0.3) {\sc\small after the merge};
        \node[narrative,opacity=0.6] (n) at (1.5,3.2) {The sky is blue.};
        \node[metadata,opacity=0.6,below=1pt of n.south west,xshift=4mm] {\mmmid{narrSkyBlue}};
        \node[narrative,right=2.5cm of n] (nn) {The sky is blue.};
        \node[metadata,below=1pt of nn.south west,xshift=4mm] {{\sc\scriptsize id=52}};
        \node[existence]    (blue) at (1.5,1.2) {Blue};
        \draw[pertains,opacity=0.6] (blue) -- (n) node[midway] (p) {};
        \draw[equates] (nn) -- (n)  node[midway,yshift=1mm] (e) {\mmmtype[equatescolor]{equates}} ;
        \node[text width= 2.3cm,font=\scriptsize,align=center] at ([yshift=-6mm]e) {$@$merged\\$@$replaced by $\longrightarrow$\\$\longleftarrow$ $@$replaces~~~~~~~~~};
        \draw[pertains] (blue) -| (nn) ;
        \node[text=red,font=\huge] at (n) {\nope};
        \node[text=red,font=\huge] at (p) {\nope};
    \end{scope}
\end{tikzpicture} 
\caption{Contribution with ID \ref*{narrSkyBlue} is merged into contribution with ID 52. }
% \label{}
\end{figure}

% ========================== same IDs ========================== %
The merge operation only affects contributions  $c$, $c'$ that have the same label and type. The end result of the merge is that only one contribution survives, which is equal to $\mathtt{m}(c,c')$. Merging doesn't modify the label nor the type of any MMM contributions. Typically, its effect is to complete the authorship set and to upgrade the \hyperref[status]{status} of a contribution.\medskip

Merging can affect "\textbf{homologous contributions}" and "non-homologous contributions". Homologous contributions are contributions  that have the same identifier. They necessarily have identical labels and types but may differ by their tag set and metadata. 
A merge of homologous contributions $c_1$ and $c_2$ 
typically happens when a user has a copy $c_1$ of a contribution locally stored and receives another copy $c_2$ of the same contribution from a peer over the distributed network. 
\medskip

Contributions $c$, $c'$ with different identifiers (non-homologues)   can also be merged as long as they have the same label and type. 
If $i < i'$ are the identifiers of $c$ and $c'$, merging  $c$ and $c'$ is called \textbf{merging} $c$ \textbf{\textit{into}}  $c'$. Generally merging  $c$ and $c'$ with identifiers $i\leq i'$ consists in the following:

\begin{itemize}
    \item Update $c'$  to $\mathtt{m}(c,c')$ which has the same identifier as $c'$.
    \item Create an \equates{} edge  between $c$ and $c'$. I recall that a bidirectional edge has three labels and three tag sets. The new \equates{} edge can be labelled or tagged "merged" (main label/tag set), "replaced by" (direction  $c$ to $c'$), and/or "replaces" (direction $c'$ to $c$). 
    \item Recursively obsolete $c$.
    \item Recursively make  non-homologous copies of all recently obsoleted edges around $c$ and redirect them towards $c'$, as in \S\ref{versioning}.
    \item Add $c'$ and new incident edges  to all pens to which $c$ and old edges belong.
\end{itemize}

As long as \mmmmark{obsolete}d contributions are in limbo,  the merging operation is commutative, associative and idempotent because applying function  $\mathtt{m}: {\bf C}\times {\bf C} \to {\bf C}$  is. The order in which merges occur and repetitions of the same merge are impactless. 
\medskip

% ======================== merging pens ====================== %

{Function $\mathtt{m}: {\bf C}\times {\bf C} \to {\bf C}$  may   be generalised to allow the merge of two  pens $p$ and $p'$ that have different contents but same concrete type. The resulting/surviving  pen is a pen whose identifier is $\textrm{max}\{i,i'\}$ and whose set of contents is the union of the contents of $p$ and the contents of $p'$. The merge of mutable pens follows since merging causes all incident edges, including $\mmmtype{pennedBy}$ edges, to be copied to the surviving pen. 
}\medskip

% ======================== different IDs ====================== %

Merging is essential to redundancy management on the MMM.
A \textbf{relaxed version of the merge} operation can be implemented to allow for the merge of two contributions with different but epistemically equivalent labels. The regular merge described above updates the surviving contribution to $\mathtt{m}(c,c')$ and in doing so may modify some contribution attributes. In contrast, because "epistemic equivalence" is subjective, the {relaxed merge} must not modify any of the two contributions that are merged, except for marking one as \mmmmark{obsolete}. %
A diversity of mechanisms may be implemented to identify potential  motives for merges (e.g. language similarity \cite{gomaa2013survey}) and submit them to the user for translation into actual merges. 
\medskip

% ============================================================== %

Because \mmmmark{obsolete} marks are local, merges also tend to be local. 
Alice may merge non-homologous contributions $c$ and $c'$ locally resulting for instance in the authorship set of $c'$ to grow. If Bob also has copies of $c$ and $c'$, Bob will not be affected by the merge. 
This is desirable as $c$ and $c'$ may be public contributions and Alice alone should not be allowed to update
public material for everyone. If Bob independently decides to merge $c$ and $c'$, Bob will adopt the same merging strategy as Alice: merging the contribution with the smallest ID into the contribution with the largest. 
If Bob doesn't have a copy of $c$ nor of $c'$, by default, Bob will not inherit obsolete contribution $c$ from Alice. 
If Bob has a copy of $c$, we may use Alice's \equates{} link  between $c$ and $c'$ as a trigger for Bob to substitute $c$ with $c'$. The more users merge  $c$ into $c'$, the greater the likelihood that $c$ ends up obsoleting from the entire distributed MMM  database. %(cf \S\ref{storage-layer}). 
\medskip

% =================== Merging landscape ======================== %
The notion of merging contributions naturally extends to a notion of \textbf{merging landscapes}.
For any two landscape $L$ and $L'$,   $\mathtt{m}^\star(L,L')$ 
is the landscape that contains exactly the union of all contributions of $L$ and all contributions of $L'$ where  contributions $c_1\in L$ have been merged with their homologous contributions $c_2\in L'$.
Let us write $L\sqsubseteq L'$ whenever for any contribution $c_1\in L$ there is a homologous contribution $c_2\in L'$. 
The set of landscapes partially ordered by  $\sqsubseteq$
forms a join-semilattice where the join of two landscapes is given by  $\mathtt{m}^\star$. The MMM is the join of all landscapes. 
If we guaranteed delivery of every contribution to every peer on the distributed MMM network (and also immutability of contributions which we mostly have for public contributions) then we could guarantee landscape convergence (locally users would eventually see the same landscape) \cite{shapiro2011}. 
Importantly however, we \textit{aren't} aiming at convergence. We {don't} want peers to see the same landscape 
\textit{despite} the distributed nature of the MMM, i.e., we don't want them to have the same local territory. On the contrary I propose to leverage the distributed nature of the MMM in order 
support a diversity of points of view on the record, and reduce the overall amount of digital content that any peer sees to just what is relevant to them\footnote{The subject of echo chambers is discussed in \S\ref{navigating} and in \cite{fumier}.}
(cf \S\ref{wayfarer}). 
Because not everyone is interested in the same material, the MMM need not be materialised at any point in time at any single node of the distributed network \cite{fumier}. 

% ============================================================== %
% ========                UPDATING                      ======== %
% ============================================================== %
\subsubsection{Updating  the Landscape}\label{updating}

Let us continue ignoring \hyperref[marks]{marks}, including the \mmmmark{obsolete} mark. The possible ways of updating a landscape $L$ are the following:
\begin{enumerate}
    \item Add a contribution to $L$ (this is the principal way of updating a landscape, cf \citedesignbias{events}).
    \item Merge  duplicate contributions (same label, same type)  as described in \S\ref{merging}.
    \item Add a tag to the tag set of a contribution in $L$.
    \item Add an authorship to the authorship set of a contribution in $L$. 
    \item Upgrade the status of a contribution in $L$. 
 \end{enumerate}
\medskip

For any of the five kinds of updates $u$ listed above, let us  write $L+u$ to denote the landscape obtained by applying $u$ to $L$. We have 
$L\sqsubseteq L+u$,
provided we consider obsolete contributions when non-homologous duplicates are merged.   Landscapes are thus monotonically non-decreasing with updates.
Updates to a user's local territory cause the territory to grow.
We have a simple case of CRDT (Conflict-Free Data Type) \cite{shapiro2011} because 
as noted above, the set of landscapes forms a semilattice ordered by $\sqsubseteq$
and the merge function $\texttt{m}^\star$ computes the join of two landscapes  (cf \S\ref{merging}). 
\bigskip

{The \href{https://www.yworks.com/products/yed}{YEd desktop tool} 
has recently been used as a demonstrator of the basic MMM editing functionalities. In particular, it has been used   to test the design of the MMM data model on information emanating from the aeronautical engineering domain. 
The set of YEd "palettes" that  have been used in that exercise  is available on Gitlab \cite{MMMgraphml}. YEd certainly isn't the only existing graph editor tool that can be adapted to act as a rudimentary MMM editor. It has the advantage of  serialising graphs into the  standard  GraphML format. And conveniently, it allows to group nodes and thus partially supports MMM pens. It does not allow to have edges act as endpoints of other edges however. We have used a temporary workaround breaking edges with a dummy node (cf  the \href{https://gitlab.com/MMM-Mat/mmm/-/blob/master/Graphml\%20files/YEd\%20Palettes/MMM\_EDGE\_BREAK\_POINTS.graphml?ref\_type=heads}{\texttt{MMM EDGE BREAK POINTS.graphml}} palette).  }

\subsection{Landscape Consuming Activities} \label{activities-consuming}% Exploring/

In section \ref{activities-editing}, we saw how to modify the MMM by adding and obsoleting content. Now, assuming the MMM is populated, we explore uses we can make of it.

% ============================================================== %
% ========                 MEASURING                    ======== %
% ============================================================== %
\subsubsection{Measuring}\label{measuring}

% Distance
A distance can be defined between any two contributions $c$ and $c'$ in the MMM, for instance as  the length of the shortest  undirected path between $c$ and $c'$. {The notion of distance between contributions in the MMM network can be refined to capture a notion of "\textbf{epistemic proximity}" between contributions. 
}

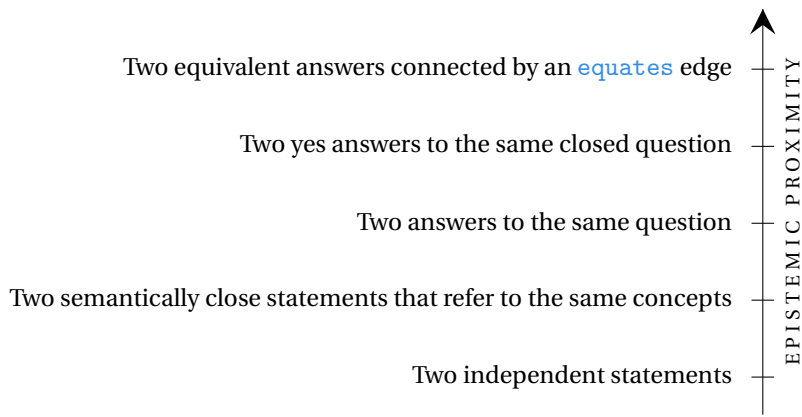
\begin{figure}[H]
    \centering %\scriptsize
    \begin{tikzpicture}
        \matrix (m) [matrix of nodes,row sep=0.5cm,column sep=2cm,
          nodes={rectangle},
          column 1/.style={anchor=base east},
        ]{

            Two equivalent answers connected by an \equates{} edge\\
            Two yes answers to the same closed question\\
            Two answers to the same question \\
            Two semantically close statements that refer to the same concepts \\
            Two independent statements  \\
        };
        \draw ([xshift=1.5mm]m-1-1.east) -- +(0:3mm);
        \draw ([xshift=1.5mm]m-2-1.east) -- +(0:3mm);
        \draw ([xshift=1.5mm]m-3-1.east) -- +(0:3mm);
        \draw ([xshift=1.5mm]m-4-1.east) -- +(0:3mm);
        \draw ([xshift=1.5mm]m-5-1.east) -- +(0:3mm);
      \draw[decoration={markings,mark=at position 1 with
      {\arrow[scale=3,>=stealth]{>}}},postaction={decorate}]    ([xshift=3mm,yshift=-5mm]m-5-1.east) -- ([xshift=3mm,yshift=8mm]m-1-1.east) node[yshift=-4mm,midway,sloped] {\sc e\,p\,i\,s\,t\,e\,m\,i\,c\, \,p\,r\,o\,x\,i\,m\,i\,t\,y};
      \end{tikzpicture}\hfill\,        
    \caption{Suggested scale of epistemic proximity for two statements.} % of statements $s_1$ and $s_2$. This is a suggested scale for a notion of epistemic distance to be defined formally.}
    \label{epistemicproximity}
\end{figure}

% Depth
Naively, the depth \textit{underneath} a contribution $c$ can  be defined as the length of the longest acyclic directed  path incoming $c$. {A notion of \textbf{absolute contribution depth} can be introduced to characterise contributions against a referent contribution depth 0 -- the depth of the most vacuous (/abstract)
question possible (e.g. 
%"What happened before time?" / "What if..."  
"What is the nature of existence?").}
% maturity
The 'maturity' of a contribution may be measured in terms of the number of annotations to it, especially the number of \mmmtype{nuances}, \mmmtype{details},  (answered) \mmmtype{questions} to it, and recursively, the maturity of those annotations. % possibly compare number of details with number of nuances
% reliability
The 'reliability' of a contribution $c$ can be defined in terms of maturity, in terms of ratio of \mmmtype{details} and \mmmtype{nuances},  and in terms of the depth of $c$ (length of directed paths outgoing $c$ leading to mature contributions).
\furtherwork{Other, finer metrics can be defined using the graph theoretic properties of landmarks and of areas of the landscape  in order to further qualify elements of the landscape.} This form of "epistemic topography" may allow nuancing  primitive binary qualification of information (e.g. correct/incorrect, true/false, consensual/not consensual, there/not there) replacing it by richer more profound qualification (cf \SMref{binary}).   \medskip

The number of red-flagging edges incident on a contribution $c$ (i.e., the number of \equates{} edges connecting $c$ with $\bot$), and the number of authors of each public red-flagging edge can also be counted to quantify the quality of $c$ or of neighbours of $c$. \medskip

Application code may rely on metrics and thresholds to decide when to trigger the publication or display of a MMM contribution (e.g. on an external feed). 

% ============================================================== %
% ========               ZOOM                           ======== %
% ============================================================== %
\subsubsection{Zooming in/out}\label{zooming}

An exploitable property of the MMM format is that most links are "vertical" links (unidirectional) as opposed to "horizontal" (adirectional or bidirectional) meaning that in some respects they express the idea that what is expressed at their start point is more defined, more narrow or more precise
than what is expressed at their endpoint which is more abstract, more compendious or more indiscriminate.
Unidirectional/Vertical edges tend go from more specific to more general. This feature can be used to implement "epistemic zooming in and out". 
Of course, in the MMM, directed paths can be circular, and I expect circularity will be common. This is not a problem. The "epistemic zooming" proposition is merely practical. It is to allow some   interactive filtering out of content, locally (as opposed to revealing some profound property of information). \medskip

{A fundamental functionality that we may want MMM editors to support is "contribution collapse": all contributions on an acyclic directed path to a given contribution are hidden and displayed on demand. The YEd graph editor  mentioned before partially supports this functionality assuming YEd groups are used to represent MMM nodes (cf  \href{https://gitlab.com/MMM-Mat/mmm/-/blob/master/Graphml\%20files/collapsingNodes.svg?ref\_type=heads}{\texttt{collapsingNodes.graphml}} \cite{MMMgraphml}). }

% ============================================================== %
% ========                FILTER                        ======== %
% ============================================================== %
\subsubsection{Filtering  Highlighting}\label{filtering}

The diversity of metrics that can be defined to qualify landscapes (cf \S\ref{measuring}) can be used to define a diversity of custom filters. These would allow users to experience the same landscape from different points of view. I propose that   a collection of pre-defined  adjustable filters be provided to users. 
Different {areas} of a landscape, and even different single contributions, may be managed by different filtering rules. 
\furtherwork{Further research is  needed to define and test possible default filter rules and determine the degree of stringency needed.  A default filter rule could for instance reject any contribution  that hasn't been challenged $x$ times and nuanced $y$ times by $z$ different users, and/or whose annotations are shallow in the sense that the depth underneath them is less than $d\in\mathbb{N}$. In this case, default values for $x,y,z$ and $d$ need to be determined based on an understanding of the possible local and global consequences of systematically rejecting more or less content  with this rule.} 
\medskip

Filter rules may also be used to trigger the local marking of contributions (cf \S\ref{marks}) -- e.g. as \mmmmark{hidden} or \mmmmark{dim}. Front-end code can use the marks to appropriately display the landscape to the user. \medskip

Rules can also conversely be defined to highlight certain contributions. We may for instance want to highlight contributions that have high \hyperref[measuring]{measures} of quality, or that are marked as \mmmmark{rewarded}.

% ============================================================== %
% ========              AGGREGATING                     ======== %
% ============================================================== %
\subsubsection{Aggregating}\label{aggregating}

Aggregation on the MMM means grouping together annotations that serve the same purpose -- e.g. grouping together all yes answers to a given \mmmtype{question} $q$ which can be identified as contributions linked to $q$ by an \mmmtype{answers} edge tagged $@$yes. Aggregation can support redundancy management mechanisms. Comparing aggregable contributions can allow identifying overlaps between those contributions. UIs can be designed to preempt  the recording of new contributions by the user that overlap with existing content, or to assist the user in narrowing down the value they can add. UIs can also be designed to systematically hide repetitive content identified as epistemically very close or equivalent to content not hidden. This way, sheer quantity and repetition of a piece of information may not directly translate into visibility of this piece of information. 
\medskip

Aggregation can be leveraged to further improve the landscape readability showing   less objects at once. 
Neighbouring contributions  may be grouped together  into "macro-meta contributions" by the graphical user interface (GUI), based on their subtype and on their position in the landscape.  For instance all \mmmtype{definition}s of the same term, or all \mmmtype{answers} to the same question 
may be displayed by the GUI as a single macro-meta node  on which the user needs to click to access the details of each definition or answer. Macro-meta contributions are part of the view, not the data model.

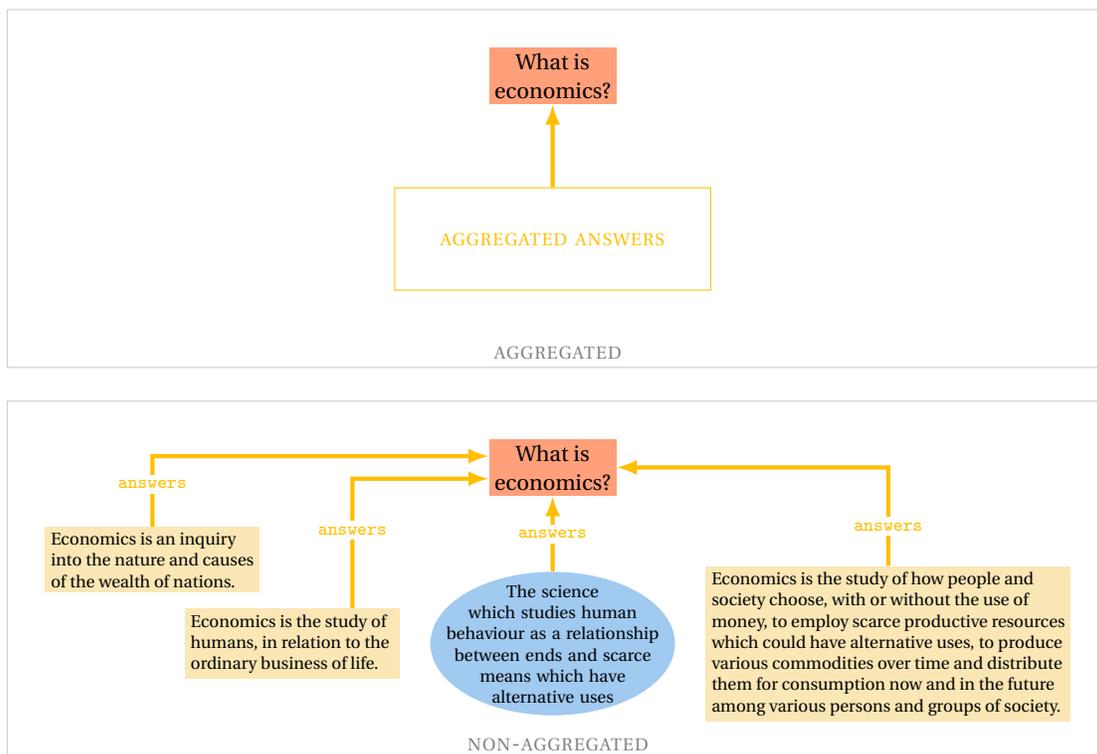
\begin{figure}[H]
    \centering\scriptsize
    \begin{tikzpicture}  
        % Question
        \node[question,font=\small] (q) at (1.5,3.2) {What is\\ economics?};
        % Lionel RObbin's definition:
        \node[opacity=0,existence,inner sep=0pt,below=1cm of q,text width=2.3cm,text height=13.5mm] (LR) {};
        % Paul Samuelson's definition 
        \node[opacity=0,narrative,right=.4cm of LR,text width=4.7cm] (PS) {Economics is the study of how people and society choose, with or without the use of money, to employ scarce productive resources which could have alternative uses, to produce various commodities over time and distribute them for consumption now and in the future among various persons and groups of society.};
        % Adama Smith's definition
        \node[opacity=0,narrative,left=0.4cm of LR,text width=27mm] (AM) {Economics is the study of humans, in relation to the ordinary business of life.};
        % Alfred Marshall's definition:
        \node[opacity=0,narrative,above left=2mm and -1cm of AM,text width=27mm] (AS) {Economics is an inquiry into the nature and causes of the wealth of nations.};
        % MACRO META NODE
        \node[draw=answerscolor,answerscolor,inner sep=6mm,below=1.1cm of q,font=\small,text width=3cm] (mm) {\sc aggregated answers};
        \draw[answers] (mm) -- (q) {};
        % node[pos=0.3,fill=white]  {\mmmtype[answerscolor]{answers}} ;
        % ANSWERS LINKS
        %
        \node [draw=gray!40, inner sep=5mm, fit={(q) (AS) (PS)}] (tour) {};
        \node[text=gray] at ([yshift=2mm]tour.south) {\sc\small aggregated};  
        \end{tikzpicture}  \\[4mm]

        \begin{tikzpicture}  
            % Question
            \node[question,font=\small] (q) at (1.5,3.2) {What is\\ economics?};
            % Lionel RObbin's definition:
            \node[existence,inner sep=0pt,below=1cm of q,text width=2.3cm,text height=13.5mm] (LR) {};
            \node[text width=6cm,align=center] (LRi)  at (LR) {The science\\ which studies human\\ behaviour as a relationship\\ between ends and scarce\\ means which have\\ alternative uses};
            % Paul Samuelson's definition 
            \node[narrative,right=.4cm of LR,text width=4.7cm] (PS) {Economics is the study of how people and society choose, with or without the use of money, to employ scarce productive resources which could have alternative uses, to produce various commodities over time and distribute them for consumption now and in the future among various persons and groups of society.};
            % Adama Smith's definition
            \node[narrative,left=0.4cm of LR,text width=27mm] (AM) {Economics is the study of humans, in relation to the ordinary business of life.};
            % Alfred Marshall's definition:
            \node[narrative,above left=2mm and -1cm of AM,text width=27mm] (AS) {Economics is an inquiry into the nature and causes of the wealth of nations.};
            % ANSWERS LINKS
            %
            \draw[answers] (LR) -- (q)  node[midway,fill=white]  {\mmmtype[answerscolor]{answers}} ;
            \draw[answers] (PS) |- (q)  node[pos=0.2,fill=white] {\mmmtype[answerscolor]{answers}} ;
            \draw[answers] (AS) |- ([yshift=1.5mm]q.west)  node[pos=0.3,fill=white] {\mmmtype[answerscolor]{answers}} ;
            \draw[answers] ([xshift=0.8cm]AM.north) |- ([yshift=-1.5mm]q.west)  node[pos=0.3,fill=white]  {\mmmtype[answerscolor]{answers}} ;
            \node [draw=gray!40, inner sep=5mm, fit={(q) (AS) (PS)}] (tour) {};
            \node[text=gray] at ([yshift=2mm]tour.south) {\sc\small non-aggregated};  
            \end{tikzpicture}

    \caption{Aggregation of all documented answers to a question, in one  {macro-meta-node}.}
    \label{macro-meta-node}
\end{figure}

% ============================================================== %
% ========               FINDING                        ======== %
% ============================================================== %
\subsubsection{Navigating, Exploring, Searching and Finding}\label{finding} \label{navigating} \label{searching}%  and {Exploring}/Searching

On a MMM landscape, \textbf{navigating} or \textbf{exploring}  means travelling down a \hyperref[path]{path}. 
% \textbf{Navigating} means finding a way to a 
\medskip

Let $c$ and $c'$ be two MMM contributions. Let $a$ be a landscape \hyperref[area]{area}. Let $u$ be a user and let $T_u$ denote  $u$'s \hyperref[territory]{territory}. 
We say that $c$ is \textbf{findable} or \textbf{visible} \textit{from} contribution $c'$ (resp. {from} area $a$) if there is a \hyperref[path]{path} between $c$ and $c'$ (resp. between $c$ and any landmark $c'\in a$). We say that $c$ is findable \textit{by} user $u$ if   $c$ is findable from $T_u$.
\furtherwork{Further consideration of path  properties  (e.g. directedness) may help refine the notion of findability.}  \medskip

Based on appropriate notions of findability and relevant \hyperref[filtering]{filters}, various landscape exploration strategies may be defined.  
{I call "\textbf{wayfarer exploration strategies}" strategies that explore MMM landscapes by following paths in the MMM network. %\medskip
Alternative "\textbf{parachuting exploration strategies}" can be used to explore the MMM. In contrast to wayfarer strategies, parachuting strategies ignore the MMM network topology and the MMM data model semantics. They explore MMM contributions in more traditional lexical ways using natural language processing techniques to infer semantic similarities between contributions.
\medskip

Search over the MMM  matches textual 
search queries to MMM landscape areas. 
% APPROXIMATE & EXACT QUERIES
The search query may be \textit{selective}  -- for instance when the user is looking for 
a contribution with a given identifier, or for \mmmtype{question}s recorded between two given dates, or for contributions that have more than 3 authors and are tagged "$@$genetic regulation". In this case, a parachutist strategy is enough. 
The query may also be \textit{approximate} with an exploratory intention behind it. 
In this case, a wayfarer approach might be better suited  to locate epistemically relevant content. 
Given a textual search query $q$ formulated by the user, a wayfarer search strategy (i) picks a MMM landmark (or area) $s$ as  starting point of the search, (2) performs a wayfarer exploration of the landscape following paths (of a certain capped length) outgoing $s$, (3)  circumscribes a relevant  target area $a$ of the landscape that is to be outputted as the search result, and possibly (4) derives an epistemic  link in the MMM landscape between $q$, $s$ and $a$. Graph theoretic properties combined with properties of the MMM data model (cf \S\ref{measuring}) will be useful to formalise the details of these steps. I leave this as an   open problem for now.
\furtherwork{A subsequent article will propose ontology-based  exploration strategies 
to support searching the landscape from a configurable point of view.}}
\medskip

Defining a relevant wayfarer exploration may entail specifying the semantics of  the MMM data model. 
% SEMANTICS
The semantics of the MMM data model have  deliberately been left vague and informal\footnote{We have been using the term "epistemics" rather than "semantics".}. Varying specifications of these semantics are possible. Different specifications may induce  different notions of "consistency" for instance. 
One may specify that two  \mmmtype{existence} nodes are consistent as long as they aren't connected by "contradicting" paths: one path being  a sequence of \equates{} edges, another path being  composed only of \equates{} edges except for one \mmmtype{differsFrom} edge.
A variation of the semantics relevant to scientific researchers might restrict this definition to  \equates{} edges that are tagged with a certain tag -- e.g.  "$@$biological equivalence" or "$@$logical equivalence". 
Specifying the MMM data model semantics allows to define the purpose of a wayfarer  exploration (e.g. checking the consistency of  an area of landscape).
Wayfarer exploration is then as close as we get to reasoning on the MMM.  The semantics determines how  the information conveyed by MMM contribution types should be interpreted, how they should inform decisions and in particular how they should orient the exploration. 
A semantics may specify that \equates{} edges  tagged  "$@$naturalLanguageTranslation" or more specifically    "$@$EN $\to$ FR", don't count. Available language translations would then  be ignored by the wayfarer exploration. 
%(cf \S\ref{filtering}).
\medskip

The wayfarer approach favours discovery of information that is epistemically related to information that has already been discovered. This might favour preparedness of the user to new information. The user's local territory grows gradually to include contributions that are epistemically close to the contributions she  already understands (output area $a$ of a search is probably directly related to input query $q$). 
\label{wayfarer}
\textit{In itself} wayfarer discovery  might also favour epistemic isolation reminiscent of the actual Web's  
echo chambers. Epistemic proximity is however not semantic proximity. Contradictory answers to the same question are for example very close epistemically. So are the  different interpretations of the same question. 
Arguably, humans  can't systematically avoid being directed by confirmation bias  towards the same statements supporting the same views.
But if they 
come across questions on the MMM (e.g. "How do vaccines work?"), there is no obvious natural way 
for them to systematically preempt exposure to the different answers that are  contributed by individuals with different mindsets. The different answers are likely to be \hyperref[implanting]{implanted} on  the same MMM landmark, namely the question they answer.  Epistemic locality does not necessarily entail social locality as very different people can take interest in the same questions.  
\medskip

Furthermore, wayfarer exploration not only leverages the connectedness of the overall MMM network it may also contribute to enhancing it. 
By making users 
aware of information that is already documented, it can help preempt the documentation of redundant content and encourage the building of short \hyperref[bridging]{epistemic bridges} between previously epistemically distant contributions.
\medskip

% ============================================================== %
% ========                                              ======== %
% ============================================================== %
\subsubsection{Safe Reformulating and Translating}\label{translating}

{
One idea or piece of information  can be expressed in multiple ways. Different expressions of the same idea don't necessarily speak to the same people, if only because people understand different languages. Connectedness of the overall MMM network increases the likelihood that there is a path between the different expressions of the same idea: one expression   is \hyperref[finding]{findable from}  another expression. People who understand idea $i$ via expression $E$ can be made aware of the annotations concerning $i$ contributed by people who better understand an alternative expression $E'$ of $i$. 
In the MMM, the epistemic connection between contributions $E$ and $E'$ being explicitly documented, it can be used to ensure "safe passage" between $E$ and $E'$. Arguably, as the formulation of a piece of information changes, the information itself changes, possibly degrades. This may have repercussions on people's understanding of the information. "Safe passage" means that the new formulation is accessed with awareness of ensuing  semantic changes. 
The people who translate or vulgarise information to give other people access to it are sometimes not the  experts who can gauge the extent of the semantic drift caused by the reformulation. But if the drift is documented in the MMM (e.g. with a labelled \equates{} or \mmmtype{differsFrom} edge between the two formulations) by someone who can discern it, then it becomes discussable like any other piece of information. Reformulation on the MMM is like any other ordinary act of deriving information out of pre-existing information. It is a documentable process that can be challenged and justified explicitly. 
\medskip

In addition to promoting epistemic democracy, multiple connected formulations of the same idea  is \textit{a desirable form of redundancy} that 
supports mitigation of \textit{a useless sort of redundancy}.
The useless sort of redundancy  occurs between 
two expressions that are semantically too similar to be humanly   distinguishable. One of the expressions can be removed without risk of reducing, now or in the future, the number of people potentially able to understand the idea. Desirable redundancy might not be relevant to \textit{every} user, but globally it contributes to the connectedness of the MMM network which in turns helps compare MMM contributions and identify useless overlaps. 
}

% ============================================================== %
% ========                                              ======== %
% ============================================================== %
\subsubsection{Epistemic Time Travelling}  \label{timetravelling}

The system architecture that we propose in the supplementary material \ref{implementation} allows for "epistemic time travel". On the MMM, epistemic time travel consists in playing back and forth the history of changes to a landscape.

% ============================================================== %
% ========                                              ======== %
% ============================================================== %
\subsubsection{"Citing the future"}\label{citing}

On the MMM, citation links are conveyed by MMM {edges} (or paths) linking a citing contribution to a cited contribution. Though an appropriate choice of edge  types  and possibly through the documentation of edge {labels}, MMM citation links can be made to convey   \hyperref[glue]{epistemic glue}. 
In traditional documents like scientific articles, citation links   direct the reader to a resource from  the past. In contrast, MMM links direct the reader to a durable location in the landscape. 
The area around this location may evolve over time as contributions (nuances, supporting details, questions, \textit{etc.}) are \hyperref[implanting]{implanted} on it. But the cited contribution location  perdures. % in the future.
Suppose that  article $a$ published in \the\year{} refers to current "\textit{RNA vaccination}" techniques and cites the latest scientific publication on that subject, namely  article $a'$. Both $a$ and $a'$ are implanted in the MMM. Ten years later,  Alice reads $a$. By then, $a$ is still somewhat relevant but  the contents of $a'$ are long outdated. Contribution $a'$ is now deep under a decade of thorough \hyperref[annotating]{annotations} conveying  new understanding of "\textit{RNA vaccination}" and new technical propositions.  The link from $a$ to $a'$ in the MMM doesn't just point Alice in the direction of the old outdated article $a'$, it points her towards a relevant area of the MMM that has been continually updated. 
\medskip

Note that %in the example mentioned in \S\ref{citing}, 
understanding an old article $a$ properly might require more than updated information on the subject  of the article $a'$ that is cited by $a$. 
It might require understanding the historical context in which the link between $a$ and $a'$ was made. 
Epistemic time travel (cf \S\ref{timetravelling}) allows to access the state of the landscape back then and play the succession of events that lead to outdating the contents of $a'$. Thus while $a'$ becomes outdated, the link (epistemic glue) between $a'$ and $a$ continues to convey information of actual worth, for at least as long as $a$ doesn't also outdate. It is the purpose of central "refrigeration mechanisms" (MMM archival) to save some contents such as contribution $a$ from global obsolescence (disappearance from the MMM). Refrigeration mechanisms need to be defined to implement choices of what contributions deserve to be archived and for how long, depending possibly on  (the dynamics of) the contributions'  implantation.

% ============================================================== %
% ========                FILE HIERARCHY                ======== %
% ============================================================== %
\def\FHFIG{
    \begin{figure}[H]
        \hfill \begin{tikzpicture}  
                \node[pen] (w) at (0,0) {Work};
                \node[pen,below=7mm of w] (r)  {Research Work};
                \node[pen,left=10mm of r] (re)  {Researchers-Entrepreneurs};
                \node[pen, right=30mm of r] (t)  {Teaching};
                \draw[pennedin] (re) |- (w);
                \draw[pennedin] (r) -- (w);
                \draw[pennedin] (t) |- (w);
                 \node[existence,inner sep=1pt,below right=1cm and 7mm of re.west,anchor=west,font=\small]  (membersCVS)  {{\tt member\_list.cvs}};
                 \node[pen, below=7mm of membersCVS.west, anchor=west] (talksRE)  {Past Talks};
                \draw[pennedin] (membersCVS) -| ([xshift=4mm]re.south west);
                \draw[pennedin] (talksRE) -| ([xshift=2mm]re.south west);
                \node[existence,inner sep=1pt,font=\small,below right=7mm and 5mm of talksRE.west,anchor=west,]  (speakersCVS)  {\tt guest\_speakers.odt};
                \draw[pennedin] (speakersCVS) -| ([xshift=2mm]talksRE.south west);
                % TEACHING
                \node[pen,below right=1cm and 0mm of t,anchor=west] (ac)  {Algorithmic\\ Complexity};
                \node[pen,below=10mm of ac.west, anchor=west] (gt)  {Graph Theory};
                \node[pen,below=7mm of gt.west, anchor=west] (logic)  {Logic};
                \node[pen,below=7mm of logic.west, anchor=west] (rw)  {Rewriting};
                \draw[pennedin] (ac) -| ([xshift=2mm]t.south);
                \draw[pennedin] (gt) -| (t);
                \draw[pennedin] (logic) -| ([xshift=-2mm]t.south);
                \draw[pennedin] (rw) -| ([xshift=-4mm]t.south);
                % RESEARCH WORK
                \node[pen,below right=1cm and 10mm of r.west,anchor=west] (DL)  {Description Logic};
                \draw[pennedin] (DL) -| ([xshift=8mm]r.south west);
                \node[pen,below right=8mm and 5mm of DL.west,anchor=west] (bibDL)  {Bibliography DL};
                \draw[pennedin] (bibDL) -| ([xshift=2mm]DL.south west);
                \node[existence,inner sep=1pt,font=\small,below right=8mm and 5mm of bibDL.west,anchor=west,]  (article2014)  {\tt article2014.pdf};
                \draw[pennedin] (article2014) -| ([xshift=2mm]bibDL.south west);
                \node[pen,below right=3.4cm and 10mm of r.west,anchor=west] (GP)  {Grant Proposals};
                \draw[pennedin] (GP) -| ([xshift=6mm]r.south west);
                \node[existence,inner sep=1pt,font=\small,below right=7mm and 8mm of GP.west,anchor=west,]  (cnrs)  {\tt cnrs.pdf};
                \node[existence,inner sep=1pt,font=\small,below =6mm of cnrs.west,anchor=west,]  (dfg)  {\tt dfg.pdf};
                \draw[pennedin] (cnrs) -| ([xshift=4mm]GP.south west);
                \draw[pennedin] (dfg) -| ([xshift=2mm]GP.south west);
                \node[pen,below right=5.5cm and 10mm of r.west,anchor=west] (RN)  {Regulation Networks};
                \draw[pennedin] (RN) -| ([xshift=4mm]r.south west);
                \node[pen,below right=8mm and 10mm of RN.west,anchor=west] (bib)  {Bibliography};
                \draw[pennedin] (bib) -| ([xshift=6mm]RN.south west);
                \node[existence,inner sep=1pt,font=\small,below right=8mm and 10mm of bib.west,anchor=west,]  (review)  {\tt 2011review.pdf};
                \node[existence,inner sep=1pt,font=\small,below =8mm of review.west,anchor=west,]  (2017)  {\tt article2017.pdf};
                \node[existence,inner sep=1pt,font=\small,below =8mm of 2017.west,anchor=west,]  (2020)  {\tt article2020.pdf};
                \draw[pennedin] (review) -| ([xshift=6mm]bib.south west);
                \draw[pennedin] (2017) -| ([xshift=4mm]bib.south west);
                \draw[pennedin] (2020) -| ([xshift=2mm]bib.south west);
                % BANs
                \node[pen,below right=4cm and 10mm of RN.west,anchor=west] (BAN)  {Boolean Automata Networks};
                \draw[pennedin] (BAN) -| ([xshift=4mm]RN.south west);
                \node[existence,inner sep=1pt,font=\small,below right=8mm and 12mm of BAN.west,anchor=west,]  (zip)  {\tt 2008--2012.zip};
                \node[existence,inner sep=1pt,font=\small,below =8mm of zip.west,anchor=west,]  (contrex)  {\tt counterexample.png};
                \node[existence,inner sep=1pt,font=\small,below =8mm of contrex.west,anchor=west,]  (DI)  {\tt dark\_information.pdf};
                \node[existence,inner sep=1pt,font=\small,below =8mm of DI.west,anchor=west,]  (pos)  {\tt PositiveCycleSize4.graphml};
                \node[existence,inner sep=1pt,font=\small,below left =8mm and 6mm of pos.west,anchor=west,]  (svsas)  {\tt SynchronismVSasynchronism.pdf};
                \draw[pennedin] (zip) -| ([xshift=10mm]BAN.south west);
                \draw[pennedin] (contrex) -| ([xshift=8mm]BAN.south west);
                \draw[pennedin] (DI) -| ([xshift=6mm]BAN.south west);
                \draw[pennedin] (pos) -| ([xshift=4mm]BAN.south west);
                \draw[pennedin] (svsas) -| ([xshift=2mm]BAN.south west);
                \node[existence,inner sep=1pt,font=\small,below right=8.7cm and 10mm of RN.west,anchor=west] (pres)  {\tt presentation.pdf};
                \draw[pennedin] (pres) -| ([xshift=2mm]RN.south west);
                \node[existence,above right=8mm and 26mm of bib] (bib2)  {Bibliography};
                \draw[pertains,bend left=10] ([xshift=8mm]bibDL.south) to[out=-40,in=180] (bib2);
                \draw[pertains,bend left=10] (bib) to[out=-10,in=-160] (bib2);
        \end{tikzpicture} \\[-9.8cm]
        \includegraphics[scale=0.5,trim=1pt 0 0 0, clip]{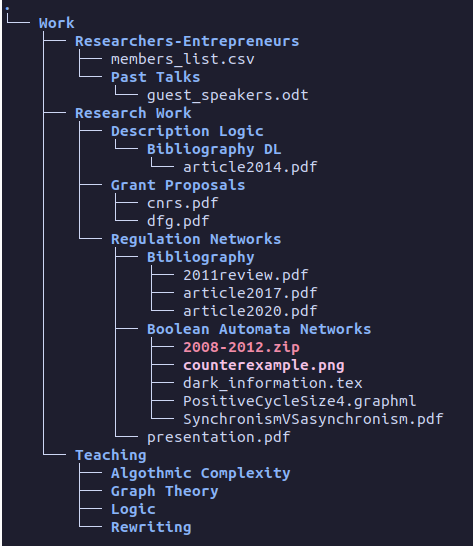}
            \caption{Bottom left: a simplified version of a part of the hierarchy of files and folders that a researcher could have on one of their devices, as presented by the Tree program. Above right: 
            the corresponding MMM landscape area. % for the corresponding file hierarchy.  
            I recall that all visual representations of MMM formatted information in this article are arbitrary. Precisely, a UI could in this case provide a visual representation of the MMM to the user identical to the terminal output of the Tree program. 
            File and folder paths can be documented as  tags in contributions' tag sets. \mmmtype{pennedIn} edges can be tagged "$@$file contained in" or "$@$sub-directory of". Other MMM contributions can be added. The diversity of MMM edge types can be leveraged to complete the file system's hierarchical organisation and play a role similar to rich symbolic links.  
            }
        \label{fig-FH}
        \end{figure}
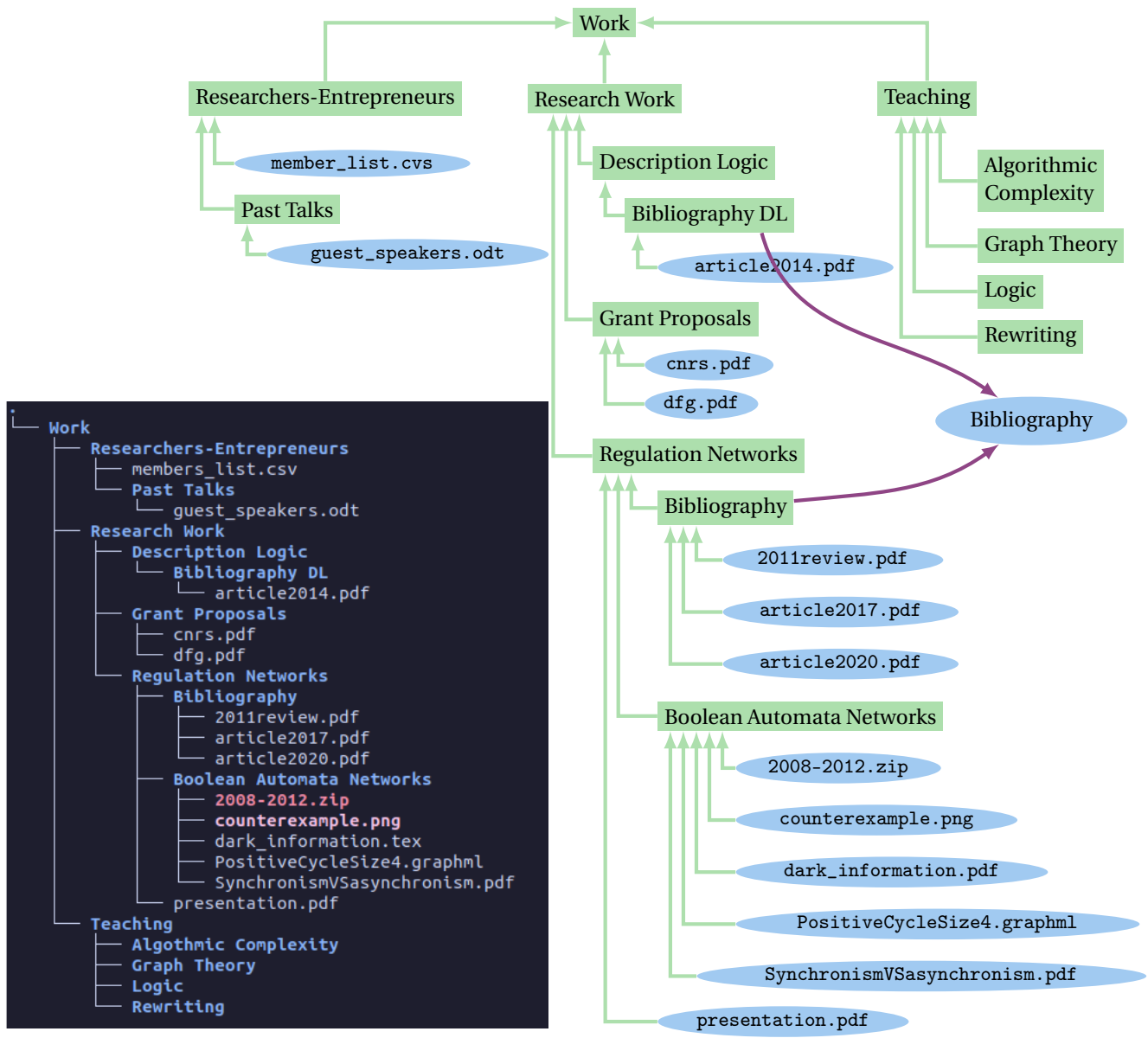
}

\subsubsection{File and Folder Organising}\label{FH}

MMM contributions (e.g. \mmmtype{existence} nodes) can be used to represent bibliographical references to   documents that exist outside the MMM. Similarly, MMM contributions can be used to represent resources like files and folders existing in a file hierarchy. Mapping (parts of) a device's local file hierarchy into the user's local MMM territory would allow the user to seamlessly  navigate 
between their MMM notes and their local file hierarchy (see Fig. \ref{fig-FH} and \SMref{socks}.). All contributions resulting from this mapping should be marked  with a distinct mark, say \mmmmark{FH}. And all contributions marked  \mmmmark{FH} should be considered  private and unpublishable.

\FHFIG

\subsection{Landscape Sharing Activities}\label{activities-sharing}

In this section we discuss the social dimension of the MMM proposal.\medskip

In the sequel let us assume the following context. 
There exists software interfaces for human users to interact  with the MMM. The software are locally installed on users' devices. Each user has their own local \hyperref[territory]{territory} as mentioned in \S\ref{territory}. A user's local territory is stored on the user's device(s). Devices (belonging to the same or to different users) connect to each other (over the internet or a LAN) to exchange MMM contributions. 
I say these devices are \textit{MMM hosts}, participating in the distributed MMM network. I sometimes refer to users  as \textit{peers} in this network. I don't assume that MMM hosts operate as servers. Details of the implementation of the distributed MMM network and  the software involved are given in the supplementary material~\ref{implementation}.

\begin{figure}[H]
    \centering
    \includegraphics[scale=0.4,trim={3cm 0 3cm 0},clip]{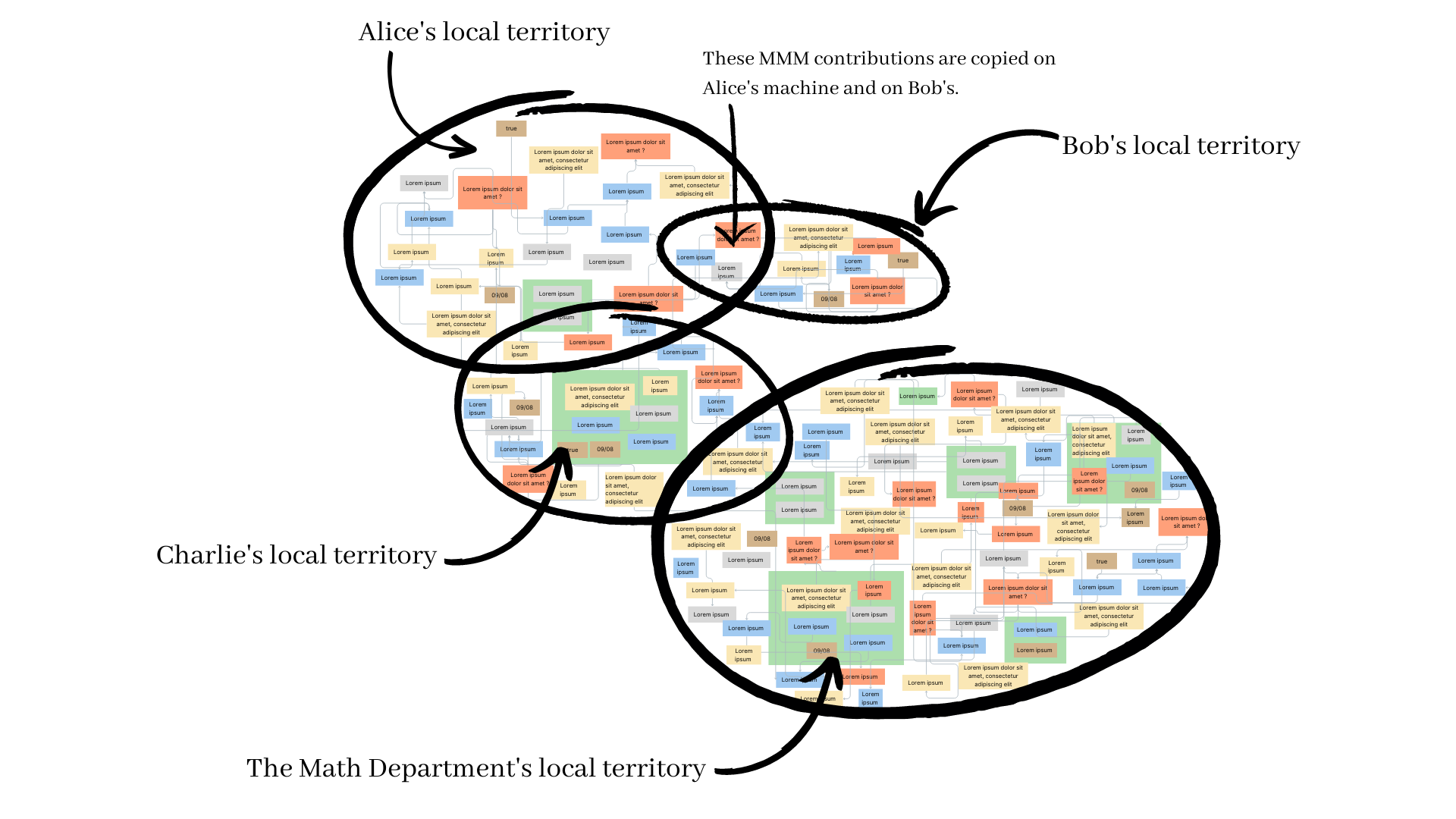}
    \caption{The MMM is the reunion of all MMM contributions stored on the local territories of users. As users share MMM contributions with each other, replication of the material occurs. Some contributions are copied on multiple hosts.}
\end{figure}

% ============================================================== %
% ========                REWARDING                     ======== %
% ============================================================== %
\subsubsection{Rewarding}\label{trickling} %\label{rewarding}

Suppose that Alice publishes contribution $c_A$ (e.g. a description of how she feeds the lab mice). Then Bob publishes contribution $c_B$ as an annotation of $c_A$ (maybe a specification of the problems Bob encounters in carrying out an experiment using Alice's mice). Later Charles publishes $c_C$ which, among many other things, builds on $c_B$. And it turns out that Charles ends up greatly rewarded for $c_C$ (maybe $c_C$ gets published in a prestigious academic journal, maybe it is the object of a Nobel Prize). I propose to use the collectively documented train of thought  $c_A\longrightarrow c_B\longrightarrow c_C$ to formally and proportionately acknowledge Alice's and Bob's participation in the work that eventually produced $c_C$.  
Charles (or the prestigious publisher or the Nobel Foundation) may locally mark $c_C$ as \rewardmark{}.
The  \rewardmark{} mark can be parametrised with (1)  data specifying or referring to the details of the reward (e.g. "Nobel Prize"), (2) data specifying the MMM distance to the rewarded contribution ($0$ in the case of $c_C$), and (3) the identifier of the rewarded contribution if the distance is greater than $0$. 
The distributed system through word of mouth (successive local shares, cf \S\ref{sharing})
propagates  \rewardmark{} marks. Charles' reward can "trickle" down to Bob  and then Alice. Every tricking step increments the distance recorded as parameter to the \rewardmark{} mark.
 On reception of a homologous copy ${c'_A}$ of Alice's contribution $c_A$, a comparison can be made\footnote{By the concierge module, cf \SMref{logic-layer}.} between the \rewardmark{} marks of both copies. The minimum distance to $c_C$ can be kept as parameter of the \rewardmark{} mark.  I call this mechanism  "\textbf{trickling reward}".
The topic of acknowledgement and reward is  discussed in \cite{fumier}.

% ============================================================== %
% ========              COLLABORATING                   ======== %
% ============================================================== %
\subsubsection{(Slow) Collaborating}\label{collaborating}
The MMM is a collective document: users  can  offer answers to the same \mmmtype{question}s, they can \mmmtype{nuance} each others' statements, they can add to each others' \hyperref[pens]{mutable pens}, they can connect  each other's contributions using appropriate  MMM edges \textit{etc}. \medskip

% ============================================================== %

Consider a traditional  document $d$ decomposed into area $a_d$ of  the MMM 
($a_d$ is the "MMMified" version of $d$, {cf \S\ref{documenting}}).  Several users may work simultaneously on  $a_d$, \hyperref[annotating]{annotating} contributions that are already in $a_d$ and episodically sharing with each other their annotations. 
Collaborators don't have to keep copies of every contribution in $a_d$. \textbf{Floating and dangling edges are possible.} Alice may  keep a copy of the edge $e$ going from her contribution $c_A$ to Bob's  contribution $c_B$, without storing a local copy of $c_B$. Alice might regard the label and type of $e$ as enough information about $c_B$. 
\medskip

The MMM solution isn't optimised for fast communication. 
It is to natively support "\textbf{slow-first collaboration}" which is when collaborators can work without having fast editor access at any moment to what their collaborators are doing (real time  read access remains a possibility). 
Slow-first collaboration is sometimes enough and sometimes preferable (cf \SMref{slowcollaboration}). 
\medskip

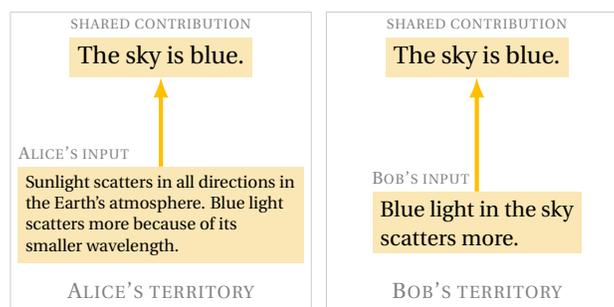
\begin{figure}[H]

    \begin{center}
        \begin{tikzpicture}  
            \node[narrative] (n) at (2,3.4) {The sky is blue.};
            \node[narrative,font=\scriptsize,text width=3.6cm] (nnn) at (2,1.3) {Sunlight scatters in all directions in the Earth's atmosphere. Blue light scatters more because of its smaller wavelength.};
            \draw[details] (nnn) -- (n)  ; % node[midway,yshift=-4pt] { \mmmtype[detailscolor]{supports}} ;
            \draw[help lines,step=40mm,gray!40] (0,0) grid (4,4);
            \node[text=gray] at (2,0.3) {\sc\small Alice's territory};
            
            \def\myshift#1{\raisebox{1ex}}
    \path [postaction={decorate,decoration={text along path,text=red,text align=center,text={|\scshape\scriptsize\color{gray}\myshift|shared contribution}}}]      (n.north west) to  (n.north east);
    \path [postaction={decorate,decoration={text along path,text=red,text align=center,text={|\scshape\scriptsize\color{gray}\myshift|Alice's input}}}]      (nnn.north west) to  ([xshift=-4mm]nnn.north);
    
            \begin{scope}[xshift=4.2cm]
            \draw[gray!40] (0,0) rectangle ++(4,4);
            \node[text=gray] at (2,0.3) {\sc\small Bob's territory};
            \node[narrative] (n) at (2,3.4) {The sky is blue.};
            \node[narrative,font=\small]   (nn) at (2,1.2) {Blue light in the sky\\ scatters more.};
                \draw[details] (nn) -- (n) ;
                \def\myshift#1{\raisebox{1ex}}
                \path [postaction={decorate,decoration={text along path,text=red,text align=center,text={|\scshape\scriptsize\color{gray}\myshift|shared contribution}}}]      (n.north west) to  (n.north east);
                \path [postaction={decorate,decoration={text along path,text=red,text align=center,text={|\scshape\scriptsize\color{gray}\myshift|Bob's input}}}]      (nn.north west) to  ([xshift=-1mm]nn.north); 
            \end{scope}
    
        \end{tikzpicture} 
        \end{center}
        \caption{Automatic assisting mechanisms for redundancy management can leverage the properties of the area surrounding Alice's contribution $c_A$ and Bob's contribution $c_B$ in order to detect similarity between $c_A$ and $c_B$. Application code make encourage Alice and Bob  to merge $c_A$ and $c_B$, to obsolete one of them or  to document an explicit relationship between them. {\todo{++ communication work in 2 separate contributions contributed independently by 2 users  + fast collaboration in a RTC shared contribution.}}}
        \label{fig-collaboration-merge}
    \end{figure}

Users edit \textit{different} contributions and then link them if relevant. 
To support real-time collaboration \textit{within} contributions (collaborators concurrently editing the same contributions), will require nesting finer-grained CRDTs  in the native coarse-grain MMM CRDT-like structure
\cite{kleppmann2017conflict}.
\medskip

Contributions are atomic units  in the MMM just like characters are atomic units in traditional digital documents (cf Table \ref{googletable}). Let us expand on a recommendation made in \S\ref{drafting} which frontend code can implement:

\begin{bestpractices}{} Just like you don't half-way type a character in a digital document, avoid half-way documenting a contribution in the MMM.
When you are  done editing  contribution $c$ mark it so as  to indicate that $c$ is finished (which is different from meaning $c$ is \textit{definite} because of \hyperref[obsoleting]{obsoleting} and \hyperref[versioning]{versioning} mechanisms described in  \S\ref{drafting}), e.g. mark it as  \mmmmark{synchronisable}. 
Only allow \mmmmark{synchronisable} contributions to be accessed from other devices. % at the end of a work session
\end{bestpractices}

% ============================================================== %
% ========                 GOOGLE                       ======== %
% ============================================================== %
\begin{center}
    \begin{table}[H]
        \begin{tabular}{|p{0.47\textwidth}|p{0.47\textwidth}|}
            \hline
            \rowcolor{black!5} \multicolumn{1}{|c|}{\textbf{Google Docs}} & \multicolumn{1}{c|}{\textbf{MMM}} 
            \\\hline
            A Google Docs  necessarily belongs to the holder of the Google account that it was created from. The owner and the editors of a Google Docs have more rights than its commentators. &
            Users don't have statuses, only contributions do. Any user can annotate  a contribution that she has access to. Alice  may annotate Bob's contribution, Bob may reject Alice's request to share her annotation with him, and Alice may persist her annotation and share it with Charlie even if it displeases Bob.\\\hline
            The finest granularity of collaboration is possible. 
            Editors can 
            see the effect of each other's keystrokes in real-time. Their keystrokes can interact with each other when they simultaneously edit the same area of the document. & 
            Fast fine-grained collaboration isn't native to the MMM. The MMM only supports coarse-grained collaboration  where collaborators don't simultaneously  edit the same atomic unit of information.
            \\\hline
            Coarse-grained collaboration is supported. Users can comment and make suggestions to the main document. 
            Their annotations  are spatially tied  to character positions in the document. & 
            
            Annotations are \textit{epistemically} tied to the landscape.  {This is a simple case of CRDT because  changes to the landscape may be applied without knowledge of the original context in which the changes were made.}
            \\\hline
            %%%
             Versioning allows viewing past states of the document and to compare versions of the document after changes have been applied to it. & Changes are 
            %natively 
            encapsulated in contributions and logged (cf \S\ref{event-sourcing}). The MMM system supports fine-grained "epistemic time travel" discussed in \S\ref{cold}. \\\hline
        \end{tabular}
        \caption{Collaboration with Google Docs and collaboration with the MMM.}
        \label{googletable}
    \end{table}
    \end{center}

% ============================================================== %
% ========                TOPICS                        ======== %
% ============================================================== %
\subsubsection{Defining Topics}\label{topics}

In the MMM, topics -- e.g.  "holiday ideas" or "AI" --   are typically documented in the user's territory as the labels of {\mmmtype{existence} nodes}.
They can also be documented as the labels of any other kind of MMM contribution including  \mmmtype{question} nodes, \mmmtype{narrative} nodes, pens and edges.  When Alice records a new contribution $c$ in her \hyperref[territory]{territory}, she \hyperref[implantation]{implants} $c$  (she links $c$  to other contributions in her territory). If this materialises a \hyperref[path]{semantic path}  between $c$  and the \mmmtype{existence} node labelled "holiday ideas", then $c$ can be considered related to the topic "holiday ideas" 
If the implantation of $c$  in Alice's territory materialises a path between $c$  and the \mmmtype{existence} node labelled "AI", then $c$  will be regarded as related to the topic of "AI". \medskip

We define the \textbf{topic anchor} to be the MMM contribution whose label -- e.g.  "holiday ideas" or "AI" --  gives the topic name. \medskip

{
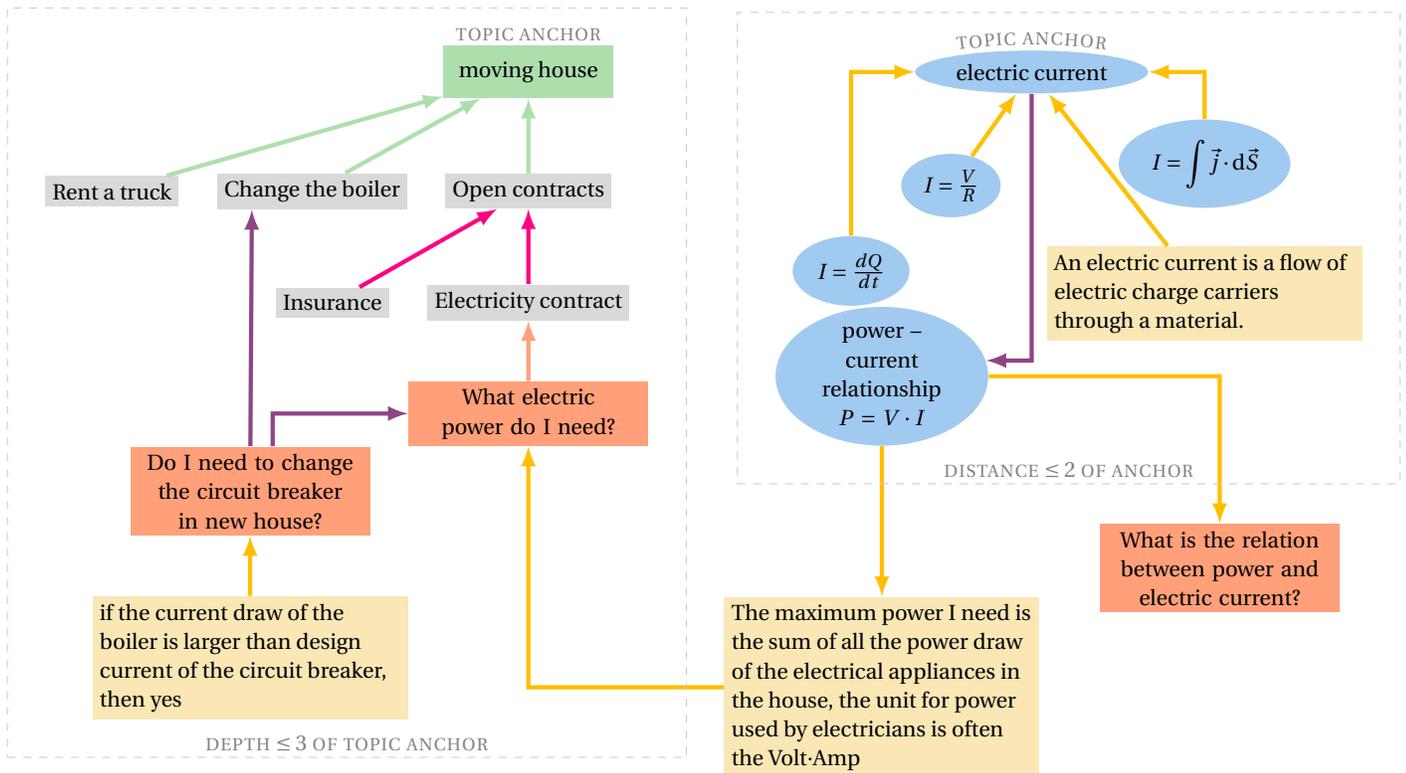
\begin{figure}[H]
    \centering\small
    \begin{tikzpicture}  
        % TOPIC 1: DEMENAGEMENT
        \node[pen,inner sep= 2mm] (t1) at (0,0) {moving house};
        \node[action,below=1cm of t1] (contracts)  {Open contracts};
        \node[action,below=1cm of contracts] (elec) {Electricity contract};
        \node[action,left=5mm of elec] (ins) {Insurance};
        \node[action,left=5mm of contracts] (boil) {Change the boiler};
        \node[action,left=5mm of boil] (truck) {Rent a truck};
        \node[question,below=8mm of elec,text width=3cm] (qc) {What electric power do I need?};
        \node[question,below left=0mm and 5mm of qc,text width=3cm] (qd) {Do I need to change the circuit breaker in new house?}; % the  load capacity of 
        \node[narrative,below right=20mm and 10mm of qc,text width=4cm] (ac)  {The maximum power I need is the sum of all the power draw of the electrical appliances in the house, the unit for power used by electricians is often the Volt\(\cdot\)Amp};
        \node[narrative,below=8mm of qd,text width=4cm] (ad)  {if the current draw of the boiler is larger than design current of the circuit breaker, then yes};
        \draw[pennedin] (contracts) -- (t1);
        \draw[pennedin] (boil) -- (t1);
        \draw[pennedin] (truck) -- (t1);
        \draw[instantiates] (elec) -- (contracts);
        \draw[instantiates] (ins) -- (contracts);
        \draw[questions] (qc) -- (elec);
        \draw[pertains] ([xshift=3mm]qd.north) |- (qc);
        \draw[pertains] (qd) -- ([xshift=-8mm]boil.south);
        \draw[answers] (ac) -| (qc);
        \draw[answers] (ad) -- (qd);
        %
        % TOPIC 2: ELECTRICITY
        \node[existence,right=40mm of t1] (t2) {electric current};
        \node[existence,below left=1cm and -5mm of t2] (D1)  {$I=\frac{V}{R}$};
        \node[narrative,below right=5mm and 8mm of D1,text width=4cm] (D2)  {An electric current is a flow of electric charge carriers through a material.};
        \node[existence,below left=5mm and 3mm of D1] (D3)  {$I=\frac{dQ}{dt}$};
        \node[existence,above =5mm  of D2] (D4)  { \(\displaystyle I = \int \vec j \cdot \mathrm d \vec S \)};
        \node[question,below right=5.8cm and -2mm of t2,text width=3cm] (Q4)  {What is the relation between power and electric current?};
        \node[existence,above=2cm of ac,text width= 2cm,inner sep=0mm] (PI)  {power -- current\\ relationship \\ \(P = V\cdot I\)};
        \draw[answers] (D1) -- (t2);
        \draw[answers] (D2) -- (t2);
        \draw[answers] (D3) |- (t2);
        \draw[answers] (D4) |- (t2);
        \draw[answers] (PI) -| (Q4);
        \draw[pertains] (t2) |- ([yshift=3mm]PI);
        \draw[details] (PI) -- (ac);
        % %
        \node [draw=gray!40, dashed, inner sep=5mm, fit={(t1) (truck) (qc) (qd) (ad)}] (ex1) {};
        \node[text=gray] at ([yshift=2mm]ex1.south) {\sc\small depth $\leq 3$ of topic anchor};  
        \node [draw=gray!40, dashed, inner sep=5mm, fit={(t2) (D1) (D2) (D3) (D4) (PI)}] (ex2) {};
        \node[text=gray] at ([yshift=2mm]ex2.south) {\sc\small  distance $\leq 2$ of anchor};  

\def\myshift#1{\raisebox{3ex}}
\path [postaction={decorate,decoration={text along path,text=red,text align=center,text={|\scshape\color{gray}\myshift|topic anchor}}}]      (t1.west) to  (t1.east);
\def\myshift#1{\raisebox{1.5ex}}
\path [postaction={decorate,decoration={text along path,text=red,text align=center,text={|\scshape\color{gray}\myshift|topic anchor}}}]      (t2.west) to [bend left=10] (t2.east);

        \end{tikzpicture}  

    \caption{Two topics: "moving house" and "electric current". The topic \textit{anchors} are respectively the pen and the \mmmtype{existence} node in which each topic name is documented. Topic scopes are captured in topic \textit{extents} and can overlap.
    {In this example, if the topic {extents} are restricted to the dashed rectangles, then no contribution yet falls into the scope of both topics at once. }
    }
    \label{fig-topics}
\end{figure}}

The \textbf{topic extent}   defines an area of the landscape surrounding the topic anchor. This area contains landmarks that  are considered to fall within the scope of the topic. 
The area can be defined in terms of distance or depth -- e.g. contribution $c$ is relevant to the topic if $c$ is at a distance $d\leq 3$ of the topic anchor $a$, or at a depth $d\leq a$ underneath $a$. 
Features of the {MMM format} may be exploited to refine the delineation of topic extents (cf \S\ref{measuring}). 
For instance, a user might want to filter out of the topic's extent, contributions connected via \mmmtype{relate} or \mmmtype{relatesTo} edges. Or she might not be interested in \mmmtype{question}s whose answers fall outside the topic extent or whose answers are shallow. 
\medskip

Formally a MMM topic is a couple $T=(a,e)$ where $a$ is the topic anchor and $e$ is the topic extent.\medskip

Let $T=(a,e)$ be a topic whose  extent $e$ circumscribes an area of radius $n\in\mathbb{N}$ around contribution $a$. Let $c$ be a contribution in that area at a distance  $m\leq n$ of $a$. Let $e_c$ be the area of radius $n-m\in\mathbb{N}$ around contribution $c$. We say that topic $T_c=(c,e_c)$ is \textbf{inherited} from topic $T$. More generally, a topic $T_c=(c,e_c)$ inherited from topic $T=(a,e)$ is such that contribution $c$ belongs to the area circumscribed by $e$, and  this area contains the area circumscribed by $e_c$.\medskip

MMM topics are for \textbf{sharing and synchronising  MMM contributions} on a need-to-know basis. The "need" is captured in advance in the topic extent.

% ============================================================== %
% ========              SHARING                         ======== %
% ============================================================== %
\subsubsection{Sharing Contributions}\label{sharing}

% ============================================================== %
% ========                RECEIVING                     ======== %
% ============================================================== %
\label{sending}

Users can send MMM contributions to each other, individually or by batches. Contributions that Alice has shared with Bob or received from Bob are marked with a parametrised \sharemark{} mark. \medskip

When Alice receives a contribution $c$ from Bob, $c$ is initially marked as \mmmmark{new} on Alice's territory. A customisable amount of systematic \hyperref[filtering]{filtering} can be applied to \mmmmark{new} contributions on delivery (cf \S\ref{concierge}, \S\ref{filterer} and \S\ref{topographer} in \S\ref{logic-layer}). Alice  might only want to see contributions that are already well \hyperref[implanting]{implanted} in the global MMM or in the sender's local  territory. If the  \mmmmark{new} contribution $c$ is not automatically filtered out, it remains for Alice to reject it or to accept it. If Alice rejects $c$, $c$ is deleted (not obsoleted) from her territory. If Alice accepts $c$, then the \mmmmark{new} mark on $c$ is removed. If there already is a homologous copy of $c$ on Alice's territory, it is \hyperref[merging]{merged} with $c$.
{Sharing maintains the CRDT like properties of the set of landscapes mentioned in \S\ref{updating}.
When Alice accepts a new contribution, her local territory grows 
according to the partial order $\sqsubseteq$ on landscapes defined in \S\ref{merging}.} % (see also \S\ref{updating}).}%
\medskip

% ============================================================== %
% ========               CONTRACTS                      ======== %
% ============================================================== %
I propose that \textbf{share contracts} be associated with MMM contributions when they are shared. 
The default share contract forbids the recipient of a MMM contribution from communicating the source's address to a third party. Only the source host can relax the contract. 
Possibly, {in a GDPR-compliant way,} the contract lists alternative hosts who have given their permission to be known as alternative hosts of the MMM material to share.
The contract may contain some copyright clauses  restricting what the recipient host can do with a shared contribution \cite{fumier}. It may formalise a non-disclosure agreement.
\furtherwork{Research work is needed to define ways of enforcing contracts. Alice must not be able to change the contract she has with Bob. The software she uses to connect with Bob  must not violate the contract. And/or peers with whom Alice connects shouldn't accept to interact with Alice in ways that violate the contracts  applying to the data Alice has.}
\medskip

% ============================================================== %
% ========                SUBSCRIBING                   ======== %
% ============================================================== %
\subsubsection{Subscribing to Topics}\label{subscribing}

Users can subscribe to topics. The anchors  of the topics that  they subscribe to must  exist on their local territories. To subscribe to topic $T=(c,e)$, 
the user must compose a  "\textbf{subscription request}" and send it to one or several MMM hosts. The user not only specifies what information they  are interested in acquiring, they also specify  who they want to get the information from. 
The subscription request is a message with the following data:
\begin{enumerate}
    \item A MMM topic $T=(c,e)$ whose anchor $c$ can be found on the user's local territory. 
    \item How often and until when the user wishes to receive $T$-related material.  
    \item A host from which  the user wishes to get $T$-related material / the host to which the subscription request is sent.  This host must have the anchor contribution $c$ on their local territory as well if they are to serve $T$-related material to the subscriber.
    \item A subscription contract specifying if the subscription can be forwarded by the recipient to an alternative $T$-serving host, and by the sender to an alternative $T$-interested subscriber\footnotemark. 
\end{enumerate}

\footnotetext{Suppose Alice is subscribed to  Bob's $T$-related contributions. The subscription contract  might  allow Bob to forward Alice's subscription over to Charlie whom Bob gets his $T$-related information from -- assuming Bob's subscription contract with Charlie allows it. And the Alice-Bob contract might specify that Alice can forward the data she has on Bob to Eve so that Eve can  receive   $T$-related contributions directly from Bob. }

Serving a subscription consists in sharing $T$-related material found on one's territory to a subscribed peer (cf \S\ref{sending}). 
\furtherwork{Further work is needed to  ensure that  serving subscription material is {efficient}. Work is in particular needed to determine the respective responsibilities of client and server in identifying subscription material in the landscape.   Identifying the contributions $c'$ that fall into a topic $T$'s scope requires running through the landscape and possibly computing inherited topics.} 
To facilitate the task of serving MMM material to his subscribers, Bob might want to constrain the  topic extents that he is willing to serve, e.g. to "\textit{one-size-fits-all-subscribers}" penned areas. He might leave most of the measuring and filtering work to his subscribers. \medskip

As suggested in \S\ref{sending}, users can send each other unrequested MMM contributions. I propose to deal with these spontaneous exchanges of MMM material as \textbf{subscription \textit{invitations}}. To share contribution $c$ with Bob, Alice invites Bob to subscribe to $T=(c,e)$ where  $e$ is empty if Alice wants to share nothing else than $c$ with Bob. The extent $e$ can alternatively define an area of radius $1$ around $c$ if Alice wants to share $c$ and immediate annotations to $c$. When he receives Alice's invitation to subscribe to her $T$-related material, Bob can modify certain parameters of the proposed subscription. He can for instance modify the extent of the topic and reduce the frequency of $T$-related news he will be receiving from Alice. I suggest that conversely, when Alice shares contribution $c$ with Bob, especially if Alice is the author of $c$, then  Alice automatically subscribes to Bob's copy of $c$ so that she episodically receives from Bob updates and annotations concerning $c$.
\medskip

Alice's  copy of the anchor contribution $c$ is marked as  \subscribemark{}, and possibly, so are also  other contributions that fall into the topic's scope $e$. 
The \subscribemark{} mark is parametrised with some of the subscription parameters. 
%Bob sends  $T$-related contributions  to Alice in response to her request. 
\medskip

% ============================================================== %
% ========                PODS                          ======== %
% ============================================================== %

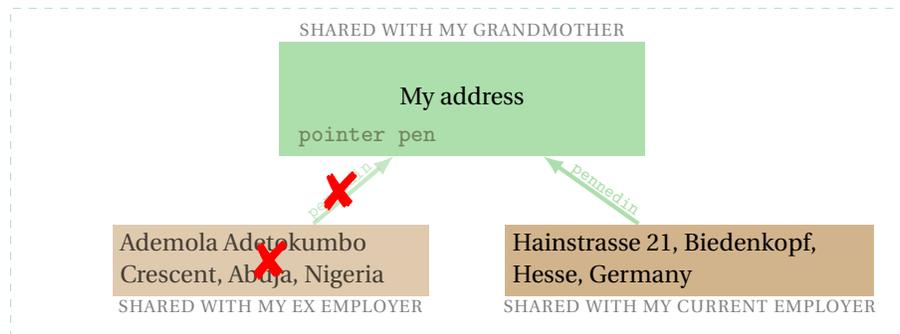
\begin{figure}[H]
    \centering
    \begin{tikzpicture}\draw[pencolor,dashed, opacity=0.9] (0,0) rectangle ++(12,4.4);
        \node[pen,inner sep=6mm] (adr) at (6,3.2) {
            \hspace{1cm}My address\hspace{1cm}};
        \node[datavalue,below left=0.9cm and -20mm of adr,text width=4cm,opacity=0.7] (Ni) {Ademola Adetokumbo Crescent, Abuja, Nigeria};
        \node[datavalue,right=10mm of Ni,text width=4.7cm] (De) {Hainstrasse 21,
        	Biedenkopf,	Hesse, Germany};
     \path [draw,pennedin,opacity=0.7] (Ni) to      node[sloped,yshift=3pt,pos=0.4] (Niedge) {\mmmtype[pencolor]{pennedin}}     (adr);
        \path [draw,pennedin] (De) to      node[sloped,yshift=3pt,pos=0.4]  {\mmmtype[pencolor]{pennedin}}     (adr);
\node[below left=-5mm and -22mm of adr.south west,font=\small] {\color{pencolorSTR}\mmmtype[pencolorSTR]{pointer pen}};
\node[text=red,font=\huge] at (Ni) {\nope};
\node[text=red,font=\huge] at (Niedge) {\nope};

\def\myshift#1{\raisebox{-5ex}}
\path [postaction={decorate,decoration={text along path,text=red,text align=center,text={|\scshape\small\color{gray}\myshift|shared with my ex employer}}}]      (Ni.west) to  (Ni.east);
\path [postaction={decorate,decoration={text along path,text=red,text align=center,text={|\scshape\color{gray}\small\myshift|shared with my current employer}}}]      (De.west) to  (De.east);
\def\myshift#1{\raisebox{6ex}}
\path [postaction={decorate,decoration={text along path,text=red,text align=center,text={|\scshape\color{gray}\small\myshift|shared with my grandmother}}}]      (adr.west) to  (adr.east);

    \end{tikzpicture}
    
    \caption{Sharing information by value or by pointer. Contributions marked with {\color{red}\nope} are \mmmmark{obsolete}d contributions.  Although the  MMM format is not primarily designed to store and manage mutable data like phone numbers and addresses, mutable 
    MMM contributions (cf \S\ref{mutable}) can nonetheless marginally support the sharing of mutable data by pointer or by value.
     N.B.: The assumption is that the MMM node storing Alice's address  is part of Alice's local territory. \furtherworkcaption{Combining  MMM concepts with concepts from the \href{https://solidproject.org/}{Solid project} \cite{sambra2016solid}, a proposal for personal data management will be made in a follow-up article.}}
    \label{editmutablecontribution}
\end{figure}
\furtherworkmark{}

% ============================================================== %
% ========              SON                             ======== %
% ============================================================== %

Subscribing to a topic on the MMM is comparable to joining a Semantic Overlay Network (SON) \cite{crespo2002semantic,crespo2004semantic} Connections between peers are determined by semantics. SONs are clusters of peers that share an interest for a concept {(e.g. minimal tech music)}  and have resources related to that concept. Resources and peers are assigned to concepts and concepts  are themselves organised into a predefined taxonomy of concepts. 
The taxonomy is leveraged to forward queries to the right peers and  optimise search performance.  
In the MMM system, the set of all peers subscribed to topic $T=(c,e)$
is reminiscent of a SON, even if contribution $c$ is not part of a common predefined hierarchy of concepts and might not represent a concept at all ($c$ might be a \mmmtype{question} or something else). We don't necessarily have the potential for a peer-to-peer connection between any two peers in the cluster.

% ============================================================== %
% ========                SYNC                          ======== %
% ============================================================== %
\subsubsection{Synchronising Across Devices}\label{syncing}\label{synchronising}

Synchronising an area $a$ of the landscape with another device one owns is similar to sharing it with a peer. 
% MARKS
The local copy of contribution $c$ on device $d_0$  is marked with the \syncmark{} mark parametrised with the list of other devises that $c$ is copied to. 
In contrast to sharing, synchronising usually doesn't ignore the house-keeping marks.  Typically, if a device has contribution $c$ marked as \mmmmark{highlighted}, other devices of the same user will also have $c$ marked as \mmmmark{highlighted}.
\medskip

Relay servers can be used to streamline cross-device synchronisation. Data transiting through these thin servers should be end-to-end encrypted.

% ============================================================== %
% ========                PUBLISHING                    ======== %
% ============================================================== %
\subsubsection{Publishing Contributions}\label{publishing}

I propose to promote a strong notion of publication. 
\begin{principle}{Irrevocable Publicness}\label{irrevocable}
    Information that is published can't be unpublished nor altered by any number of individuals, not even by the author or publisher of the information. 
\end{principle}

A strong notion of publication requires an  accompanying  paradigmatic shift towards tolerance for miscommunication and misinformation. This is discussed in  \SMref{errors}.
\medskip

% ============================================================== %

Publishing a contribution to the MMM starts with upgrading its \hyperref[status]{status} to \mmmmark{public}. 
Importantly, if Alice didn't create $c$ herself, if she got $c$ from Bob, then the propagation contract she has with Bob might forbid Alice from publishing $c$.  
Contributions marked as \mmmmark{public} may remain unseen by everyone except their creator. But the point of making a contribution public is to share it. Sharing a \mmmmark{public} contribution happens as described in \S\ref{sharing}. Like other contributions,
{public} contributions propagate  through word of mouth (successive shares). So they don't necessarily propagate. A public entity like a university participating in the distributed MMM network may  reject all \mmmmark{public} contributions it receives unless the contributions  are from its affiliated researchers  and  are highly \hyperref[implanting]{implanted} in the university's local territory.
\medskip

% ============================================================== %

From the moment a \mmmmark{public} contribution is shared, the publication is virtually irrevocable. This is because (1) the  \mmmmark{public} status of a contribution can't be downgraded and (2) sharing is irrevocable. \medskip

% ============================================================== %

\begin{bestpractices}{} 
use \hyperref[versioning]{versioning} and \hyperref[obsoleting]{obsoleting} mechanisms to share a correction (eg a typo correction) for a contribution that has already propagated over the distributed network. 
\end{bestpractices}

% ============================================================== %

The author of a \mmmmark{public} contribution  $c$ shares equal \textbf{ownership} of \textbf{and control} over $c$ with anyone who has a copy of $c$. 
Anyone is free to make a local copy of $c$, share $c$ and  annotate $c$. 
No-one (not even the author) can directly modify the label and type  of $c$ (cf Table \ref{table-updates} below). However, anyone can modify the epistemic environment around $c$ by (publicly) commenting on $c$, supporting it, detailing it, nuancing it, questioning it, red-flagging it, relating it to other information \textit{etc}. So anyone can potentially sway the way others interpret $c$. 
As mentioned before, sheer quantity and repetition of information  in relation to $c$ does not necessarily translate into visibility of this information on the MMM however (cf \S\ref{aggregating} and \SMref{implantation}).
So no single user or group of users has the exclusive power to definitely sway the interpretation of $c$.\medskip

Users can't delete public contributions from the MMM once they have been shared. They can only delete  their own local copies.
I recall that the MMM is not suited for the documentation of all kinds of content. It is especially ill-suited for content that \textit{can't} be disputed such as feelings. It is better suited for analytical  information that has some collective value such as scientific contributions.
All the current owners of a local copy of a \mmmmark{public} contribution  $c$ could mark $c$ as \mmmmark{obsolete} and $c$ could eventually entirely disappear from the MMM. 
Contributions that disappear shortly after they propagate  might not  be deemed of the same quality\footnote{This doesn't necessarily refer to a \textit{binary} quality of information such as good/bad. See \S\ref{measuring} and \SMref{binary}. } as persisting pervasive contributions. 
I propose that \furtherwork{collective obsolescence be leveraged to inform the design of smart archiving mechanisms 
capable of identifying  digital public information that is worth  archiving {(cf \SMref{cold})}. }
\medskip

% ============================================================== %
% ========              UPDATE TABLE                    ======== %
% ============================================================== %

\begin{table}[H]
    \begin{center}
\begin{tabular}{|p{7cm}|>{\centering\arraybackslash}p{3cm}|>{\centering\arraybackslash}p{3cm}|}
    \hline
    \multicolumn{3}{|c|}{\textbf{Allowed landscape modifications depending on contribution status}}\\\hline
    \multirow{2}{*}{\centerline{\textbf{Landscape modification:}}}
     & \multicolumn{2}{c|}{\textbf{Status of the concerned contribution:}}\\
    & \mmmmark{Private} & \mmmmark{Public} \\\hline
    Add a new contribution & \yes  & \yes \\\hline
    \rowcolor{black!25} Add an authorship    & \yes & \yes \\\hline
    Add a tag    & \yes & \yes \\\hline
    Add, modify % the parameters of
    remove a mark   & \yes & \yes \\\hline
       \rowcolor{black!7} Remove an author or authorship & \yes & \nope \\\hline
    \rowcolor{black!7} Remove a tag & \yes & \nope \\\hline
    \rowcolor{black!7} Change the endpoints of an edge & \yes   & \nope \\\hline
    \rowcolor{black!7} Add a label to an edge  & \yes &\nope \\\hline
    \rowcolor{black!7}Add a label to a pen & \yes & \nope \\\hline
    \rowcolor{black!7}Change the label of a contribution  & \yes  & \nope \\\hline
    \rowcolor{black!7}Change the concrete type of a contribution & \yes  & \nope \\\hline
    \rowcolor{black!7}Add an author to an authorship list &\yes & \nope \\\hline
    \rowcolor{black!25} Upgrade the status  & \yes  & -- \\\hline
    Downgrade the status  & \nope  & \nope \\\hline
    Change the id of a contribution  & \nope  & \nope \\\hline
\end{tabular}
\caption{The different modifications that a user can apply to a MMM landscape, depending on the \hyperref[status]{status} of the  contribution involved in the modification.
\yes{} stands for possible and \nope{} stands for impossible. Landscape modifications  only apply to a user's local territory. If the concerned contribution is sent to peers, the change may propagate through \hyperref[merging]{merges}.
For instance, Alice may add an authorship mentioning herself in the authorship set of contribution $c$. Even if $c$'s status is \mmmmark{public}, the change only affects Alice's local copy of $c$ at first, until $c$ is sent to another user Bob and Bob accepts it.
\furtherworkcaption{Some changes mandatorily propagate whenever there is an opportunity for them to -- cf rows in dark grey. 
Some changes such as changes relative to house-keeping marks, only propagate to a user's devices. 
Share contracts may settle how/if modifications 
affecting   private contributions propagate. 
Restrictions on the conditions for modifying and propagating tags are to be determined.} 
}
\label{table-updates}
\end{center}
\end{table}
\furtherworkmark{}

\section{Conclusion}
I  introduced  the {MMM data model} and a notion of  epistemic {landscape} based on it. I defined {the MMM} as the reunion of all epistemic landscapes. I detailed the different types of  MMM {contributions} and their attributes, which are involved in the MMM. 
I presented landscape based activities centered around editing landscapes, consuming information and sharing information. 
\medskip

Like the Semantic Web's underlying formalisms, the MMM's networked structure allows pieces of information to be connected to each other so that they gain in precision from context, and they can be reused and meaningfully processed -- by humans in the case of the MMM. %, rather than by machines. 
The MMM data model is assigned \textit{loose} semantics   because like the original Web\footnote{The CERN Web \cite{TBLWeb}, not so much the Web centered around social media.}, the MMM is meant to accommodate a diversity of use cases involving  humans at work contributing to scientific research and other   evolving informational fields.  \medskip

I proposed to introduce rich \hyperref[measuring]{metrics} expanding our traditional
definitions of informational quality. MMM metrics   leverage the non-binary epistemic qualities of information captured in the MMM data model. They also {account for} the contextual "{implantation}" of information reflected by the MMM network topology.
\medskip

I formalised the notion of \hyperref[implanting]{implantation}  and I evoked  the incentives for authors to implant their contributions well.
Implantation is central to my proposal.
It is to promote connectedness of the MMM network, making every piece of information more likely to be  \hyperref[finding]{findable} from an arbitrary location in the MMM landscape.
Connectedness % And connectedness of the overall MMM network 
can facilitate global redundancy mitigation. The supplementary material \ref{implementation} proposes to further support it through automatic connection suggesting mechanisms   (see in particular  the \hyperref[parachutist]{lawless parachutist software component}). 
\medskip

The supplementary material \ref{implementation} proposes a  technological infrastructure to support the MMM. The MMM is organically distributed among peers who store  the parts of the MMM network that are relevant to them  and share parts that might be relevant to others.
Globally, MMM contributions  get replicated as they get shared  and as peers decide to keep local copies for themselves. 
Locally, %on users' "\hyperref[territory]{epistemic territories}",  the 
globally unique  identifiers %of MMM contributions 
avoid  duplication. I propose to encourage multiple UIs in order to accommodate the diversity of epistemic cultures. I defined a notion of \hyperref[territory]{local epistemic territory} characteristic of a human user.  MMM contributions inputted through the various UIs used by a user are all to be funnelled to that user's territory. 
\medskip

On the MMM, all granularities of informational resources can be documented, identified and referred to, including fine-grained informational resources like single ideas and coarse-grained informational resources like the entire contents of an article. 
Information consumers can \textit{precisely} \hyperref[topics]{select} the pieces of information that they consume and keep copies of. Their local epistemic territories can grow \hyperref[filtering]{without getting littered} by irrelevant atomic pieces of information. Useless archival and useless exchanges of information between users can be {mitigated}. \medskip

The "refrigeration" process mentioned in the supplementary material \ref{implementation}  is a possible MMM based alternative to \href{https://archive.org/web/}{the Internet Archive's Wayback Machine} \cite{bowyer2021wayback}. Valuable pieces of MMM contributions can be archived enmeshed as they are with the epistemic MMM landscape. 
Collective  obsolescence ensures that any atomic piece of information that isn't of use to anyone, eventually %spontaneously 
disappears from the record and archiving concentrates on information deemed relevant.
\medskip

This proposal is primarily geared towards  digital content sobriety. Enhanced epistemic democracy is expected to follow.  The focus is  on what people are \textit{not} interested in. My hypothesis is that to be well-informed one must have enough focus and time to  reason and to analyse information \cite{fumier}. One must not suffer from information overload. 
I propose to equip people with enhanced means of disregarding information, without necessarily missing out on relevant content. Alice may not need any details on  how electricity works. Bob may not be interested in having an example of fungal dissemination mechanism. With our proposed MMM solution, Alice and Bob  can be made aware of the availability of these resources without having to come into contact with them even if their current interests are taking them in the immediate epistemic vicinity of those resources. 
Assuming as I do that there is a huge amount of information that a person does \textit{not} want to have access to, and that different people are uninterested in different areas and different \hyperref[zooming]{levels of information}, allows us to regard each host of the MMM network as a point of view on the MMM record participating in the distributed pruning of information. \medskip

The MMM proposal is the result of a compilation of ideas contributed by the scientific and entrepreneur communities  over several years of discussions. 
Further collective  participation is welcome to help address  remaining  questions in the aim of materialising the MMM. % materialise the MMM. 
% ============================================================== %
% ========            RESEARCH QUESTIONS                ======== %
% ============================================================== %
\furtherwork{Indeed, a number of research questions spanning over multiple domains still require attention. 
Some are essential administration questions:
\textit{How should \hyperref[IDs]{MMM identifiers} be defined? How can their definition provide an indexing of MMM contributions that facilitates  \hyperref[finding]{search} over the MMM? How can  \hyperref[sharing]{share contracts} be enforced? What should they regulate? Should data identifying authors be hashed into the MMM identifiers of the contributions they author in order to support authentication of shared MMM contributions? If so, should the \hyperref[merging]{merge operation} be adapted to persist the hashed data?}
The local territories of different users, in particular different users of the same machine, may overlap.
\textit{Can we give granular access rights to MMM contributions so that the overlaps  don't have to be duplicated on the disk?} Requirements listed in \S\ref{MMMreqs} limit future possibilities of modifying  the \hyperref[main-attributes]{main attributes} of MMM contributions. However, the basic \hyperref[metadata-attributes]{metadata attributes} can and may need to be adapted to provide satisfactory answers to some of these non-epistemic questions. 
Other questions requiring further work are concerned by the epistemic organisation of MMM information: 
\textit{How can we nest fine-grain CRDTs in MMM contributions to support \hyperref[collaborating]{fast paced collaboration}? How is MMM \hyperref[subscribing]{subscription material} identified and by whom, the serving host or the recipient host? What relevant \hyperref[measuring]{MMM based metrics}  can  be defined? What   relevant \hyperref[filtering]{MMM based filters} can  be defined? What are possible global repercussions of using certain filters systematically and universally? What \hyperref[gluebot]{learning mechanisms} can be implemented on the MMM to enhance connectedness and promote other desirable global (topological) qualities of the MMM record? }
The architectural proposal made in the Supplementary Material proposes to persist MMM data in a \hyperref[graphDB]{graph database}. \textit{What should be the design of this database to optimize the different landscape-based activities -- provided the MMM network is not exactly a graph because of pens and because of edges acting as nodes? What \hyperref[qMMML]{query language} should we rely on?} 
The supplementary material also mentions the possibility of documenting into the MMM  the design choices and semantics of external data \textit{schemas} 
as well as possible known relations between data schemas. Because of the flexibility of the MMM data model, there is flexibility in the way external data models can be mapped to the MMM data model. \textit{What external data models are worth mapping to the MMM data model, why and how should they be mapped?} 
Perhaps most generative of questions is the interface between the MMM and formal ontologies.
A preliminary RDF--MMM mapping is available on Gitlab \cite{MMMRDF}.  As the MMM data model is not equivalent to RDF, different mappings are possible to serve different purposes in relation to the following questions: \textit{How can formal ontologies thread the MMM and provide support for "long distance" MMM exploration and search? Conversely, can updated, possibly informally expressed  knowledge documented on the MMM assist the design, evaluation, completion, alignement and generally the evolution of formal models? How can the MMM data model and its interface with standard ontology and knowledge graph formalisms be automatically leveraged to those ends? Are there kinds of information of interest in ontology engineering that are computationally easier to get from epistemic glue in the MMM than from formal inferences on represented knowledge?} Answers to the latter questions would specify some additional incentives for populating the MMM.  
In a follow-up article I propose to formalise a knowledge graph called the "\textbf{Socio-Economic Overlay Network}" (SEON) and its interface with the MMM. The SEON is to relay the MMM on informational content like metadata that is less amenable to discussion than typical MMM information. The questions listed above need to be specified, and the list is not exhaustive nor definite.} \medskip

In addition to the research questions the MMM poses a number of implementation challenges. 
\furtherwork{The supplementary material \ref{implementation} sketches a technological solution abiding by local-first principles \cite{kleppmann2019local}. It would be relevant to examine possible synergies and overlaps with ongoing initiatives and existing technologies.}  The technological solution proposed for the MMM encourages multiplicity of frontend interfaces with a standard MMM backend. Opportunities to interface existing tools with this backend are key. \furtherwork{ The supplementary material \ref{implementation}   proposes a distributed architecture to support the MMM. The appropriate nature and conditions of network connections between hosts of this network  need to be determined. \textit{What protocols should they rely on? When can direct peer-to-peer connections be implemented? When are relay servers relevant?}}

% ============================================================== %
% ========               BIBLIOGRAPHY                   ======== %
% ============================================================== %
\addcontentsline{toc}{section}{References}
\bibliography{bib.bib} 
\label{articleLastPage}

\appendix

% ============================================================== %
% ========               IMPLEMENTATION                 ======== %
% ============================================================== %
\clearpage

\setcounter{page}{1}
\cfoot{\thesection~ \thepage\  /\ \getpagerefnumber{articleLastPagepartA}}

\thispagestyle{empty}

\refstepcounter{section}
\addcontentsline{toc}{section}{\thesection: Implementation: The Standard MMM Application}

\begin{center}
 \episodetitleS{Supplementary Material \thesection:}{The Standard MMM Application}\\[5mm]
\author{M. Noual, G. Bouzige} 
\end{center}
\bigskip

The main article defines the MMM data model for documenting and annotating any granularity of informational content. In this supplementary material we complete the MMM proposal with an outline of the architecture of a software solution to support the distributed MMM network.\medskip

\label{implementation}

% ============================================================== %
% ========              MMM APPLICATION                 ======== %
% ============================================================== %
\label{layers}

To manage MMM landscapes and support the landscape based activities presented in part I, we propose a software solution which we call the \textbf{\MMMapp}. Following  Design Bias \ref{experts} and the principle of "slow first collaboration" presented in  \S\ref{collaborating}, the \MMMapp{} is to be a locally installed software primarily operating \href{https://offlinefirst.org/}{offline} \cite{offlinefirst}.  
% Components can work alone.
When online, instances of the \MMMapp{} can connect to each other and exchange MMM formatted data. Together they constitute the nodes of  a distributed information network. We call them MMM hosts. %
\medskip

% ============================================================== %
% ========                 MMM APP LAYERS               ======== %
% ============================================================== %
We propose that the \MMMapp{} be organised into four layers:

\begin{figure}[H]
    \centering\includegraphics[scale=0.2,trim={2cm 0 3cm 0},clip]{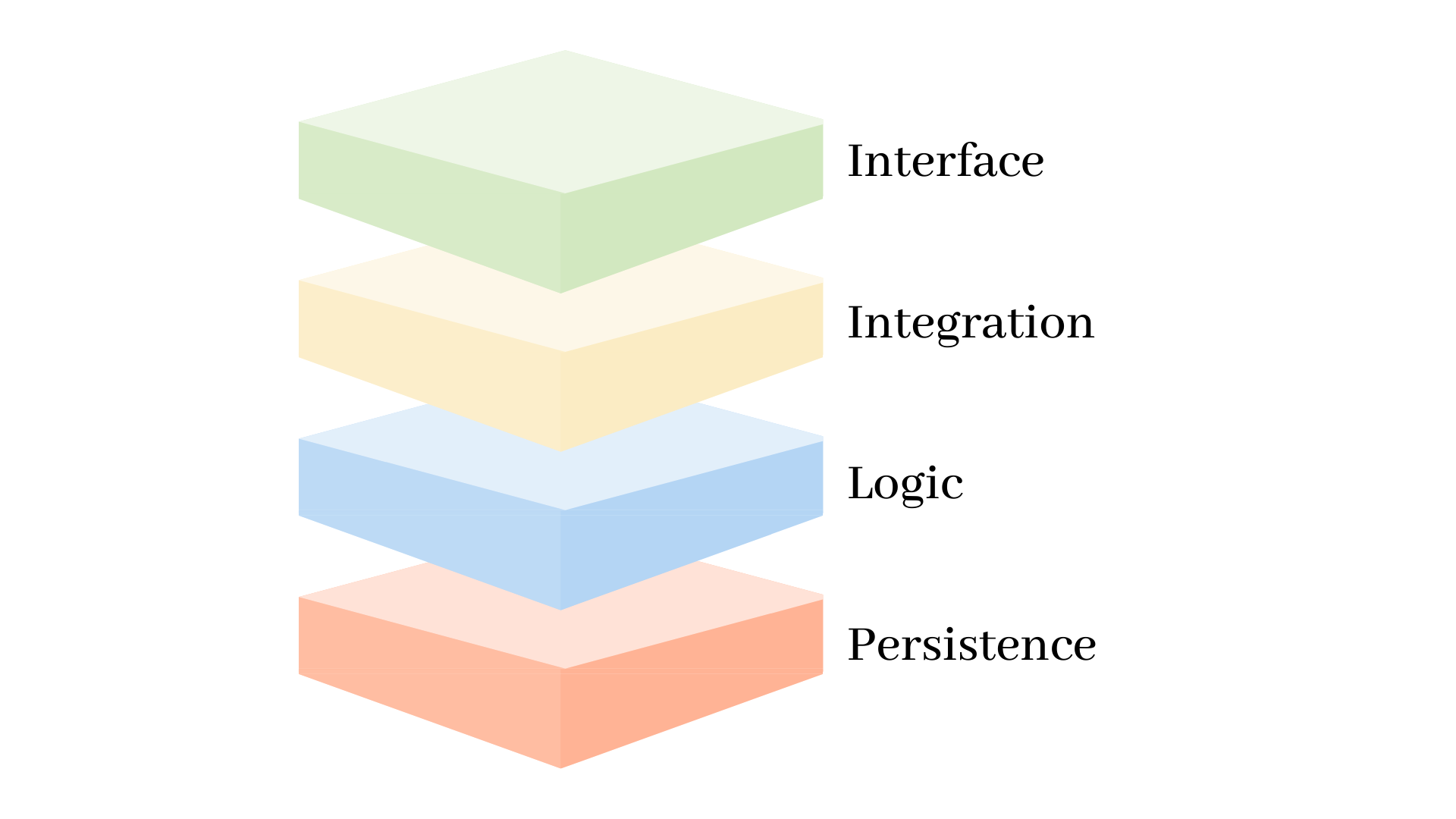}
    \caption{{The four layers of the \MMMapp}}
\end{figure}

Connections between MMM hosts are  made by the bottom two layers, namely the persistence layer and the logic layer, using {some form of request-response message-passing mechanism.} % like remote procedure calls (RPCs) 
{Those bottom two layers  implement a message bus for the MMM-speaking distributed ecosystem.}  
\medskip

We now present each of the four layers of the \MMMapp{} separately.

% ============================================================== %
% ========                     STORAGE                  ======== %
% ============================================================== %
% ============================================================== %
% ========                     STORAGE                  ======== %
% ============================================================== %
\subsection{The persistence Layer}\label{storage-layer}\label{storage}

The persistence layer stores MMM formatted data. More  precisely it stores the user's \hyperref[territory]{local territory} or a part of it on the machine's local disk, making the data accessible offline. For a scientific researcher, the local territory that needs to be stored might consist in MMM formatted research work notes, bibliographical material,  administrative information, \textit{etc}, i.e., an  amount of \textit{text} data similar to  what an individual usually stores on their devices, on university servers and on the cloud.
\medskip

% ============================================================== %
% ========                    REMOTE                    ======== %
% ============================================================== %
Off-site  backup  is a possibility, for the user's entire local territory or just for parts of it. \medskip

The persistence layer is comprised of an event log, a graph database and  a database management system.

% ============================================================== %
% ========                  EVENT SOURCING              ======== %
% ============================================================== %
\subsubsection{Event Sourcing}\label{event-sourcing}
\citedesignbias{events}'s event perspective on information  makes an event-sourcing architecture natural \cite{ES}. 
\citedesignbias{irrevocable}'s  \hyperref[irrevocable]{irrevocable publicness}  further facilitates the idea of storing data in a \textit{mostly} {append-only} event log.
We say "\textit{mostly}" because of obsolescence (cf \S\ref{obsoleting}) and archival mechanisms (cf \S\ref{cold}).
An append-only log is best suited for our default collaboration case presented in \S\ref{collaborating}. Events correspond to the creation or the reception of well-formed MMM contributions. And MMM contributions  are shared like  \mmmmark{public} MMM contributions whose main attributes are immutable (cf Table \ref{table-updates}). \furtherwork{The append-only feature of the event log may locally need to be relaxed in order to support real-time collaborative editing of the main attributes of MMM contributions (cf \S\ref{collaborating}).} 
\medskip

Events are logged in the order of their creation or delivery. The event logs of different users   represent  different local timelines with different ordering of events  and overall different sets of events. {As with CRDTs \cite{shapiro2011}, the different local  timelines don't need to converge} (and unlike CRDTs, we are also not aiming at the convergence of landscape states). 
\medskip

Locally, when a new contribution $c$ is created, by the user or following an in-coming share request, an event is added to the local event log. The event includes the JSON-MMM serialisation of  $c$ \cite{MMMJSON}. 
When an existing contribution $c$ is modified (one or several of its mutable attributes is modified), the modification needs to be saved. This requires finding $c$ in the event log and modifying the corresponding event. If $c$  was created a long time ago, searching for it in the log by its identifier is not efficient. It is preferable to search for it by the timestamp providing the date at which $c$ was logged. But the modification of $c$ might have come from the distributed network. In this case, the timestamp is unknown. The next section proposes a solution delaying the updating of the event log.\medskip

The event log can be used to see how the user's local MMM territory  evolved over time. The sequence of events that applied to the territory can be played back and forth. 
Note that we have assumed that no new event is created when an existing contribution is modified.  The number of events in the event log is equal to the number of contributions in the user's territory.  
This means that when we use the event log for \hyperref[timetravelling]{time traveling} we don't see how contributions were modified over time. Relaxing our definition of events might be relevant to enable a finer {time travelling} experience.
\medskip

As we will see in \S\ref{integration-layer} when we present the interface layer of the \MMMapp{}, a user may interact with her local MMM territory through multiple user interfaces (UIs). Different UIs  can be designed with different purposes in mind. They can be for note-taking, for reference management, for collecting scientific intelligence, for communicating \textit{etc}. Content that the user saves to her local territory through UI 1 (e.g. her university's electronic laboratory notebook) could be thematically unrelated (linked by a very small number of MMM paths) to content that the user saves to her local territory through UI 2 (e.g. a MMM based cooking app she has on her phone). 
For  log lookup  efficiency, it might be interesting to decompose the event log  into several smaller event logs corresponding to thematically independent landscape areas. To keep logs manageable in size, non-actionable content (e.g.  \mmmmark{obsolete} contributions that are not destined to be shared) may be removed from some logs. 

% ============================================================== %
% ========                    GRAPH DB                  ======== %
% ============================================================== %
 \subsubsection{The Graph Database}\label{graphDB}

In a situation where Alice shares MMM data with Bob, a MMM edge $e$ created by Alice between contributions $c$ and $c'$ may be received by Bob before Bob receives the endpoints $c$ and $c'$ of $e$. Searching for the endpoints of the edge in Bob's local event log might therefore fail to produce a result. To avoid unnecessarily searching the event log for contributions that aren't there, 
we propose to double the event log with a graph database (GDB) representing the current landscape state (a.k.a. the {application state}).
\medskip

The event log allows to natively query the MMM landscape as a historical record. The design of the GDB will allow to natively query it {as a graph}. 
Formally, MMM landscapes are not standard graphs because of \hyperref[pens]{pens} and because of  \hyperref[edges]{edges that can have other edges as endpoints}.  Further research is needed to determine the exact design 
of the MMM GDB. Ideally, \hyperref[measuring]{epistemic proximity} of contributions in the MMM landscape (like an edge and its endpoints) will be reflected as proximity in the GDB.
An indexing scheme (relative to \hyperref[IDs]{MMM identifiers}) will need to be defined to efficiently retrieve contributions that are epistemically close (e.g. like a pen and its contents) but not contiguous in the GDB layout. 
\medskip

% INITIAL STATE 
Initially a user's local MMM GDB  may start in a blank state representing  the "empty" landscape (which contains  nothing more than the mandatory \hyperref[pit]{pit landmark $\bot$}). Alternatively, the GDB can start from a non-empty  landscape, possibly copied from a remote device. 
% APPLY CHANGES
As successive events are applied to the landscape, the state of the GDB changes.  
Changes are applied in the order the software becomes aware of them. %
\medskip

% MULTIPLE STATES
As with the event log, it might be interesting to split the GDB  into several smaller GDBs to which disjoint sets of events are applied. Each sub-GDB would store a contiguous (thematic) area of the user's territory that the user tends to access independently of areas stored is the other sub-GDBs.
\medskip

% LAZY MERGE STRATEGY
On reception of a contribution $c$ following a share request from a remote MMM host, $c$ is loaded into the local GDB. \textit{Where to position $c$ in the local GDB} is a question we leave open. Whatever information we have on $c$ and on the context in which it was shared (e.g. $c$'s MMM identifier, the semantics of its label, known adjacent  contributions,  the source host of the share request) can be leveraged to help make an appropriate choice of positioning. \hyperref[implanting]{Implantation} being central to our proposal, it is relevant to implicate the user in that choice. Note that possibly a local homologous copy $c_l$ of $c$ (with the same identifier) already exists in the local persistence layer. In that case, the old and the new copies  need to be \hyperref[merging]{merged}. A possible lazy strategy is save $c$ into the event log and into the GDB as if it were a brand-new contribution, and to count on \hyperref[bridging]{epistemic bridging}. Because $c$ and $c_l$ have the exact same meaning, paths are likely to  materialise between them. The distance between them in the user's local territory is likely to eventually become very small. A future landscape exploration  may  constitute an opportunity to notice that  $c$ and $c_l$ are homologues and to merge them on the fly.

% ============================================================== %
% ========                   COLD & WARM                ======== %
% ============================================================== %
\subsubsection{Cold and Warm Storage}\label{cold}\label{refrigeration}

We propose that users mark contributions in their local territory as \mmmmark{refrigerated} (meaning archived) when they no longer have an immediate use for them but don't want to  \mmmmark {obsolete} them (delete them). 
Past events corresponding to the creation
of now \mmmmark{refrigerated}  contributions  can be deleted from the GDB as long as they are persisted in the user's local event log.  
\medskip

% WARM & COLD
Refrigerated contributions are called \textbf{cold}  contributions. Non-refrigerated contributions are called \textbf{warm}  contributions. A warm landscape is one that is composed only of warm contributions (and $\bot$). 
The \textbf{warm MMM} is the distributed set of warm local territories. It holds  MMM formatted information that is of \textit{actual} interest to at least one  MMM user.
\medskip

% SNAPSHOTS
We propose to keep periodical "snapshots"  of the local GDB. The snapshots  represent historical states of the user's territory. Given a snapshot $s$ made at date $t_1$ in the past, the state that the landscape was in at  later date $t_2>t_1$ can be reconstructed by applying to $s$ the sequence of events logged between  $t_1$ and $t_2$. \hyperref[timetravelling]{Time travelling} back in the past (removing contributions from a GDB state in the reverse order they were logged) is also possible. \medskip

GDB snapshots (which typically contain contributions that are still warm) and events in the log corresponding to cold contributions are what we refer to as the cold storage of the \MMMapp.
% COLD
The cold storage stores archived material that the user need not often access. It can be the part of persistence layer that is relegated to remote stores.

% ============================================================== %
% ========            DB MANAGEMENT -- QMMML            ======== %
% ============================================================== %
\subsubsection{The  Database  Management System}\label{qMMML}

The persistence layer of the \MMMapp{} needs a database  management system (DBMS) so that upper layers can efficiently 
access and modify persisted MMM data. The DBMS needs to know how to navigate between the event log and the GDB, querying one or the other or both appropriately. 
In our proposal (cf \S\ref{documenting}), we drop the traditional paradigm whereby (collections of) well-circumscribed closed individual documents are the principal entities to be queried. We favour the following view instead: there is a \textit{single} open collective document structured as a network of contributions, namely the MMM, and  it is the principal entity to  be queried. In practice, \MMMapp{s} often work offline. So queries are often restricted to local territories or even smaller thematic areas of the global MMM landscape. But unlike traditional documents, MMM areas are not fundamental to the way the data is stored and organised (cf \S\ref{area}). Search areas  may only be delineated at run time. %  when the query is formulated. 
\medskip

MMM data is primarily written and read locally. When a connection is available, the DBMS can also query remote MMM hosts. When a connection with a remote host can't be established, the DBMS  decides whether to queue the query and process it when the connection returns, or to drop it altogether.
    \medskip

A query language QMMML needs to be defined or derived from existing query languages  such as {Cypher} \cite{francis2018cypher}.

% ============================================================== %
% ========                     LOGIC                    ======== %
% ============================================================== %
% ============================================================== %
% ========                     LOGIC                    ======== %
% ============================================================== %
\subsection{The Logic Layer}\label{logic-layer}

The logic layer is the central part of the \MMMapp{}. It contains  generic 
software components in charge of implementing   MMM logic. 
Most components of the logic layer are default modules, common to all instances of the \MMMapp{}.
Some however (like the gluebot, cf \S\ref{gluebot}) are non-essential and optional.

% ============================================================== %
% ========                     KERNEL                   ======== %
% ============================================================== %
\subsubsection{The Orchestrator}

The orchestrator module  orchestrates other modules  of the \MMMapp, especially other modules of the logic layer operating in the background. It relays commands and data between modules and layers.
\medskip 

% ============================================================== %
% ========                   EDITING LOGIC              ======== %
% ============================================================== %
\subsubsection{The MMM Editor Module}
The MMM editor  module is layered over the persistence layer's DBMS. It executes QMMML queries to support   complex MMM logic and edit the landscape stored in the persistence layer. It implements the landscape editing activities of Section \ref{activities-editing}. 
The editor module  works offline.

% ============================================================== %
% ========                 PARACHUTIST                  ======== %
% ============================================================== %
\subsubsection{The Lawless Parachutist} \label{parachutist}
The lawless parachutist module is one of two explorer modules of the \MMMapp{}. The other is the lawful wayfarer introduced below. %, cf \S\ref{wayfarer-module} below. 
The lawless parachutist explores MMM formatted information using traditional natural language processing (NLP)
ignoring the graph topology of the MMM landscape. 
It detects similarities between MMM contributions 
that are not yet reflected by epistemic connections in the MMM landscape.  %  MMM paths.
This is key to enhancing the connectedness of the MMM landscape which is itself key to our proposal. 
\medskip

% CONNECTION
In case of  a connection, 
the parachutist's exploration may extend beyond the local territory of the user and over to territories of known remote MMM hosts. Further work is needed to specify permitted crawling strategies 
-- e.g. in terms of the maximum number of permitted device hops, 
which hops are most relevant, how to discover more hosts towards which relevant hops can be made and whether connections with remote hosts can be executed asynchronously in the background. 
\medskip

% ============================================================== %
% ========          MENTIONING / TACIT LINKS            ======== %
% ============================================================== %
The parachutist  supports the documentation of "\textbf{tacit links}". A user documents  contribution $c$  in the MMM. The label of $c$ involves concept $\textbf{C}$. The parachutist explores the landscape and discovers an \mmmtype{existence} node $n$ labelled $\textbf{C}$. A \pertains{}  edge
from $n$ to $c$ is suggested to the user who validates it or rejects it.  If they validate it,
the edge is created. 

% ============================================================== %
% ========                 WAYFARER                     ======== %
% ============================================================== %
\subsubsection{The Lawful Wayfarer} \label{wayfarer-module}
The lawful wayfarer module is the second explorer module of the \MMMapp{}.  It is the "thinker" of \MMMapp{}: it implements the configurable wayfarer exploration strategies presented in \S\ref{finding}. 
It specialises in visiting the landscape stored in the GDB of the persistence layer.  Contrary to the parachutist, it only explores the landscape by following MMM \hyperref[area]{paths}.
% CONNEXION    RPCs
Like the parachutist, it may extend its exploration to remote territories. It does so when it encounters an edge  straddling  the local territorial border: one of the edge's endpoints is in the user's local territory, the other is the territory of a remote, connected MMM host. 
This situation may result in MMM contributions  being copied to the user's local territory without  involving \hyperref[sharing]{explicit manual sharing} from a peer.
If the host of the edge's endpoint is not connected at the time of the wayfarer's encounter with the edge, then the wayfarer possibly submits  an asynchronous connection request to the concierge module (the concierge module is presented below in \S\ref{concierge}).

% ============================================================== %
% ========                    GLUEBOT                   ======== %
% ============================================================== %
\subsubsection{The naive gluebot} \label{gluebot}
The naive "gluebot" is an optional module that assists the collective documentation of the landscape by humans. Its purpose is to   increase the connectedness of the landscape.
It does this cautiously by contributing {question}s \textit{in between} two existing contributions $c$, $c'$  deemed comparable with one another by the explorer modules. %have identified as 
% $c$, $c'$ identified  by the parachutist or wayfarer as similar or worth comparing with one another
It uses large language models to express questions in human intelligible language, asking \textit{how} targeted contributions $c$, $c'$  epistemically relate to each other. The gluebot documents its questions  in \mmmtype{question} nodes which it  ties to $c$ and $c'$ using  \pertains{} or \mmmtype{questions} edges. 

% ============================================================== %
% ========               TOPOGRAPHER                    ======== %
% ============================================================== %
\subsubsection{The Topographer} \label{topographer}
The topographer is the standard MMM analytics module. It  stores a bank of fundamental \hyperref[measuring]{metrics} (cf \S\ref{measuring}). 
It calls the explorer modules to implement exploration strategies  optimised  for measuring MMM data according to those metrics. 
As most metrics involve graph theoretic properties of the landscape and properties of the MMM data structure, the topographer especially relies on the wayfarer explorer to "walk around" landmarks and circumscribed areas in order to compute the targeted measures on them. 

% ============================================================== %
% ========                 FILTERER                     ======== %
% ============================================================== %
\subsubsection{The Filterer}\label{filterer}
The filterer module stores a collection of default and user defined filters. 
It calls on the topographer to assist the user in defining her own custom filters and the conditions in which to apply those filters. The filterer can also be called to constrain wayfarer explorations.

% ============================================================== %
% ========                 SEARCH ENGINE                ======== %
% ============================================================== %
\subsubsection{The MMM Search Engine}\label{search-engine}

The MMM search engine implements search strategies  mentioned in \S\ref{finding}. In the case of a selective query, the parachutist is typically invoked to search the event log lawlessly  (heedless of MMM topology). In the case of an approximate exploratory query,  the search engine calls the wayfarer to locate epistemically relevant content. 
In addition to topological properties of the epistemic MMM network, topological properties of the infrastructural distributed network of MMM hosts can be exploited as additional information to refine search results. 
\medskip

The same search may not yield the same results at all times because of possible unavailability of remote MMM hosts.

% ============================================================== %
% ========                 IMPLANTER                    ======== %
% ============================================================== %
\subsubsection{The Planter} \label{planter}

The planter module is in charge promoting \hyperref[implanting]{implantation} (cf \S\ref{implanting}). It runs while  the user is in the process of documenting a new contribution $c$. It calls on the search engine using information in $c$ to formulate a search query, possibly combined with contextual information uncovered by the wayfarer (information conveyed by neighbouring contributions of the landscape in case $c$ is already partially implanted). 
It searches for pre-existing contributions $c'$ in the landscape that $c$ could relevantly be connected to. It guesses the appropriate concrete MMM edge types to use for that.  To help preempt the documentation of redundant contributions by the user, the UI makes the user aware of the planter's suggestions.

\begin{bestpractices}{} If the planter finds a pre-existing contribution (or area) $c'$  that expresses an idea very similar to the one that you are presently documenting in  contribution $c$, then reduce the information you place on $c$ to its bear minimum (while ensuring that the area $\{c',c\}$ ends up conveying all the semantic load you were about to place on $c$ alone). If relevant, give up on documenting $c$ altogether. 
\end{bestpractices}

As an example suppose Alice is documenting the story of her building a passive house in \mmmtype{narrative} node $c$. A simple run of the planter module in parachutist mode identifies a remote \mmmtype{existence} node $c'$ labelled "Passive house". The planter makes the suggestion that Alice connect $c'$ to $c$ with a \pertains{} edge.  Alice accepts the suggestion, the edge is created. The planter then invokes the wayfarer to search the area around $c'$  for further similarities with $c$. Alice continues to type her story in $c$. She details how she thermally insulated the house. The planter uses the additional detail to narrow the wayfarer search around $c'$. An existing \mmmtype{question} node asking "What are the different kinds of thermal insulation for house walls?" is identified, as well as a \mmmtype{definition} of the term "Mineral wool". Again, the planter makes suggestions to Alice to connect $c$ with these  pre-existing contributions. Alice accepts one and rejects the other. \textit{etc}\medskip

The planter may associate a confidence score to its suggestions. 
If  a confident suggestion between $c$ and $c'$ is rejected by the user, 
the planter may invoke the gluebot to make an indirect connection between $c$ and $c'$. 
\medskip

% ============================================================== %
% ========                    CONCIERGE                 ======== %
% ============================================================== %
\subsubsection{The Concierge} \label{concierge}
The concierge module handles interactions with  remote MMM hosts in the background.  
It supports  the sharing activities of Section \ref{activities-sharing}.
It sends and receives \hyperref[sharing]{share requests} and  \hyperref[subscribing]{subscription requests}, and manages incoming MMM material. 
% CONNEXION
The concierge supports \textbf{synchronous sharing} when  the sending and the receiving MMM hosts are simultaneously online. This can work for certain situations like online meetings, collective note-taking during a conference talk, and  in case of out-of-band rendezvousing. 
\textbf{Asynchronous} collaboration requires buffering messages and  using relay servers  to play a role similar  to email servers. 
Relay servers are relevant to substantiate communities of interest. They can  also serve as collective (thematic) backup. 
Asynchronous sharing is relevant to our proposal because of slow-first collaboration (cf \S\ref{collaborating})  and because delay in the reception of contributions favours stringent filtering and thereby  mitigates unnecessary replication of fine-grained pieces of information. The longer one waits before receiving contribution $c$, the more likely it is that $c$ has been annotated by one or several peers, and so the  more  {information available there is about $c$}, the more precisely the qualities of $c$ can be \hyperref[measuring]{measured}.  This  increases the chances of either (i) automatically and  appropriately  \hyperref[filtering]{filtering} out $c$ if it is sub par or useless to the receiving peer, or (ii) strongly \hyperref[implanting]{implanting} $c$ in the user's local territory otherwise.

% ============================================================== %
% ========           CONCIERGE's HOUSE-KEEPING          ======== %
% ============================================================== %
The concierge also handles local house-keeping in the background. 
It orders the persistence of \mmmmark{new} material accepted by the user, (lazily) checking its newness and merging with local homologues (cf  \S\ref{graphDB})
and possibly invoking the filterer and   the planter modules. 
It orders the deletion of \mmmmark{obsolete} contributions at the end of the limbo period.

% ============================================================== %
% ========                     INTEGRATION              ======== %
% ============================================================== %

% ============================================================== %
% ========                     INTEGRATION              ======== %
% ============================================================== %
\subsection{The Integration Layer}\label{integration-layer}

\textbf{The integration layer} is an {optional} layer that contains software modules needed to  map external {standard}
data formats onto the MMM format. 
The integration layer stores a collection of {standard} mappings between standard data schemas (e.g. RDF \cite{manola2004rdf,MMMRDF}) and JSON-MMM \cite{MMMJSON}. % the MMM format. 
Modules of the integration layer use the mappings to perform ETL-like integration (Extract, Transform, Load) \cite{denney2016validating}. Incoming data is systematically  consolidated before it is loaded in the persistence layer.  
The integration layer may in particular dedicate a module to the MMM file hierarchy functionality mentioned in \S\ref{FH}. 
\medskip

Some modules of the integration layer also conversely process MMM formatted data from the persistence layer, and ready it for output and processing by non-MMM applications.
This may include the parametrisable conversion of MMM formatted content into a variety of human friendly file formats  such as CSV, PDF, EML and HTML.  
One module may be dedicated to the conversion of  (i) MMM formatted information contained in selected areas of the landscape, into (ii) human friendly natural language text. The module can use 
% Selected areas of the MMM landscape can be compiled into traditional (text) documents. 
language  generation \cite{li2022pretrained,zhang2022survey}   to weave the information expressed in MMM contributions coherently with the semantics of MMM types and landscape topology. This module may be used by a UI component to  display MMM search results in purely verbal form, or as part of a MMM fed QA chatbot. 
\label{pretty-print}\label{rendering}
In \S\ref{drafting} we mentioned that the MMM can be used for drafting. The epistemic structuring of MMM  material  
presents an opportunity to propose automatic composing of standard documents from (collectively taken) MMM notes. We might consider using circumscribed, appropriately filtered MMM areas as input to produce technical reports, slide decks, articles,  standard documentation \textit{etc} as outputs. 
\medskip

The integration layer {is only \textit{marginally} meant to  support} 
syntactic interoperability between  data schemas. The MMM proposal is geared towards supporting human experts at work \textit{manually} documenting textual information. Automatic population of the MMM is therefore not central to our proposal, with the exception of formalised ontological content which is to  be the object of a follow-up article. For most data schemas, the advantage of translating data into MMM format and storing it as such in the MMM persistence layer isn't clear. 
The MMM may however be more relevant  for human-intelligible schema \textit{documentation} (by human schema designers) and for recording  semantic and epistemic glue between different schemas. The MMM may serve as a \textbf{collective networked  catalogue} inventorying database schemas and what we know and think of them. This  may support an \textit{incremental} from of semantic interoperability between data schemas 
%{(cf \SMref{incremental})} 
favouring the gradual emergence and updating of standards and schema convergence 
in the long run.

% ============================================================== %
% ========                    INTERFACE                 ======== %
% ============================================================== %
% ============================================================== %
% ========                    INTERFACE                 ======== %
% ============================================================== %
\subsection{The Interface Layer}\label{interface-layer}

\textbf{The \interfacelayer{}} 
{together with the integration layer} materialises the boundary between the MMM world supported by the \MMMapp{}
and the rest of the world comprised of  third-party tools, data and users  that don't speak the MMM language.   
The \interfacelayer{} is comprised of style sheets,  modules implementing specific interface logic, and default (user)  interface components. 

% ============================================================== %
% ========             INTERFACE MODULES                ======== %
% ============================================================== %

% ============================================================== %
% ========                RENDERER                      ======== %
% ============================================================== %
\subsubsection{The Rendering Engine}\label{renderer}
The rendering engine is a module of the \interfacelayer{} that parses MMM formatted data and renders MMM landmarks and landscapes on the user's screen. It combines the MMM data with local style sheets (possibly involving \hyperref[marks]{MMM marks} such as \mmmmark{dim} and \mmmmark{highlighted}) and applies a visual model to display MMM objects within a UI component.

% ============================================================== %
% ========                BROWSER ENGINE                ======== %
% ============================================================== %
\subsubsection{The Browser Engine}\label{browser-engine}
The MMM browser engine (BE) interfaces with the UI and the rendering machine. Provided with some session data as well as events that the UI has detected from the user's interaction with the UI, the BE determines which MMM data to display.
It fetches  the data from the lower layers of the \MMMapp{}, and then calls the rendering engine to display it. The BE supports the essential MMM browsing actions such as \hyperref[zooming]{zooming in and out},  \hyperref[navigating]{navigating forward and backward}, \hyperref[filtering]{filtering}, refreshing the landscape after the loading of \mmmmark{new} content by \hyperref[concierge]{the concierge}.
% FILE BROWSER/MANAGER
It can also support a file navigation mode: a module of the integration layer (cf  \S\ref{integration-layer}) converts data from the local file hierarchy into MMM format (cf \S\ref{FH}), and the BE  presents this MMM data to the user using a 
specific style sheet  {emulating} a traditional file navigator. 

\bigskip

% ============================================================== %
% ========                 INTERFACES                   ======== %
% ============================================================== %
\citedesignbias{experts} (\textit{Experts at work know best}) requires that UIs for the MMM accommodate the diversity of epistemic cultures of MMM users including  domain experts. It should be possible for different users to navigate and edit the MMM using interfaces that best suit the specifics of  their informational work. 
Ideally, it should even be possible for them to use the familiar existing software they are already using for documentation, publication,  information navigation, organisation, and even communication  -- e.g. note-taking apps, electronic laboratory notebooks, reference management tools, community platforms, social media platforms, web browsers, file browsers, email. 
Users would populate  the MMM and benefit from its  \hyperref[smart]{smartly} networked structure \textit{through} those existing tools.
And the  digital content they  manipulate with those different tools would funnel to their local territory. 
\bigskip

Next, we make some suggestions of   MMM interfaces.

% ============================================================== %
% ========                      CLI                     ======== %
% ============================================================== %
\subsubsection{The standard MMM CLI}\label{CLI}
The standard MMM CLI (command line interface) supports most MMM actions presented in Section \ref{activities} and is delivered by default with the \MMMapp. 
It relies on   modules of lower levels of the \MMMapp{}.

% ============================================================== %
% ========                      GUI                     ======== %
% ============================================================== %
\subsubsection{The MMM browser}\label{GUI}
The MMM browser is the  standard MMM GUI (graphical user interface) for the \MMMapp,  delivered by default with the \MMMapp. 
It also implements the activities of Section \ref{activities} by relying on the rendering and browsing engines. % (see above). 

% ============================================================== %
% ========                   STATIC JSON                ======== %
% ============================================================== %
\subsubsection{Out-Of-Band Sharing}
The MMM CLI and  GUI support the native \hyperref[activities-sharing]{sharing} and \hyperref[storage-layer]{storing} mechanisms of the logic and persistence layers. Out-of-band sharing and storing of MMM content is also possible. 
The CLI and the GUI  allow the user to load JSON-MMM  files. They also allow  taking
static snapshots of selected landscape areas. The  areas are serialised in JSON-MMM  files \cite{MMMJSON} that can be saved outside the persistence layer and optionally 
exported into other file formats and shared by email (calling on \hyperref[integration-layer]{integration modules}). 
\medskip

% ============================================================== %
% ========               DYNAMIC SHARING                ======== %
% ============================================================== %
We may also want to support \textit{dynamic} out-of-band sharing of a landscape area $A$ so that  future MMM annotations to $A$ will  also be shared.
Without resorting to a relay server, dynamic sharing only works when the MMM host of  $A$ is online. 
% ============================================================== %
% ========                      DYNAMIC                 ======== %
% ============================================================== %
A university may have a server to host its researchers' personal webpages.
Researchers may want to display 
some contributions of their MMM territories that are {public} and relative to their current research work. {To that end,  MMM landscape areas may be embedded in webpages  similarly to the ways in which OpenStreetMap maps are embedded in webpages} \cite{haklay2008openstreetmap,OSMmedium}.
\medskip

% ============================================================== %
% ========                      URI                     ======== %
% ============================================================== %
Subscription requests and invitations normally handled by the logic layer's concierge may be copied with their parameters in files that be can be shared out-of-band. These files may be   opened by a  MMM interface which will transfer the files to the integration layer of the \MMMapp{} and then to the concierge in the logic layer. 

\bigskip

% ============================================================== %
% ========                    APIs                      ======== %
% ============================================================== %
\subsubsection{APIs}\label{APIs}

% ============================================================== %
% ========               VARIATIONS                     ======== %
% ============================================================== %
Different kinds of experts may want to explore the MMM according to different strategies. They will typically look for different kinds of information. 
They will have different typical interpretations and uses of native \hyperref[types]{MMM types} (consider for instance what a biologist might use the \equates{} edge for and what a linguist might use it for). Their trains-of-thought will follow different successive landmarks, and so the documentation of their work will have different patterns. 
In order to support specialised uses of the \MMMapp{}, domain-specific MMM UIs are needed.  To facilitate their development by third-party programmers, domain-specific {libraries and APIs} need to be published. 
\furtherwork{A future question to address might be: can the progressive epistemic differences between different kinds of domain experts be machine learned to be  captured in API wrappers?} 

\medskip

\begin{figure}[H]
        \centerline{\includegraphics[scale=0.33,trim=5cm 0 0cm 0,clip]{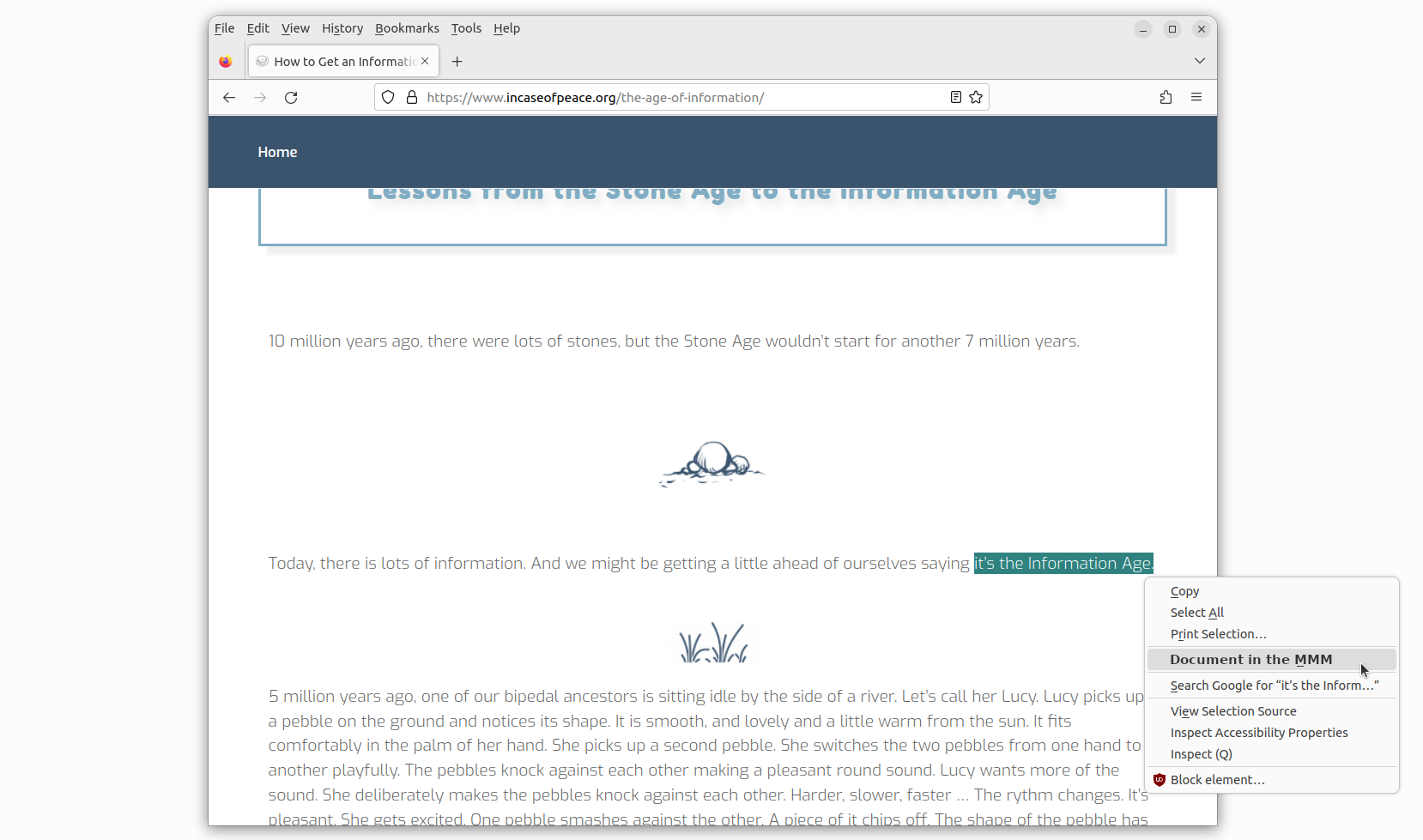}}
        \caption{A MMM plugin for a Web browser could allow a user %navigating the web 
        to right-click on textual content they find in a webpage and import it into an appropriate location in their local territory. A similar plugin  could be developed for  pdf readers. The MMM backend could be made transparent to users who are not ready for self-hosting their own local territories. To them, the functionality could  appear as a simple "annotate this" functionality. 
        }
        \label{browser-plugin}
    \end{figure}

More generally, we propose to make some of the functions of the \MMMapp{} modules available for use in third-party application code. We want to allow   third-party installed desktop/mobile  applications to retrieve content from the MMM persistence layer, to use the integration and logic layer's functionalities, and to automatically push MMM formatted content into the local MMM persistence layer.
% ============================================================== %
% ========         PACKAGES / LIBRARIES / ...??         ======== %
% ============================================================== %
We propose to provide {APIs} for frontend developers -- mostly exposing functions of the \interfacelayer{}'s modules  ({cf Fig \ref{browser-plugin}}) --, for database administrators -- mostly exposing functions of the integration layer for \hyperref[registration]{data registration} -- 
and for news feed publishers -- exposing functions to extract \hyperref[concierge]{concierge news} including new MMM publication and reward notifications  (which some users might prefer to receive through traditional channels like email). 
\medskip
% ============================================================== %
% ========                   WEB APIs                   ======== %
% ============================================================== %

MMM {APIs}, especially  backend APIs 
are primarily local APIs for offline use. They interface the locally installed \MMMapp{} with another third-party application locally   installed on the same device. This is similar to the  approach  used for EDI translators \cite{whatisEDI}.
Marginally, we might want to provide Web versions of those MMM APIs. 
A university may 
decide to execute its locally installed \MMMapp{} in server mode and publish a web API  to offer non MMM users a point of entry into the university's local academic territory.
Web APIs may   mitigate the strong local-first principles underlying our solution  and help  on-board users that are not ready for self hostage \cite{federationfallacy}. Web browsers 
that support MMM functionalities (through various plugins for instance) and cloud versions of the \MMMapp{}
may allow these users  to access and interact with their remotely stored MMM territories.
However, cloud apps and centralised remote storage should remain marginal in the MMM ecosystem. Servers should remain primarily subordinate to local clients. Clouds apps should mandatorily require of their users the provision of a local or decentralised backup for their MMM territories.
\medskip

% ============================================================== %
% ========                      PLUGINS                 ======== %
% ============================================================== %

\label{articleLastPagepartA}

% ============================================================== %
% ========               PHILOSOPHICAL VIEWS            ======== %
% ============================================================== %
\clearpage

\setcounter{page}{1}
\refstepcounter{section}
\cfoot{\thesection~ \thepage\ /\ \pageref{articleLastPagepartB}}
\addcontentsline{toc}{section}{\thesection: Philosophical Views}
\label{philo}

\begin{center}
    \episodetitleS{Supplementary Material \thesection:}{Philosophical Views}\\[5mm]
    \author{M. Noual} 
\end{center}
\bigskip

Information is a difficult term to define. The Wikipedia page on this subject was one of the first Wikipedia pages to be started in {2001}. The page has continually and considerably evolved back and forth over the decades. Arguably, generally, even more can be learned from the talk page of a Wikipedia article than from the article itself. 
In the case of \href{https://en.wikipedia.org/w/index.php?title=Information\&oldid=1174678681}{the Wikipedia article on "information"} \cite{wikiinfo}, the history of \href{https://en.wikipedia.org/wiki/Talk:Information}{discussions underlying the design of the article} is especially interesting \cite{wikitalkinfo}. Some questions have been alternatively settled and reopened, eg \textit{Is information the meaning or the conveyor of the meaning?}. {Episodically a Wikipedian would inteject that the question  \textit{What is information?} is  the most hopelessly tricky question that the 21st century has to answer. }\medskip

Fortunately, our proposal has no consensual encyclopedic intention. Below I present some convenient assumptions about information that have provided the basis for the proposal. 

% ============================================================== %
% ========                 BINARY                       ======== %
% ============================================================== %
\subsection{Information isn't binary}
\label{binary} \label{nonbinary}

In my view, traditional knowledge management places a frustrating amount of emphasis  on characterising information in binary terms: \textit{correct/incorrect, true/false, consensual/not consensual, there/not there}. Most of those binary qualifications tend to work better for data (given objects) than for information. I propose to promote a non-binary point of view on information akin to the point of view humans spontaneously take on \textit{landscapes}. Humans sometimes qualify landscapes in binary terms: \textit{beautiful/ugly, there/not there}. But often they qualify them with richer terms like \textit{hilly, coastal, desertic, verdant, bleak, rugged, wild, remote, urban, sunlit, wet} \ldots{}. Our proposal offers to promote the same kind of nuanced qualifying of information.

% ============================================================== %
% ========              MATERIAL OBJECTS                ======== %
% ============================================================== %
\subsection{Hierarchical localisation works for material objects, not for everything else as well} 
\label{socks} 
% What works for material objects of the physical world doesn't necessarily work as well for non-material  objects.

We are used to finding material objects likes \textbf{socks and books} through their locations: "\textit{top drawer / cabinet to the right of the window / master bedroom / 2nd floor /  yellow house / Ikpe Street / Port Harcourt / Nigeria / Planet Earth / Solar system%
%Local Fluff / 
\ldots{}}". \textbf{Thoughts and ideas} are not like socks and books. 
% When a human mind comes across a thought or idea it does not necessarily have an obvious locating hierarchy for it. While a book on a shelf can only be accessed by going to that shelf in whatever room/building/etc it is stored, 
One same thought can be accessed through an indefinite number of  pathways, during different conversations, different chains of thought  \textit{etc}.     In my view, \textbf{digital textual information} (DTI) \textit{expresses} thoughts and ideas. 
% I propose to consider \textbf{digital textual information} (DTI) as an expression of thoughts and ideas rather than as a representation of the "reality" in which books and socks live. 
So there is no reason to think DTI gains at being organised as socks and books. In the lines of V. Bush \cite{VB} and T.H. Nelson \cite{NelsonWeb} 
% file systems vs block and object ?
I propose to enable the \textit{organic} structuring of DTI, honouring the multitude of pathways leading to any given thought or idea. \medskip
% in the capacity of expression of thoughts and ideas. 

Importantly, this point is about hierarchical \textit{localisation}. It says nothing about  the hierarchical \textit{nature} of things. Arguably, any good piece of information is built in a deeply hierarchical way. 
% Not all pieces of information have the same status ($\neq$ value). Equally valuable pieces of information may operate at different levels of abstraction and hold different roles. Some are questions, some are answers, some are definitions, \textit{etc}. Locally, \textit{i.e.}, between two given pieces of information it very often makes sense to specify a hierarchical relationship.  

% ============================================================== %
% ========               METAINFORMATION                ======== %
% ============================================================== %
\subsection{There is no such thing as meta-information} 
\label{metainformation}\label{metadata} 

Getting informed takes time. 
%There is no shortcut. 
Knowing of the existence of Newton's Principia doesn't inform in the same way that reading the Principia does. Knowing 'who are the authors of the Pythagorian theorem' is a different kind of information than knowing 'how the theorem works'. Both are still information, just not the same.\medskip 
    
In the world of data, meta-data is of critical importance. Providing meta-data to describe your data gives your data better chances of getting found and serving others. 
% Making data reusable according to the FAIR data reusability principle, is also a way of sparing human time and effort.  To ensure reusability, the FAIR guidelines \cite{FAIR} encourage the good description of data by meta-data.
Data needs meta-data -- \textit{i.e.}, information about data -- in order  to be informative and useful. Meta-data is usually recommended for  datasets and physical objects that are impossible to confuse and replace with  the descriptions I make of them. It is less clear however that a description of the Pythagorian theorem is impossible to confuse with the Pythagorian theorem itself. Both have textual forms.\medskip
    
Data is different from  information 
%(cf point \ref{alive} and \ref{binary}) 
and there is no  reason to treat the two  the same way. \medskip
%I will deal with the case of data later in Section \ref{data}. 

Our proposal focusses on information. It makes no distinction between basic and meta information. Either something informs or it doesn't. Either it is information or it isn't. Information about something is still information, even if that something is information itself. A lot of information (arguably all information) is information about other information. If there were a point in calling somthing "meta-information", it would have to be because a piece of meta-information X has a dinstinctly special way of informing on the basic piece of information Y it is about. It could not merely complete Y like ordinary contextual information. Unlike data which has no power to inform without contextual information, information obviously has. Contextual information (eg who Newton was, the date at which the Principia was published, what people at the time thought of distant action, the expression of Kepler's laws) is information that may or may not help understand Newton's laws of movement. 
Reading  a modern reformulation of the Principia, or a summary of it is different from reading the original text. Some people previously acquainted with the Principia might remember the whole thing just by reading a summary. Other people will read the summary and understand much less of the Principia. 
This doesn't mean that the information conveyed by   the summary is  meta. It is just not the same set of information.
% than what is conveyed in the three volumes of the Principia.  
Ordinarily, different pieces of information, and different expressions of them, inform in different ways. 
It is not clear what way of informing called "meta-informing" has been identified as so special that it needs to be emphasised above  other ways of informing. 
%through the term "meta". 
It might be tempting to use the social status of Newton as a shortcut to evaluating the Principia's worth. But it is  absurd to equate an evaluation of the Principia without understanding the Principia to an evaluation of the Principia \textit{based} on an understanding of it. There is no information that can exempt from from putting in the time to understand a different information. 
There is no information \textit{about} the Pythagorian theorem     that can act as  an economical placeholder for appreciating the theorem. There     is no way of condensing the proof into something short and easy to understand without any mathematical reasoning. There is no shortcut to being informed. Outside of the realm of data where it is synonymous with "meta-data", the term "meta-information" has no purpose.

% ============================================================== %
% ========               FINDABILITY                    ======== %
% ============================================================== %
\subsection{Findability is overrated} 
\label{findability}

The FAIR data principles advocate for "findability" 
%and "accessible under well defined conditions" 
\cite{dunning1970fair,wilkinsonFAIR2016}.
% gofair https://www.go-fair.org/fair-principles/
In the context of data that risks remaining trapped underused in a 
% an obscure 
scientific silo, this is a priority. And meticulous meta-data provision is key. The situation is different in the context of unstructured text that presents the additional risk of wasting the time of busy humans. As mentioned before, ingesting information takes time. Arguably, humans might be better off not ingesting more information than they can use and process. 
Determining if a piece of information $I$ is relevant to a given human individual $H$ is very challenging. It requires  expertise in both  $I$  and  $H$ to answer questions like the following:\medskip

\begin{itemize}
    \item Is $I$ intelligible to $H$? 
    \item Is $I$ new to $H$? 
    \item What information is $H$  already familiar with? 
    \item Can I be better than $H$ at knowing what $H$'s expertise is?  
    \item How much (effort, time, concentration, \textit{etc}) will it cost $H$ to ingest $I$?
    \item How much will it cost $H$ to decide $I$ is  new and useful to them -- or to reject $I$? 
    \item What will $H$ not do as a result of doing that?  
    \item Would $H$'s time be better employed elsewhere?
    % Is the information that $H$ already has insufficient? 
    % Can $H$ make good use of $I$? 
    \item Will $H$ be able to do something more or something more worthwhile 
    %or better 
    than what they  can already do?
    %without ingesting $I$? 
    \item What can $H$  already do?
    \item Who looses/gains more by having $H$ ingest and make use of $I$ 
    %take the time to understand whether they can make use of $I$  
    rather than  perform the tasks $H$ can already do (fairly well, fairly efficiently) without $I$? 
    \item How much does $H$ gain in confidence in their existing expertise? 
    \item What if $H$ tries to injest $I$ and ends up thinking they understand $I$ but actually don't? 
    \item What if $H$ misuses $I$ or transfers a degraded version of it to a friend? 
    \item Can I do better than $H$ at predicting $H$'s understanding of $I$? 
    \item What if $I$ is not new to $H$ but $H$ takes the mere fact that  $I$  is repeated to them  as confirmation that $H$'s belief in $I$ is righteous?
    \item[] \hspace{0.6cm}\vdots
\end{itemize}

There are many reasons why a piece of information should \textit{not} be findable to you. Many of these reasons are not solely contained in the information's lexical and semantic properties.
% whether you are an idiot or whether you are an expert. 
% --- $I$ is unintelligible to you. The risk is that  you waste time trying to decrypt $I$  to no avail. Or worse, you end up wrongly thinking you understand $I$  and transfer a degraded version of it. 
% --- The ideas conveyed by $I$   are not new to you. You waste time discovering that. Worse you take the mere fact that  $I$  has been repeated to you as confirmation of what you already believe in. 
% --- You were busy doing some valuable work that you would have managed to complete with or without $I$ . Finding $I$  depletes your focus without necessarily making you gain any time or confidence in your existing expertise.  
Sometimes, simply, the information and expertise you already have is worth  relying on. 
% instead of asking for more "parachuted"  information. 
%looking to new relevant information they  can find. 
The ability to ignore new information is as essential as the ability to discover information.~
\medskip

% Arguably  making information more findable isn't necessarily a good idea if it is not preceded by significant effort to support the application of expertise and knowledge that people already have.\medskip

Actual search engines and recommendation tools can't answer the majority of the questions listed above. 
% are not able to evaluate how worthwhile your current work is, how precious your time is, and how long it would take you to ingest  one of their suggestions.
They often use collected personal user data in order to personalise search results.
This approach produces the wrong kind of non-findability. 
It is suitable to ensure customer satisfaction for customers  who expect and demand  informational familiarity and control \cite{googleguidelines}. But in general, the filtering mechanisms it relies on, based on users' past, are  inappropriate for fair information retrieval, e.g. in exploratory scientific research contexts \cite{garrett2009echo,papacharissi2002virtual,rowland2011filter,twainecho}.  
Dmitri Brereton calls for alternative, \textit{decentralised} (as in local) kinds of search customisation \cite{nextGoogle}.\medskip

Findability might be the wrong angle to address the digital information overload problem. Non-findability might be a more appropriate angle to address fundamental contextualisation questions. Offering performant alternatives to filter bubbles requires devising filtering mechanisms capable of ensuring  that information is \textit{not} findable to a user for the sole reason that in itself -- lexically, semantically, statistically -- it seems relevant to the user's query. And indeed, the level of contextualisation suggested by the questions listed above and the level of expertise required to answer them makes a decentralised (local) solution necessary.

% ============================================================== %
% ========                                              ======== %
% ============================================================== %
\subsection{Marketing can be domesticated} % educated civilised broken % Keywords are dumb
% Keywords and other implantation techniques Spreading the rumour
\label{implantation}

I have emphasised the notion of implantation in our proposal and mentioned how it relates to visibility of information on our proposed information landscape. Implanting a piece of information in an information space means  anchoring it to that space by documenting links between the piece of information and known landmarks of the space. The links to the landmarks will allow traffic to be relayed over to the piece of information. Good implantation creates meaningful links and multiplies the  pathways to the piece of information in the information space. It makes it more likely for different individuals to stumble across the piece of information in different circumstances, when they are not necessarily looking for it. 
\medskip

Implantation happens on traditional information spaces. For academics, implantation involves citations.  For collaborators it may involve positioning a file in a shared file hierarchy. For the SEO industry, it involves keywords and hyperlinks.  For open science advocates it involves metadata. 
\medskip

An obvious question is: what happens when implantation takes over other information producing activities because of its gratifying relation to visibility? -- when information producers become specialists in implantation and  in producing easily implantable content? More interestingly, can I devise an implantation method that  still provides visibility but whose excessive practice doesn't harm the record? i.e., can marketing of information be made indissociable from desirable information practices? Our proposal assumes that an epistemic network  provides a positive answer to that. 
% Good implantation is an alternative to good indexation. 

% ============================================================== %
% ========                                              ======== %
% ============================================================== %
\subsection{Different wording makes different Information}%{Information is more like hunger than socks} 
\label{message}

In the history of the composing of Wikipedia's page on "information" \cite{wikiinfo}, a recurrent point of discussion and dissensus has been  \textit{whether information is the message or the meaning of the message}. In this proposal, I choose to confuse the message with its meaning. I assume that except when the differences are  insignificant  (e.g. typos), different messages, even when they are different wordings of the same idea,  are different pieces of information. A piece of information is thus inseparable from its expression. The reason for this choice is the following. Say $\mathcal{I}$ is an idea, and $W_1$ and $W_2$ are two  different wordings of $\mathcal{I}$. The difference between $W_1$ and $W_2$ is said to be insignificant if replacing all occurrences of $W_1$ by $W_2$ (or vice versa) will lead to no human understanding $\mathcal{I}$ any less or any differently. On the contrary, if  $W_1$ and $W_2$ are significantly different, then a human who understands $\mathcal{I}$  through $W_1$ might not understand the same thing through $W_2$ or might not understand anything at all. $W_1$ and $W_2$ are not equivalent in that they don't produce the same effect on humans' understanding. Incidentally, the (great) overlap in meaning there might be between two different expressions $W_1$ and $W_2$ is referred to in this proposal as the "good sort of redundancy". It is a relation between two expressions  $W_1$ and $W_2$  that allows to bridge knowledge built on the basis of an understanding of $W_1$ and knowledge built on the basis of an understanding of $W_2$.

% ============================================================== %
% ========                                              ======== %
% ============================================================== %
\subsection{Information serves understanding}
\label{understanding}

I am especially interested in promoting human understanding. I consider information as a means for both expressing 
% an amount of 
the understanding of a human and triggering the experience of understanding in another human. 
I recognise that understanding like hunger is a personal and dynamic experience. It can manifest itself diversely, including in textual form. Our proposal offers to equip ourselves with systems  to compare, qualify and use understanding in its various textual expressions. \medskip

% ============================================================== %
% ========                                              ======== %
% ============================================================== %
\subsection{Information can express something  without representing anything}%{Information is more like hunger than socks} 
\label{expression}

Information is often seen as being \textit{about} something: information \textit{about} a person, information \textit{about} an object, information \textit{about} a natural phenomenon. Information \textit{represents} something of the "real world" (as opposed to the world of information).
%the person, the object, the natural phenomenon.
 %  Information does not just \textit{represent} objects of the "real world". 
%  Information is sometimes seen as a  symbolic representation of something (typically a representation of something "real" that is not information):
% that is not information: 
% information is \textit{about} a person, an object, a natural phenomenon. 
It is a sort of abstraction of what it informs on, capturing some of its features in a different form. Arguably there are pieces of  information that don't  represent nor abstract anything \textit{per se}, or at least they don't represent any object of the "real world" different from themselves. They are \textit{expressions} (manifestations, formulations) rather than representations.
 The Pythagorian theorem for instance is information. 
 % It exists without reference to anything outside of the mathematical world. 
 %It doesn't matter whether the mathematical world is the real world or not. The Pythagorian theorem 
 It \textit{expresses} an understanding of how certain triangles work (it doesn't represent the triangles \textit{per se}).\medskip

In this proposal, I am primarily interested in supporting the process of information, the (re)formulating of understanding. Supporting  the  end product of information, the symbolic form that results from the process of information is a secondary aim. In other terms I am more interested in providing an upgrade on the concept of whiteboard than an upgrade on the concept of  book. 
\medskip

I choose to equate \textit{information} with \textit{expression} and  confuse it with the process itself -- % equating it with \textit{expression},
% choose to equate \textit{information} with expression:  
that is, I choose to confuse information/expression with the process of giving of a form to something  (e.g. a thought, an idea, a feeling) typically in the aim of sharing something that doesn't yet have a form with someone. With this definition, information isn't necessarily  giving a \textit{different} form to something that already has one.  This implies that models (representations) are not 
ubiquitous in the world of information. 
\medskip

Note that models are expected to have certain properties that expressions aren't necessarily expected to have  -- eg 
   consistency with recognised features of the represented object\footnote{"A map is not the territory it represents, but, if correct, it has a similar structure to the territory [\ldots]." \cite{themapisnot}},
   (ontological) non-contradiction, circumscription (not everything is represented),
   and certain forms of completeness (eg non ambiguity) which imply a  degree of immutability of the representation. Modifications of a model become necessary when I realise the model is imperfectly representing what it is meant to represent. 
   Its underlying ontological commitments are possibly wrong. %perhaps not the right ones.  
   In contrast, it is sometimes enough for an expression  to be pleasing, in a certain context, to someone. % or satisfying.
   The reason why I change our expressions, isn't necessarily due to their imperfections. A perfect expression usually  becomes obsolete when the context changes. 
   \medskip
 
   Logicians distinguish between models and theories. Theories capture the pith of our understanding of what an object is and how it works (eg arithmetic). Contrary to the models  that instantiate them (there may be several models for one theory, cf natural numbers), logical theories don't need to be non-contradictory nor complete.
    They don't even need to be instantiable, i.e., there doesn't need to be a sensible object  that works in the way the theory describes. Despite all this, theories, even those lacking the expected properties of good representations, are useful objects \cite{girard1950pourquoi}. In particular, they are essential for automatic inference and they play a central role in the Semantic Web   \cite{Baader2005}.
    \medskip

    Our proposal is not to document information as part of a great unified model. The MMM is not a model but a medium (like A4 paper) for documenting information. It is  compatible with the documentation of all models (which can contradict each other), the documentation of theories, of isolated thoughts, questions \textit{etc}.

% ============================================================== %
% ========                                              ======== %
% ============================================================== %
    \subsection{Opinion looks like information} 
    \label{opinion}

    % Duck test 
    "\textit{If it looks like a duck, swims like a duck, \ldots{}
    }"\footnote{See Wikipedia's entry for the \href{https://en.wikipedia.org/wiki/Duck_test}{duck test} \cite{wikiduck}.}
    % and quacks like a duck, then it probably is a duck.
    \medskip
    
    At times of high media concentration and instrumentalisation of information, it is important to acknowledge the ostentatious ressemblance between opinion and information, especially when it is expressed by  articulate individuals. So-called "meta-information" (eg provenance of the information) is clearly not foolproof. And it certainly is not proof. Debate over what is opinion and what is information doesn't generally reduce the air time given to opinion. 
    Using meta-information as the basis of the debate shifts the exercise of distinguishing opinion from information onto an exercise of comparing a common feature: both opinion and information are assumed to have meta-information and the  meta-information of opinion is assumed to be  comparable with the meta-information of information. This 
    %  by looking into a common feature that arguably 
     makes the difference between opinion and information even more tenuous. 
    There is an alternative way to address our fear of confusing opinion with information. It is the core business of scientist researchers. It consists in
        % distinguish impression from information by 
    methodically and logically challenging the opinion/information and determining how well and under what conditions does it hold. \medskip

    Our proposal offers to manage  opinion scientifically, like any other information. Arguably characterising content in rich non-binary terms (cf \S\ref{nonbinary}) is more important than  adjudicating on what is opinion and what is not.

% ============================================================== %
% ========                                              ======== %
% ============================================================== %
\subsection{The map \textit{is} a territory} 
     \label{themap}

     Usually, a more formal object  that I have some control over is used to \textit{model} a less formal object that eludes us more. 
     The more formal object (a.k.a. the model) is an \textit{interesting} model if it is not just a placeholder for the less formal object it represents (a.k.a. the modellee): the model can be justified, discussed,  and challenged. It can be studied and understood as a stand alone object.
     \medskip

     Naively, one might assume a model relates \textit{analogically} to the modellee. But many things look alike without one of them having the capacity to inform on the other. Usually the model and the modellee are objects of  different nature (eg  Boolean Automata Networks versus genetic regulation) living in  different realms (eg the realm of maths and the biological reality). There is no reason to expect  the model/modellee relationship  to be a trivial one. 
     In our experience, a lot happens in the "no man's land" in between the different realms in which model and modellee live. Assumptions are tacitly made as to  how the model's features relate to  the modellee's features and how the informing performance occurs.
    Poor understanding of the model/modellee relationship typically leads to confusing information that is derived from the model with information  that  springs out of  the no man's land, initially placed there by the modeller himself. In other terms,   the model is made to say things it doesn't have the capacity to say. 
     Gaining an understanding of the formal model  independently of its relationship  with the modellee is a prerequisite to circumscribing, exploring and sorting out information in the no man's land.\medskip

     The existence of analogies between properties of the model and features  of the modellee aren't enough to guarantee one object has the capacity to reliably inform on the other. It also is no reason to expect that the model is the \textit{only} useful representation of the modellee. 
     Two models (eg  Newtonian mechanics and Lagrangian mechanics,  {wavelike and particular descriptions of electrons}) can simultaneously inform on the same modellee. One territory can be covered by multiple maps. 
     Different maps/models will naturally have different relationships  to the object they represent. 
     They formalise different points of view on the territory/modellee.
     Korzybski's famous formulation "The map is not the territory" amounts to "the viewpoint is not the object viewed". Arguably, the latter formulation makes it more obvious that there can be multiple models (viewpoints) representing the same object and  that both (1) model/modellee relationships and (2) relationships between diverging models, are interesting, contextual,  non-trivial relationships that are traditionally overlooked and often still informal, but worth studying. \medskip

    Logicians and model theorists have taught us  to be wary of the relationships models entertain with other objects \cite{Gödel1931,Klenk1976}. 
    These relationships can hide unexpected information that depart  from spontaneous  interpretations and naive analogies. Gaining an independent understanding of formal objects is a prerequisite to making full and safe use of them to inform on less formal objects.   
    Method in studying formal objects and their relationships to each other is essential to ensuring reliable information. One part of the scientific method most well known,  relates real world observations with formalisations. It concerns the empirical sciences which  derive information from interacting with the real world.    The formal sciences derive information from  other information. Their complementary part of the scientific method applies to formal objects and their relationships. It is more recent and discrete. Arguably, the so-called "information age" creates an urgent need to promote method and rigour in manipulating information and maintaining a     neat, uncluttered, manageable version of the record.
    \medskip

    Our chosen definition of information (cf \S\ref{expression}) avoids regarding information (including descriptions of models)  as necessarily subordinate to the physical world, like the map is subordinate to the territory. The record of information is  seen as a territory in its own right. The objects it is comprised of, pieces of information including models, exist in and of themselves and are worth studying without reference to another territory such as the physical world.  \medskip

% ============================================================== %
% ========                                              ======== %
% ============================================================== %

    \subsection{Documents aren't the right level of granularity } 
    \label{documents}
    The field of information retrieval has devised generic "relevance" metrics to characterise documents retrieved as a result of a user's query \cite{COOPER197119}. To our knowledge, measures of intelligibility and novelty are less common \cite{welearn}.  They are {rudimentary  and inappropriately \textit{non-subjective}}.
    % unsuitably  generic
    Arguably, the field is challenged by its  operating on \textit{documents}.  
    It may be that 99\% of a specific document is useless to a user because it is unintelligible or not new, and 1\% of the document is precious to that user. The principle of preservation of human mental resources, would advocate for sparing that particular user the effort of considering the 99\%. Documents are not the right granularity at which to perform information retrieval, if I want to mitigate  information overload and save the focus of humans at work.

% ============================================================== %
% ========                                              ======== %
% ============================================================== %
\subsection{Humans are also structured} % Machines are structured but so are humans 
\label{machines}

"Unstructured information" is unstructured from the point of view of machines. It often isn't from the  point of view of humans. Unstructured information  is often mostly comprised  of  text. The text expresses pieces of information and how they are tied together according to a certain human logic. 
\medskip

Machines can help with knowledge management because they can automatically process information. However, for them to understand the information and behave as I want them to, the information must be stripped down to fit into some predefined formalism that necessarily has limited expressivity compared to natural language.  And the way to understand the information expressed in this formalism -- a.k.a. the semantics -- must be explicitly predefined. In the  information that I use to make machines do things for us, there  are properties of information that I must get rid of like ambiguity, high contextuality, heterogeneity, uncertainty, vagueness, contradiction.
\medskip

Humans have different needs. They also have different capacities. %are good at different things. 
Notably, humans already have the ability to understand the information that they produce (although not all equally so because of differences in expertise), without it necessarily being stripped down. \textbf{Humans already produce information for each other to understand}.  %And competent information producers already understand the information the produce. \medskip

% fig missing

To improve our management of unstructured information, one approach is to improve machines' ability to understand   unstructured information \textit{after-the-fact}, i.e., after the information has been produced by humans, and when its human producers are no longer there thinking about what they produced, and all there is left of their intellectual activity to exploit  is a collection of static {documents} they wrote.  An alternative approach is to improve and complete the means that humans have of documenting what they already understand, striking the iron while it's hot (while the understanding is still in the process of being experienced by some humans), making the most of the contextual insights that humans are already successfully developing and adapting
% and continually evolving to adapt 
to a continually changing world. This requires a new take on what a document is. Our proposal offers one. \medskip

Our proposal aims to do for humans what the Semantic Web aims to do for machines: support the connection of pieces of information together to reveal structure   that humans can   process. This  could be an easier endeavour because humans don't need the information to be cleaned up for them in advance. % They can do all the work.  
Incidentally, more thorough documentation of human understanding will certainly facilitate the  machine-centric approach.
\medskip

A universal model or a universal language to express all the information that humans want to express is not realistic. However a (quasi-) universal \textit{medium} is. Paper is a  (quasi-) universal medium for documenting unstructured human content. A4 paper also is. Its predefined dimensions add some constraints to the documentation. The academic journal article format (AJAF) is another example. It further  constrains the documentation of human content by imposing that a title, an abstract \textit{etc} be provided. In standardising the structure of articles, these constraints support portability of the information documented in articles. A mathematician and a biologist will not be surprised to find an introduction section in each others' articles. They will have some idea of what kind of information they will find documented there. \medskip

\label{balance}
Arguably, when people think in the privacy of their own minds, they rarely say to themselves "\textit{I've just had an introduction kind of thought.}" or "\textit{\ldots{} a section 1.2 kind of thought.}". Labeling pieces of our own thoughts with indications of the roles these pieces of thoughts play in a larger system of thoughts is part of methodical thinking. But historical methods of documentation like scholarly documentation in articles 
requires a structuring of our thoughts 
that isn't reflective of the dynamics of live human thought. 
They constrain the structure of documentation with labels such as "title", "abstract", "introduction" \textit{etc} that are reflective of the linear style of historical documentation technology like the printing press. 
\medskip

People who think also obviously rarely bother labelling  their thoughts on the fly in their minds with  grammatical role labels (subject, verb, object, determiner\ldots{}) and semantic role labels (agent, experiencer, patient, stimulus, theme\ldots{}).
Still, there is structure to most human thought.
Without having decided beforehand on a specific method of thought, people usually effortlessly know when they have just had a "\textit{question kind of thought}" or a "\textit{nuancing kind of thought}". Many thoughts aren't followed by random disconnected other thoughts. There may be some logic to the sequence of thoughts. Our proposal attempts to capture and support a \textit{common denominator} structure of structured human thought. 
I aim at 
striking the right balance between asking  too much (or not the right kind of) labeling of people who thereby won't do it, and asking too little and thereby missing out on structuring information that people could have easily provided.  
The MMM format is designed to % strike that balance between 
 (i) provide {meaningful} structuring constraints as an alternative to AJAF's constraints   enabling a practical  amount of standardising of documentation, and    (ii) maintain the (quasi-) universality of the MMM documentation medium.

% ============================================================== %
% ========                                              ======== %
% ============================================================== %
\subsection{Humans can agree (but don't have to)}

Arguably the single foundation of science is that there  are things that humans can agree on, even if only momentarily.
The things humans agree on (eg models representing experiences of the world) are not foundational. The fact that humans can agree is. Nothing much would be constructible by humans if they didn't have it in them to \textit{momentarily} share a point of view on the world
and look in the same direction. %  Science wouldn't be possible without this. 
\medskip

    Two humans can agree on what exists.  For instance they both assume that a certain object  '\texttt{Bob}' exists. They can agree to call it '\texttt{Bob}' and they can agree on what  kind of object  '\texttt{Bob}' refers to, eg a '\texttt{person}' which is a type of '\texttt{mammal}'. 
    Such ontological commitments are the stuff representations of the world are made of. Representations of the world are useful because they allow for a shared vision, especially when they are well formalised like scientific  models and formal ontologies. 
    There are indisputable benefits to sharing a vision (sharing a vision of a house facilitates the construction of the house by multiple humans). 
    But systematic agreement outside of a specific project isn't necessarily beneficial to information. \medskip

    This further supports our choice to stay clear from proposing a unified model, and instead, less ambitiously propose a generic \textit{medium}, namely the MMM.

% ============================================================== %
% ========                                              ======== %
% ============================================================== %
\subsection{Contradiction is good for information}

Nothing much would happen if I all agreed with each other and if the world agreed with our representations of it. % I weren't driven to disagree. 
As the world evolves, information has to evolve too. Contradiction is a source of evolution. 
\medskip

Our solution supports the documentation and discussion of contradictions.

% ============================================================== %
% ========                                              ======== %
% ============================================================== %
\subsection{Ontological commitments don't scale well}\label{ontological-commitment}

A single point of view or model  cannot be all encompassing. 
Ontological  commitments are circumstantial and local. 
Not all new information is derived in line with old ontological commitments. 
New information is regularly produced by challenging and abandoning old ontological commitments.
This is why unlike the Semantic Web, our proposal doesn't rely on ontological commitments for interlinking informational resources. The MMM supports general epistemic interlinking of  information. The regularly destructive nature of informational and scientific progress  is also why I choose to allow ontological commitments to be explicitly documented, challenged and discussed like any other piece of information. Our proposal keeps ontological commitments local and circumstantial, so that they can evolve. \medskip

% finish paragraph

% ============================================================== %
% ========                                              ======== %
% ============================================================== %

    \subsection{There is no information without misinformation} 
    \label{errors}
    \label{quality} 
    It is impossible to produce new information without producing low quality content (cf the daily practice of scientific research). Many errors can decisively participate in the process of improving information. A contribution conveying a low quality piece of information, if dealt well (without denial), is  a step in the process of improving the record: it calls for further contributions specifying what is wrong with this piece of information and how to deal with it. It should not be removed from the record. Nor should its author necessarily  be scorned, unless they hide the error. A contribution of poor quality should be systematically exposed to improvements, and for a long enough time that concerned citizens learn from it. Visibility of an error and how it has been addressed should be entertained as long as citizens  risk  repeating the error. \medskip

Systematic low quality information is not a problem. Failure to systematically improve it is.

% ============================================================== %
% ========                                              ======== %
% ============================================================== %
    \subsection{Questions are information too}\label{questions} To be aware of what relevant questions can be asked on a given topic is to be reasonably well informed on that topic. 

\medskip

No good informing can take place without questions. 

% ============================================================== %
% ========                                              ======== %
% ============================================================== %
    \subsection{Naming things substantiates them} % substantiates them/ makes them exist
    \label{namingthings}
    Genes exist/matter because they have been given a name. There is no absolute necessity for the existence of genes imposed by the real physical world. Certainly something in that area of the physical world studied by genetics  exists. Judging by the success of genetics it deserves to be named so that I can talk about it among humans and work with it like geneticists wonderfully have. But \textit{how} I circumscribe the areas of the physical world that I want to give names to is up to us. The world doesn't care to draw a line between what constitutes a gene and what doesn't. \textit{We} do that.
    Similarly, the concept "one" exists/matters because I want it to so much that I have given it a name. The world itself doesn't care to divide itself into things that count as one and things that count as several. 
    Genomics might have gone just as far as it has  without the concept of gene, perhaps making do 
    with   the concepts of coding sequence and open reading frame. Besides,  the part of the physical world circumscribed and denoted by the  word gene hasn't even always been the same \cite{portinGenes}. 
    \medskip

    Words are a very practical step up from onomatopoeias. They conventionally delineate the observations that humans have made of the world and have cared about ever since.   Humans invented them    to speak among themselves about the surrounding physical world which matters immensely to them. The things  that humans say exist are the things that matter to humans. Often they also end up being things that matter  in the world in general because humans tend to focus, get excited about and act on the parts of the world they name (eg petrol). \medskip

   For now genes exist/matter, at least to many people. The term "gene",  its mattering, 
   the 'existence' of genes, is information -- a kind of information that can't easily be qualified of  true or false but nonetheless essential. I propose that I give ourselves the means to systematically document such ontological assumptions. Making assumptions explicit, however successful they are,  will make it easier to replace them when the day comes I have surpassed them with our science. % method

% ============================================================== %
% ========                                              ======== %
% ============================================================== %
\subsection{Information is different from communication}
% Not a communication tool
\label{SMcommunication}

I propose to consider information as something that can be communicated but doesn't have to be. This means that information work doesn't necessarily entail communication. 
\medskip

Information work refers to analytical reasoning about the internal logic and structural build of a piece of  information.  It is concerned with the semantics of information, how pieces of information fit together, how they are built from other pieces of information and woven together with logical links, how they are proven to be valid \textit{etc}. 
Communication work refers to the effort of wording a piece of information so that someone else than oneself understands it. Communication management is concerned with the conveying and delivery of information, regardless of how the information is built. \medskip

Our solution prioritises information work over communication work and management.

% ============================================================== %
% ========                                              ======== %
% ============================================================== %
\subsection{Fast communication is overrated}\label{slowcollaboration} %Slow-First Collaboration / Faster communication doesn't always work better

Not all information needs to be readily available in split seconds. For instance, the time scales at which pieces of science are produced and shared is not defined in seconds, nor rarely even in days or weeks. Communication of scientific results is often a matter of months and years. 
Real-time collaboration technology is not absolutely key to accelerating the pace of scientific production and communication. In our opinion, if anything is, it is rather socio-economic incentives like deincentivising strategic withholding of scientific results. 
Generally, information workers don't necessarily need fast access at any moment to what their collaborators are doing.
Sometimes they can even wait hours for a resource to become available because its owner is not actually online. 
Arguably, 
\textit{not} attempting to accelerate 
communication means where there is no need to,  favours the preservation of the % preserve 
solitary focus of  information producers which sometimes is more important.
\medskip

Information work may gain from real-time collaboration technology in a marginal situation: when collaborators are working on the same atomic idea together, trying to find the right words for it. They don't want to work separately to produce a wording each and choose the best one. They want to find the words together. They are working on the same wording and they want to be able to correct each others' words. Their edits to the common wording must then be synchronised in real-time. Our solution is not designed to natively support this situation which requires potential conflict resolution.\medskip

Natively, our solution supports  "epistemic separation of concerns". Collaborators are supposed to be working on different pieces of information. 
Coordination happens \textit{in between} the pieces of information. 
And when it turns out that epistemic concerns were not that separate after all, a \hyperref[merging]{merge} of collaborators' contributions can be done or \hyperref[glue]{epistemic glue} can be explicitly provided between them.  Epistemic separation of concerns is a natural and meaningful way of ensuring synchronous concurrent edits are non-blocking and still conflict free. % since they apply to different contributions. 
The point of defining an epistemic (rather than semantic) network is that concurrent  updates/contributions have  little semantic interdependence.  As long as they are well formed, individual  contributions don't risk  becoming collectively inconsistent. 
\medskip

Incidentally, epistemic separation of concerns is favourable to  \textbf{informational privacy}. Coordination between collaborators' contributions only needs to be local. Coordination occurs through the provision of  local epistemic glue (conveyed by MMM edges).  In itself it doesn't require any intermediary third-party to orchestrate the synchronisation (contrary to OT and like CRDTs \cite{crdtsfuture,kleppmanndistrib}).
A user's local territory, their world view, the information they produce and consume, can remain private \cite{nextGoogle}.

\label{articleLastPagepartB}

% ============================================================== %
% ========                 CONTENTS                     ======== %
% ============================================================== %
\clearpage
\thispagestyle{empty}\pagenumbering{gobble}
\addtocontents{toc}{\protect\thispagestyle{empty}}
\addcontentsline{toc}{section}{Contents}
\cfoot{}
\tableofcontents
\addtocontents{toc}{\protect\thispagestyle{empty}}
\thispagestyle{empty}
\pagenumbering{gobble}
\end{document}